\documentclass[aps,prc,superscriptaddress,showpacs,twocolumn,floatfix,10pt]{revtex4-2}

\usepackage{graphicx}
\usepackage{color}

\def\micro{\mu}
\def\gtorder{\mathrel{\raise.3ex\hbox{$>$}\mkern-14mu
 \lower0.6ex\hbox{$\sim$}}}
\def\ltorder{\mathrel{\raise.3ex\hbox{$<$}\mkern-14mu
 \lower0.6ex\hbox{$\sim$}}}
\def\mugegm{\mu_p G_{Ep} / G_{Mp}}
\def\gegm{G_{Ep}/G_{Mp}}
\def\ge{G_{Ep}}

\def\mugegmp{\mu_p G_{Ep} / G_{Mp}}

\def\gep{G_{Ep}}
\def\gmp{G_{Mp}}
\def\gd{G_D}

\def\etal{\textit{et al.}}

\begin{document}

\title{High precision measurements of the proton elastic electromagnetic form
factors and their ratio at $Q^2$ = 0.50, 2.64, 3.20, and 4.10~GeV$^2$}

\author{I.~A.~Qattan}  
\affiliation{Khalifa University of Science and Technology, Abu Dhabi,  P. O. Box 127788  UAE}
\affiliation{Northwestern University, Evanston, Illinois, 60208}
\author{J.~Arrington}  
\affiliation{Lawrence Berkeley National Laboratory, Berkeley, California 94720, USA}
\affiliation{Argonne National Laboratory, Argonne, Illinois, 60439}
\author{K.~Aniol}   
\affiliation{California State University, Los Angeles, Los Angeles, California, 90032}
\author{O.~K.~Baker}    
\affiliation{Hampton University, Hampton, Virginia, 23668}
\author{R.~Beams}    
\affiliation{Argonne National Laboratory, Argonne, Illinois, 60439}
\author{E.~J.~Brash}   
\affiliation{University of Regina, Regina, Saskatchewan, Canada, S4S 0A2}
\author{A.~Camsonne}   
\affiliation{Universit\'{e} Blaise Pascal Clermont-Ferrand et CNRS/IN2P3 LPC 63, 177 Aubi\`{e}re Cedex, France}
\author{J.-P.~Chen}   
\affiliation{Jefferson Laboratory, Newport News, Virginia, 23606}
\author{M.~E.~Christy}   
\affiliation{Hampton University, Hampton, Virginia, 23668}
\author{D.~Dutta}   
\affiliation{Massachusetts Institute of Technology, Cambridge, Massachusetts, 02139}
\author{R.~Ent}  
\affiliation{Jefferson Laboratory, Newport News, Virginia, 23606}
\author{D.~Gaskell}  
\affiliation{University of Colorado, Boulder, Colorado, 80309}
\author{O.~Gayou}     
\affiliation{College of William and Mary, Williamsburg, Virginia, 23187}
\author{R.~Gilman}   
\affiliation{Rutgers, The State University of New Jersey, Piscataway, New Jersey, 08855}
\affiliation{Jefferson Laboratory, Newport News, Virginia, 23606}
\author{J.-O.~Hansen}    
\affiliation{Jefferson Laboratory, Newport News, Virginia, 23606}
\author{D.~W.~Higinbotham}   
\affiliation{Jefferson Laboratory, Newport News, Virginia, 23606}
\author{R.~J.~Holt}  
\thanks{Current affiliation: California Institute of Technology, Pasadena, California 91125}
\affiliation{Argonne National Laboratory, Argonne, Illinois, 60439}
\author{G.~M.~Huber}
\affiliation{University of Regina, Regina, Saskatchewan, Canada, S4S 0A2}
\author{H.~Ibrahim}      
\affiliation{Old Dominion University, Norfolk, Virginia, 23529}
\author{L.~Jisonna}    
\affiliation{Northwestern University, Evanston, Illinois, 60208}
\author{M.~K.~Jones}  
\affiliation{Jefferson Laboratory, Newport News, Virginia, 23606}
\author{C.~E.~Keppel}   
\affiliation{Hampton University, Hampton, Virginia, 23668}
\author{E.~Kinney}    
\affiliation{University of Colorado, Boulder, Colorado, 80309}
\author{G.~J.~Kumbartzki}  
\affiliation{Rutgers, The State University of New Jersey, Piscataway, New Jersey, 08855}
\author{A.~Lung}   
\affiliation{Jefferson Laboratory, Newport News, Virginia, 23606}
\author{K.~McCormick}   
\affiliation{Rutgers, The State University of New Jersey, Piscataway, New Jersey, 08855}
\author{D.~Meekins}    
\affiliation{Jefferson Laboratory, Newport News, Virginia, 23606}
\author{R.~Michaels}  
\affiliation{Jefferson Laboratory, Newport News, Virginia, 23606}
\author{P.~Monaghan}   
\affiliation{Massachusetts Institute of Technology, Cambridge, Massachusetts, 02139}
\author{L.~Pentchev}   
\affiliation{College of William and Mary, Williamsburg, Virginia, 23187}
\author{R.~Ransome}   
\affiliation{Rutgers, The State University of New Jersey, Piscataway, New Jersey, 08855}
\author{J.~Reinhold}  
\affiliation{Florida International University, Miami, Florida, 33199}
\author{B.~Reitz}   
\affiliation{Jefferson Laboratory, Newport News, Virginia, 23606}
\author{A.~Sarty}   
\affiliation{Saint Mary's University, Halifax, Nova Scotia, Canada B3H 3C3}
\author{E.~C.~Schulte}  
\affiliation{Argonne National Laboratory, Argonne, Illinois, 60439}
\author{K.~Slifer}   
\affiliation{Temple University, Philadelphia, Pennsylvania, 19122}
\author{R.~E.~Segel}   
\affiliation{Northwestern University, Evanston, Illinois, 60208}
\author{V.~Sulkosky}   
\affiliation{College of William and Mary, Williamsburg, Virginia, 23187}
\author{M.~Yurov}  
\affiliation{Mississippi State University, Starkville, Mississippi, 39762}
\affiliation{Oak Ridge National Laboratory, Oak Ridge, Tennessee, 37830}
\author{X.~Zheng}
\thanks{Current affiliation: University of Virginia, Charlottesville, Virginia 22904}
\affiliation{Argonne National Laboratory, Argonne, Illinois, 60439}

\date{\today} 


\begin{abstract}
\begin{description}
\item[Background] The advent of high-intensity, high-polarization electron
beams led to significantly improved measurements of the ratio of the proton's
charge to electric form factors, $\gegm$. However, high-$Q^2$ measurements
of this ratio yielded significant disagreement with extractions based on
unpolarized scattering measurements, raising questions about the reliability of the
measurements and consistency of the techniques.

\item[Purpose] Jefferson Lab experiment E01-001 was designed to provide a
high-precision extraction of $\gegm$ from unpolarized cross-section
measurements using a modified version of the Rosenbluth separation technique
to allow for a more precise comparison with polarization data.

\item[Method] Rosenbluth separations involve precise measurements of the
angular dependence of the elastic e-p cross section at fixed momentum
transfer, $Q^2$. Conventional Rosenbluth separations detect the scattered electron, requiring the comparisons of measurements with very different detected electron energy and rate for electrons at different angles. Our `Super-Rosenbluth' measurement detected the struck proton, rather than the scattered electron, to extract the elastic e-p cross section. This yielded a fixed momentum for the detected particle and dramatically reduced variation of the cross section with angle, significantly reducing rate- and momentum-dependent corrections and uncertainties.

\item[Results] We measure the cross section vs angle with high relative precision,
allowing for extremely high-precision extractions of $\gegm$ at $Q^2$=
2.64, 3.20, and 4.10~GeV$^2$.  Our results are consistent with traditional
Rosenbluth extractions, but with much smaller corrections and systematic
uncertainties, comparable to the uncertainties from polarization measurements.

\item[Conclusions] Our data confirm the discrepancy between Rosenbluth and
polarization extractions of the proton form factor ratio using an improved
Rosenbluth extraction that yields smaller and less-correlated
uncertainties than typical of previous Rosenbluth
extractions.  We compare our results to calculations of two-photon exchange
effects and find that the observed discrepancy can be relatively well explained
by such effects.

\end{description}

\end{abstract}

\pacs{25.30.Bf, 13.40.Gp, 14.20.Dh}

\maketitle


\section{Introduction}

Electron scattering is a powerful tool for studying the structure of the proton. The electron is a point-like particle with no internal structure, making it a clean probe of the target structure. The incident electron scatters from the target proton by exchanging a virtual photon, and the electron-photon scattering vertex is well understood within the theory of QED.

The electron scattering cross section for elastic scattering from a point-like spin-1/2 particle is well known. The finite size of the proton, associated with the spatial charge and magnetization distributions, modifies the point-particle scattering cross section measured in elastic electron-proton scattering experiments. This allows for the extraction of the nucleon elastic charge and magnetic form factors~\cite{sachs62, kelly02, miller03, venkat11}. At large momentum transfer, the unpolarized cross section has limited sensitivity to the charge form factor, and advances in measurements utilizing polarization degrees of freedom allowed for improved extractions of $\gegm$. However, these polarization measurements were at odds with previous extractions from unpolarized scattering, making precise comparisons of the two techniques important for a detailed examination of this discrepancy.


\subsection{Elastic e-p Scattering}

In the Born (single-photon exchange) approximation, the unpolarized cross section for electron-proton elastic scattering as a function of four-momentum transfer squared ($-Q^2$) and scattering angle ($\theta_e$) is
\begin{equation} \label{eq:diffrentional2}
\frac{d\sigma}{d\Omega} = \sigma_{ns}
\Bigg[ \Big( F_1^2 + \frac{\kappa_p^2Q^2}{4M_p^2}F_2^2 \Big) +
\frac{Q^2}{2M_p^2}(F_1+\kappa_pF_2)^2\tan^2({\frac{\theta_e}{2}}) \Bigg],
\end{equation}
where $\kappa_p$ ($\kappa_n$) is the proton (neutron) anomalous magnetic moment, $\sigma_{ns}$ is the non-structure cross section and $F_1$ and $F_2$ are functions of $Q^2$ and known as Dirac and Pauli form factors, respectively. The Dirac form factor, $F_1(Q^2)$, is used to describe the helicity-conserving scattering amplitude while the Pauli form factor, $F_2(Q^2)$, describes the helicity-flip amplitude.  In the limit $Q^2 \to 0$, the structure functions yield $F_1(0) = F_2(0) = 1$. In this limit, the virtual photon becomes insensitive to the internal structure of the proton which is viewed as a point-like particle. The non-structure cross section is given by
\begin{equation}  \label{eq:nonstructure}
\sigma_{ns} = \frac{\alpha^2 \cos^2({\frac{\theta_e}{2}})}
{4E^2\sin^4({\frac{\theta_e}{2}})} \frac{E'}{E} = \Big(\frac{d\sigma}{d\Omega}\Big)_{Mott}
\Big(\frac{E'}{E}\Big),
\end{equation}
where $E$ is the energy of the incident electron and $E'$ is the energy of the scattered electron.

A linear combination of $F_1(Q^2)$ and $F_2(Q^2)$ can be used to define the Sachs form factors~\cite{sachs62}, $\gep$ and $\gmp$, the electric and magnetic form factors:
\begin{equation}
\gep = F_1 - \kappa_p \tau F_2,~~\gmp = F_1 + \kappa_p F_2,~~ \label{eq:sachs}
\end{equation}
where $\tau = Q^2/4M_p^2$. In the limit $Q^2 \to 0$, where the virtual photon becomes insensitive to the internal structure of the proton, Eq.
(\ref{eq:sachs}) reduces to the normalization conditions for the electric and magnetic form factors, respectively,
\begin{equation}
\gep(0)= 1, \gmp(0) = \mu_p,~~ \\
\end{equation}
where $\mu_p \approx 2.7928$ is the proton magnetic moment in units of the nuclear magneton~\cite{ParticleDataGroup:2024cfk}, $\mu_N = e\hbar/2M_pc$.

Finally, we can define the reduced cross section, $\sigma_R$:
\begin{equation} \label{eq:reduced}
\sigma_R = \frac{d\sigma}{d\Omega} \frac{(1+\tau)\varepsilon}{\sigma_{ns}}
= \tau \gmp^2(Q^2) + \varepsilon \gep^2(Q^2)~,
\end{equation}
where $\varepsilon$ is the virtual photon polarization parameter, $\varepsilon^{-1} = (1 + 2 (1+\tau) \tan^2({\frac{\theta_e}{2}}))$. By measuring the reduced cross section at several $\varepsilon$ values and fixed $Q^2$, a linear fit of $\sigma_R$ to $\varepsilon$ gives $\tau \gmp^2(Q^2)$ as the intercept and $\gep^2(Q^2)$ as the slope. The magnetic contribution to the cross section is enhanced at large $Q^2$ by the factor $\tau$ in Eq.~(\ref{eq:reduced}), and dominates the cross section at moderate-to-large $Q^2$ values, making it difficult to make precise extractions of $\gep$ above $Q^2 = 2-3$~GeV$^2$.

Rosenbluth extractions yield values of $\gep$ and $\gmp$ which approximately follow the dipole form:
\begin{equation}
\gep \approx \gmp/\mu_p \approx \frac{1}{(1+Q^2/M_{dip}^2)^2}~,
\end{equation}
with $M_{dip}^2=0.71$~GeV$^2$. For $\gmp$, the deviations from the dipole form are small, $\ltorder$5\%, up to $Q^2$=10~GeV$^2$, with deviations up to 30\% at $Q^2 \approx 30$~GeV$^2$. Conventional Rosenbluth extractions of $\gep$ are limited to $Q^2$ values below 10-15~GeV$^2$, with uncertainties of 10\% or more above $Q^2=2$~GeV$^2$~\cite{arrington07c, arrington11b, Punjabi:2015bba, Ye:2017gyb, Christy:2021snt}. Overall, these data are consistent with form factor scaling, $\mugegm \approx 1$, indicating that $\gep$ and $\gmp$ have nearly identical $Q^2$ dependence.


\subsection{Recoil Polarization Technique} \label{sec:recoil_polarization}

In recoil polarization experiments~\cite{dombey69, akhiezer74, arnold81}, a longitudinally polarized beam of electrons is scattered from unpolarized protons, resulting in a transfer of the polarization from the electrons to the recoil protons. In the one-photon exchange approximation, the polarization component normal to the scattering plane, $P_N$, is zero, and there are two non-zero polarization components: the longitudinal polarization, $P_l$, which is along the direction of the proton momentum, and the transverse polarization, $P_t$, which is in the scattering plane, perpendicular to the proton momentum. In the Born approximation, the ratio $\gegm$ can be extracted from the ratio of $P_t$ and $P_l$:
\begin{equation} \label{eq:ratio}
\frac{\gep}{\gmp} = -\frac{P_t}{P_l} \frac{(E+E')}{2M_p}
\tan({\frac{\theta_e}{2}}).
\end{equation}
In the JLab recoil polarization measurements~\cite{jones00, gayou01, gayou02, punjabi05, jones06, maclachlan06, ron07, paolone10, puckett10, zhan11, ron11, puckett12, Puckett:2017flj}, a focal plane polarimeter~\cite{alcorn04, Puckett:2017flj})is used to measure both the transverse and longitudinal polarization components $P_t$ and $P_l$, yielding an extraction of $\gegm$ that is independent of beam polarization and analyzing power. Because the polarization measurements are only sensitive to the ratio, these measurements can be combined with cross-section data to allow for a precise extraction of both $\gep$ and $\gmp$, even for kinematics where the cross section is dominated by one of the form factors.

\begin{figure}[!htbp]
\begin{center}
\includegraphics*[width=0.95\columnwidth, trim={2mm 5mm 1mm 3mm}, clip]{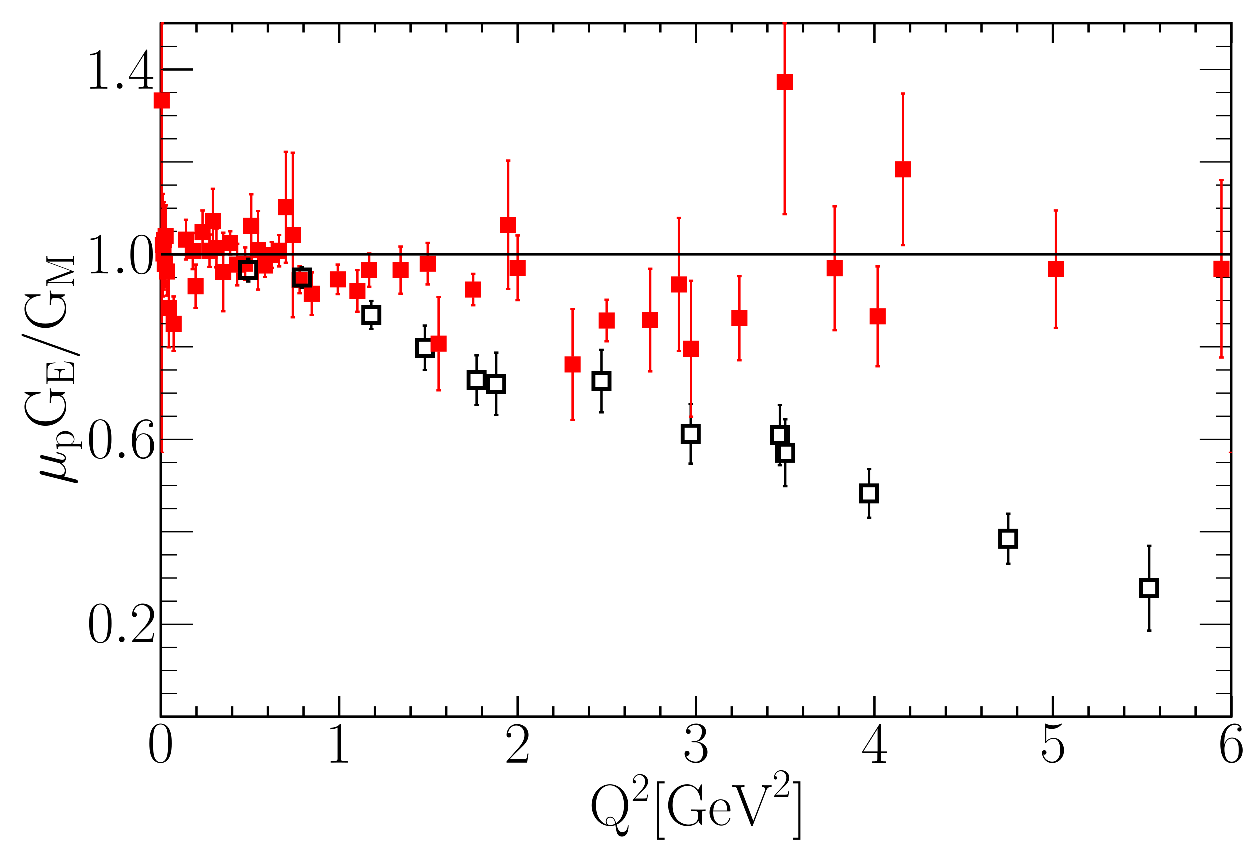}
\end{center}
\caption{(Color online) Comparison of Rosenbluth extractions (solid red squares) from the global analysis of Ref.~\cite{arrington04a} and polarization extractions~\cite{jones00, gayou01, gayou02} (hollow black squares) of $\mugegm$ at the time experiment E01-001 ran.}
\label{fig:gegm_lt_pt}
\end{figure}

Figure~\ref{fig:gegm_lt_pt} shows a comparison of a global Rosenbluth~\cite{arrington04a} extraction and world polarization~\cite{jones00, gayou01, gayou02} extractions of $\mugegmp$, based on data available in 2002. The Rosenbluth results are consistent with form factor scaling, but the polarization results decrease approximately linearly with increasing $Q^2$, and deviate significantly from the Rosenbluth data for $Q^2 > 1$~GeV$^2$.

The data from recoil polarization measurements are more precise at high $Q^2$ and should be less sensitive to systematic uncertainties than the Rosenbluth data~\cite{arrington07a, perdrisat07, arrington11b}. This, combined with the scatter between different Rosenbluth separations, led people to believe that there were experimental issues with the previous Rosenbluth extractions. A detailed examination~\cite{arrington03a} suggested that the Rosenbluth data were consistent and that only a large, common systematic effect could resolve the discrepancy. Because the polarization measurements yield only the ratio $\gegm$, it was important to understand the nature of any error in the cross-section data to properly combine the results from the two techniques to separate $\gep$ and $\gmp$.  More importantly, assuming a significant error in the cross-section measurements would have significant consequences for a large number of other experiments~\cite{arrington04a}, which normalize their results to elastic e-p scattering or require the use of the elastic cross sections~\cite{dutta03} or form factors~\cite{budd03} as input to the analysis. Thus, identifying the source of the inconsistency is important, as it will help us to understand whether there is an error in our cross-section extraction procedures and how this might impact a range of precision measurements.

\subsection{Possible sources of the discrepancy and early two-photon exchange calculations}\label{sec:earlytpe}

It was noted that the discrepancy between the Rosenbluth and polarization data could be resolved if there were a common systematic error in the cross-section measurements yielding a (5-8)\% $\varepsilon$-dependent correction~\cite{arrington03a, guichon03, arrington04a}. Several attempts were made to understand the nature of the discrepancy, with many focusing on the potential impact of missing two-photon exchange (TPE) contributions.

Guichon and Vanderhaeghen~\cite{guichon03} expressed the hadronic vertex function in terms of three independent complex amplitudes or generalized form factors that depend on both $Q^2$ and $\varepsilon$. The reduced cross section $\sigma_R$ and the recoil-polarization observables were expressed in terms of these generalized form factors, with TPE contributions assumed to yield corrections at the few-percent level. For the polarization observables, this translates into few-percent corrections to the extracted values of $\mugegm$. Rosenbluth separations are much more sensitive to small corrections to the cross section, in particular if they modify the small $\varepsilon$ dependence coming from $\gep$. For $Q^2 \gtorder 3$~GeV$^2$, the cross-section contribution from $\gep$ is at most a few percent, and even small TPE corrections could yield a comparable contribution.

A low energy hadronic model which accounts for the proton intermediate state but neglects excited intermediate states was proposed in Ref.~\cite{blunden03}. In this model, TPE corrections from elastic contributions (box and crossed-box diagrams) were included. Their results showed a $\approx$2\% $\varepsilon$ dependence with small nonlinearities at small $\varepsilon$ and insignificant $Q^2$ dependence. 

If TPE corrections are responsible for the discrepancy, then they must increase the $\varepsilon$ dependence of $\sigma_R$, yielding an apparent increase in $\gep$. They may also modify the extrapolation of $\sigma_R$ to $\varepsilon = 0$, modifying the extracted value of $\gmp$. From symmetry constraints~\cite{arrington11b}, we know that the TPE contribution must vanish at $\varepsilon=1$, and thus any $\varepsilon$-dependent TPE correction is likely to modify $\gmp$ as well as $\gep$. Therefore, it is crucial to know the $\varepsilon$ dependence of the TPE correction, in particular any nonlinearity as $\varepsilon \to 0$, to accurately extract $\gmp$.

We have focused so far on the understanding of the form factor discrepancy and TPE corrections in the mid-2000s, at the time that Jefferson Lab (JLab) experiment E01-001 made improved Rosenbluth measurements to confirm the discrepancy and to constrain TPE contributions. Since then, significant theoretical and experimental studies have been carried out in order to understand the impact of TPE corrections on electron-proton scattering observables, as summarized in Secs.~\ref{sec:TPE_theory_Phenomen} and~\ref{TPE_experimental}. In addition, there are extensive reviews of the role of the TPE effect in electron scattering~\cite{carlson07, arrington07a, arrington11b, yang13, Afanasev:2017gsk} and on their impact on the extraction of the form factors and other observables~\cite{arrington04a, kondratyuk06, borisyuk06b, carlson07, arrington07b, arrington11b, arrington11a, Arrington:2012dq, bernauer10, arrington11c, bernauer11, Qattan:2012zf, Arrington:2015ria, Qattan:2015qxa, Qattan:2018epw, Qattan:2021ysu}.


\section{Experiment Overview} \label{sec:experiment_overview}

Because of the inconsistency in the extracted values of $\gegm$ between Rosenbluth and precise recoil polarization measurements, a high-precision measurement of $\gegm$ using the Rosenbluth separation technique in the $Q^2 > 1.0$~GeV$^2$ region was important to provide a comparison between the two techniques with precision comparable to the recoil polarization measurements. It also provides a check on the possibility of additional and unaccounted-for systematic uncertainties in the Rosenbluth or recoil polarization measurements by minimizing the size of the $\varepsilon$-dependent corrections.

\begin{figure}[!htbp]
\begin{center}
\includegraphics*[width=0.95\columnwidth,height=6cm]{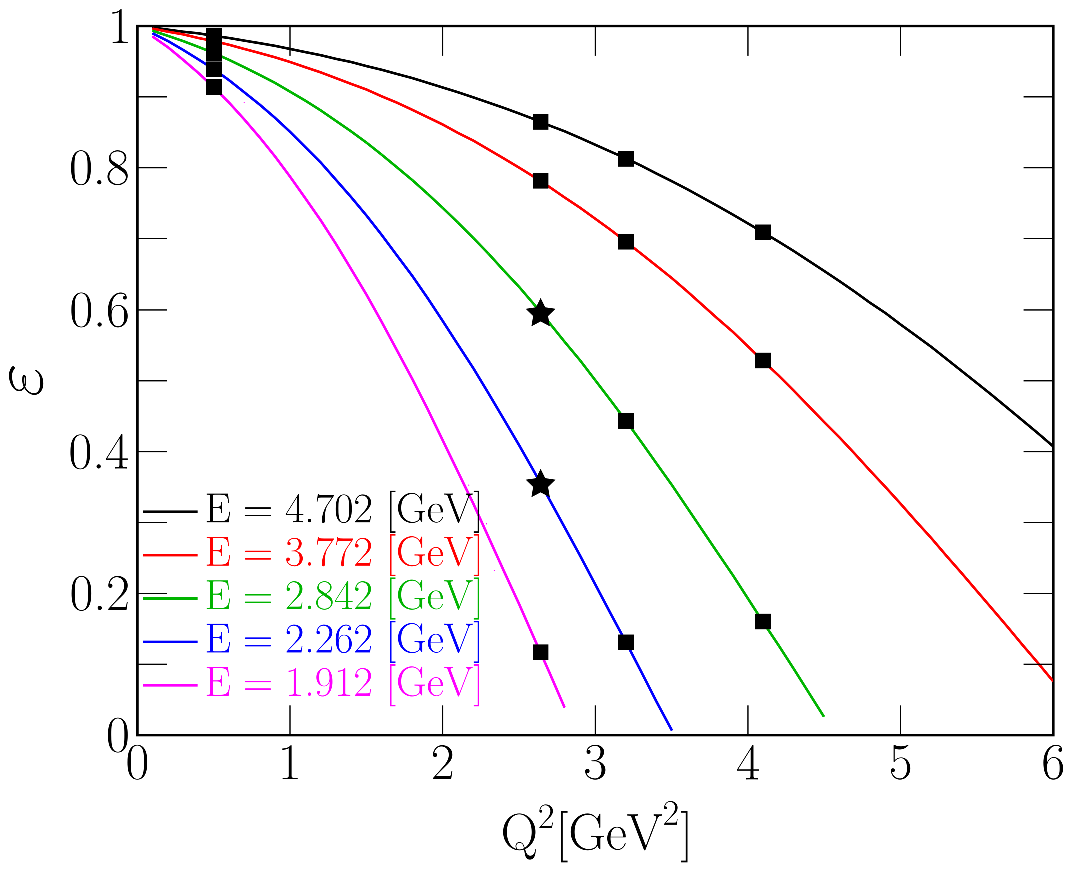}
\end{center}
\caption{(Color online) Kinematics of the E01-001 experiment. Solid lines represent different electron beam energies, while points indicate the kinematics of individual cross-section measurements. Squares indicate kinematics where only inclusive protons were measured, while stars indicate kinematics where additional electron-proton coincidence measurements were performed with electrons detected in the HRS-R.}
\label{fig:epsilon_q2}
\end{figure}

Experiment E01-001 was designed to achieve these goals. It ran in May 2002 in Hall A of the Thomas Jefferson National Accelerator Facility (JLab) in Newport News, Virginia. An incident electron beam with energy from 1.9-4.7~GeV was directed on a 4~cm unpolarized liquid hydrogen target and protons from elastic e-p scattering were detected. Measurements of the elastic e-p cross sections were made as a function of scattering angle at fixed $Q^2$ to allow for a Rosenbluth separation of the proton electric and magnetic form factors. The experiment emphasized minimizing the relative uncertainties on the cross sections at fixed $Q^2$ values to provide the most precise $\gegm$ extraction possible, even though some aspects of the analysis yield larger normalization uncertainties in the cross-section measurements that cancel out in the extraction of $\gegm$.

\begin{table}[!htbp]
\begin{center}
\caption{Nominal kinematic settings for the experiment. $E_b$ is the electron beam energy, $\theta_{L(R)}$ and $P_L(R)$ are the nominal central spectrometer angle and central momentum settings, respectively, for the HRS-L(R). $Q^2_{L(R)}$ is the $Q^2$ for the HRS-L(R), $\varepsilon_{L(R)}$ is the virtual photon polarization parameter for the HRS-L(R). The last column is the $\Delta P$ cut applied in the analysis of the elastic peak for the left arm data (see Sec.~\ref{extract_sigma}). The right arm was always set to $P_R = +0.756$~GeV, corresponding to $Q^2=0.5$~GeV$^2$, except for the coincidence kinematics where the right arm settings for electron detection are listed at the bottom of the table. The right-arm $\Delta_P$ cut was from $-$18 to +12 MeV for beam energies above 2.5 GeV, and $-$19 to +11 MeV for beam energies below 2.5 GeV. Small kinematic offsets were later determined and applied to the energy and scattering angles; see Table~\ref{tab:kinematics} for the final kinematics used in the analysis.}

\begin{tabular}{c c c c c c c} \hline \hline
$E_b$ & $\varepsilon_L$	& $\theta_L$ & $P_L$ & $\varepsilon_R$ & $\theta_R$ & $\Delta P$ cut\\
(GeV)   &               &   (deg)    &  (GeV)&              &   (deg)    & (MeV)\\
\hline
\multicolumn{7}{l}{$Q^2_L$=2.64~GeV$^2$, $Q^2_R$=0.50~GeV$^2$} \\
\hline
1.912 & ~0.117      & 12.631 & 2.149 & 0.914 & 58.288 & $-$12:+2 \\
2.262 & 0.356\dag 	& 22.166 & 2.149 & 0.939 & 60.075 & $-$12:+6 \\
2.842 & ~0.597\ddag & 29.462 & 2.149 & 0.962 & 62.029 & $-$12:+9 \\
3.772 & 0.782 	    & 35.174 & 2.149 & 0.979 & 63.876 & $-$13:+11 \\
4.702~ & ~0.865~    & ~38.261~ & ~2.149~ & ~0.986~ & ~64.978~ & $-$14:+12 \\
\hline
\multicolumn{7}{l}{$Q^2_L$=3.20~GeV$^2$, $Q^2_R$=0.50~GeV$^2$} \\
\hline
2.262 & 0.131 	 & 12.525 & 2.471 & 0.939 & 60.075 & $-$12:+3 \\
2.842 & 0.443 	 & 23.395 & 2.471 & 0.962 & 62.029 & $-$12:+8 \\
3.772 & 0.696 	 & 30.501 & 2.471 & 0.979 & 63.876 & $-$12:+12 \\
4.702 & 0.813 	 & 34.139 & 2.471 & 0.986 & 64.978 & $-$14:+14 \\
\hline
\multicolumn{7}{l}{$Q^2_L$=4.10~GeV$^2$, $Q^2_R$=0.50~GeV$^2$} \\
\hline
2.842 & 0.160 	 & 12.682 & 2.979 & 0.962 & 62.029 & $-$12:+5 \\
3.772 & 0.528 	 & 23.665 & 2.979 & 0.979 & 63.876 & $-$15:+10 \\
4.702 & 0.709 	 & 28.380 & 2.979 & 0.986 & 64.978 & $-$17:+13 \\
3.362 &~0.398*&20.257 & 2.979 &  N/A   &  N/A   &   N/A \\
\hline
\multicolumn{7}{l}{\dag~ $\theta_R=71.481$~deg, $P_R=0.356$~GeV for coincidence run}  \\
\multicolumn{7}{l}{\ddag~ $\theta_R=47.439$~deg, $P_R=0.597$~GeV for coincidence run} \\
\multicolumn{7}{l}{*~ $\theta_R=61.184$~deg, $P_R=1.177$~GeV} (coincidence run only) \\

\hline\hline
\end{tabular}
\label{kinematics}
\end{center}
\end{table}

Protons were detected simultaneously using the two nearly-identical high-resolution spectrometers (HRSs), referred to as the left and right arms~\cite{alcorn04} or HRS-L and HRS-R. The HRS-L was used to measure scattering at $Q^2$ values of 2.64, 3.20, and 4.10~GeV$^2$. Simultaneously, the HRS-R measured scattering at $Q^2$ = 0.5~GeV$^2$ to serve as a monitor of beam charge, current, and target density fluctuations. A total of 12 points were measured covering an angular range of 12.52$^o$ $< \theta_L <$ 38.26$^o$ for the left arm and 58.29$^o$ $< \theta_R <$ 64.98$^o$ for the right arm. Here, $\theta_{L(R)}$ is the nominal angle of the spectrometer with respect to the electron beam for the left (right) spectrometer. Figure~\ref{fig:epsilon_q2} and Table~\ref{kinematics} show the nominal kinematics covered in the E01-001 experiment.

For elastic scattering, the kinematics are overconstrained and the proton momentum can be determined from the outgoing angle of the struck proton. This allows us to select elastic events by comparing the measured proton momentum, $P_{\mbox{meas}}$, to the expected momentum of the scattered protons, $P_{\mbox{calc}}(\theta_p)$, calculated from two-body kinematics using the measured scattering angle of the protons $\theta_p$:
\begin{equation} \label{eq:pcal}
P_{\mbox{calc}}(\theta_p) = \frac{2E_b(M^2_p+E_bM_p)\cos\theta_p}
{M^2_p+E^2_b+2E_bM_p-E^2_b\cos^2\theta_p},
\end{equation}
where $M_p$ is the mass of the proton. The difference between detected and expected momentum for elastic scattering is:
\begin{equation} \label{eq:delta_p}
\Delta P = P_{\mbox{meas}} - P_{\mbox{calc}}(\theta_p).
\end{equation}

The protons detected in the HRS can come from a variety of scattering processes. The contribution of interest is the elastic peak, corresponding to $\Delta P=0$, and its radiative tail. In addition, there are backgrounds due to quasi-elastic scattering from the aluminum target windows and high-energy protons generated from photoreactions ($\gamma p \to \pi^0 p$ and $\gamma p \to \gamma p $). For each of the kinematics covered, data were taken with the dummy target to subtract away the endcap contribution from the spectrum. Protons associated with the photoproduction of $\pi^0$ were simulated using a calculated Bremsstrahlung spectrum and $d \sigma \over dt$ $\propto s^{-7}$, normalized to the observed background, and then subtracted away from the spectrum.

After subtracting non-elastic events, the net number of elastic events from the measurement is then compared to the number of elastic events in the e-p peak as simulated using the Monte Carlo simulation program SIMC~\cite{makinsphd,oneillphd,mohring02} under the same kinematic conditions and cuts. The ratio of the data yield to the simulated yield is applied as a correction factor to the input e-p cross section.

The background and elastic simulations are sensitive to the HRS resolution. Three coincidence kinematics were taken to allow for a clean separation of the elastic events from background events, allowing us to isolate elastic protons and test the simulated shape of the $\Delta P$ spectrum. Protons were detected in the HRS-L in coincidence with electrons in the HRS-R. The coincidence data were also used to measure the proton absorption and to identify clean samples of protons, as discussed in Secs.~\ref{proton_absorp} and~\ref{A1_eff}.


\subsection{Treatment of Systematic Uncertainties} \label{sec:systematics}

Because we are aiming for a precise extraction of a small effect, with particular interest in the angular dependence of this effect, it is important to determine which sources of systematic uncertainties may have a non-trivial angular dependence. Frequently, uncertainties are broken into two categories: normalization or scale uncertainties, which shift all data points identically, and uncorrelated or point-to-point uncertainties, which are independent for each data point. In this experiment, normalization uncertainties do not affect the extraction of $\gegm$ at all, as it is sensitive to the slope of the cross section relative to the intercept. In many previous Rosenbluth separations, uncertainties that are correlated with angle, but where the exact correlation is not known, are treated as uncorrelated (point-to-point) or else decomposed into a normalization contribution and an additional point-to-point contribution chosen to be large enough to account for the expected range of angle-dependent behavior. In our analysis, we separate uncertainties into a combination of normalization, point-to-point, and contributions that could be linearly correlated with $\varepsilon$ (referred to as ``slope uncertainties''). This is important as a linear correction would impact $\gegm$, our primary quantity of interest, more than an equivalent uncorrelated contribution. Thus, treating a linear correlation as an uncorrelated contribution would suppress the impact on $\gegm$, while also enhancing the uncertainty associated with testing the linearity of the reduced cross section. Note also that the slope uncertainty is quoted as an estimate of the $\varepsilon$ dependence over the full possible range, $0 < \varepsilon < 1$, rather than just over the measured $\varepsilon$ range. This is particularly important for the low-$Q^2$ measurement where the measured $\varepsilon$ range is extremely small. The uncertainties are evaluated separately for each $Q^2$ value, as the experiment focused on reducing the relative uncertainties for measurements at a fixed $Q^2$ value, in some cases at the cost of larger normalization uncertainties. 

\subsection{Advantages and disadvantages of Proton Detection} \label{proton_vs_electron}

Nearly all previous elastic e-p cross-section measurements were made by detecting the scattered electrons, rather than the struck protons. While detecting the struck proton instead of the scattered electron has some drawbacks in extracting the absolute cross section, it has several advantages in extracting the ratio $\gegm$.  At large $Q^2$ values, $\gep$ yields a small, $\varepsilon$-dependent contribution to the cross section, and the extraction is extremely sensitive to any effects that depend on the scattering angle at fixed $Q^2$.  When detecting electrons, the detected particle momentum and rate vary rapidly with $\varepsilon$, so any rate or angle-dependent effects can significantly modify the extraction of $\gep$.

\begin{figure}[!htbp]
\begin{center}
\includegraphics*[width=0.95\columnwidth]{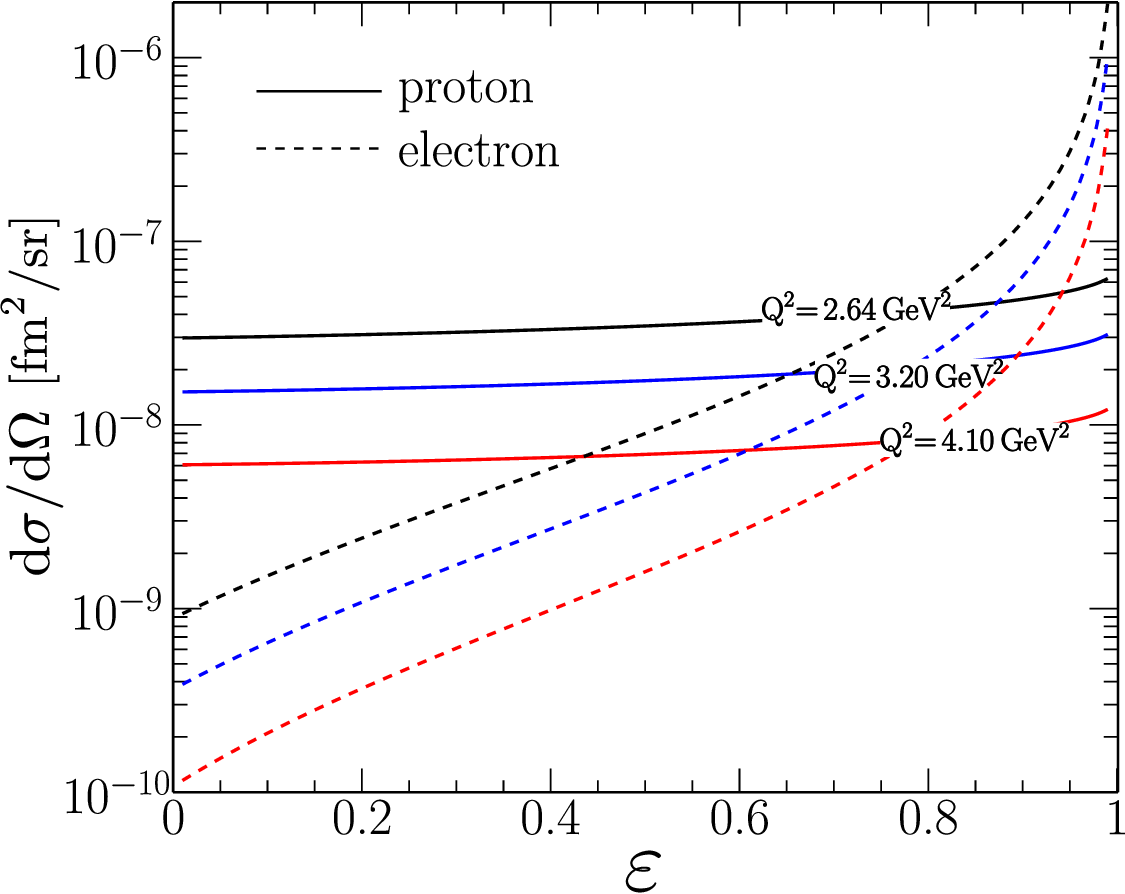}
\end{center}
\caption{(Color online) The differential Born elastic cross section for proton (electron) detection, $d\sigma / d\Omega_p(e)$ as a function of $\varepsilon$ at $Q^2$ = 2.64, 3.2, and 4.1~GeV$^2$ (black, red, and blue lines, respectively). While the underlying e-p scattering cross section is identical in both case, the Jacobian between the electron and proton solid angles enhances the proton differential cross section for small $\varepsilon$ values.} \label{fig:sigma_epsilon}
\end{figure}

When detecting the struck proton at fixed $Q^2$, the detected particle momentum is always fixed, eliminating momentum-dependent corrections in the detector as a potential source of uncertainty in $\gegm$.  It also yields a much smaller variation of rate with $\varepsilon$, as well as lower peak rates, as shown in Figure~\ref{fig:sigma_epsilon}. This significantly reduces both the size and $\varepsilon$ dependence of any rate-dependent corrections.  In addition, the electron cross section is much lower at small $\varepsilon$ values, and so many previous Rosenbluth separations are statistics limited at small $\varepsilon$ and/or used higher beam currents for these settings, leading to additional uncertainty associated with target heating or beam current readout. Finally, the cross section is less sensitive to knowledge of the beam energy and the angle of the detected particle, especially at larger $\varepsilon$ values, where the electron detection is extremely sensitive. For example, at $Q^2$=2.64~GeV$^2$ and $\varepsilon \approx 0.9$, the cross section changes by ~50\% for a 1$^o$ change in the electron scattering angle, and this sensitivity varies rapidly with $\varepsilon$. For proton detection, the highest sensitivity is below 20\% per degree and varies more slowly with $\varepsilon$. The impact of these effects is illustrated in more detail in Fig. 2 of Ref.~\cite{Arrington:2021alx} which compares electron and proton detection for a variety of quantities from 0.5-4.5~GeV$^2$.

\begin{figure}[!htbp]
\begin{center}
\includegraphics*[width=0.95\columnwidth]{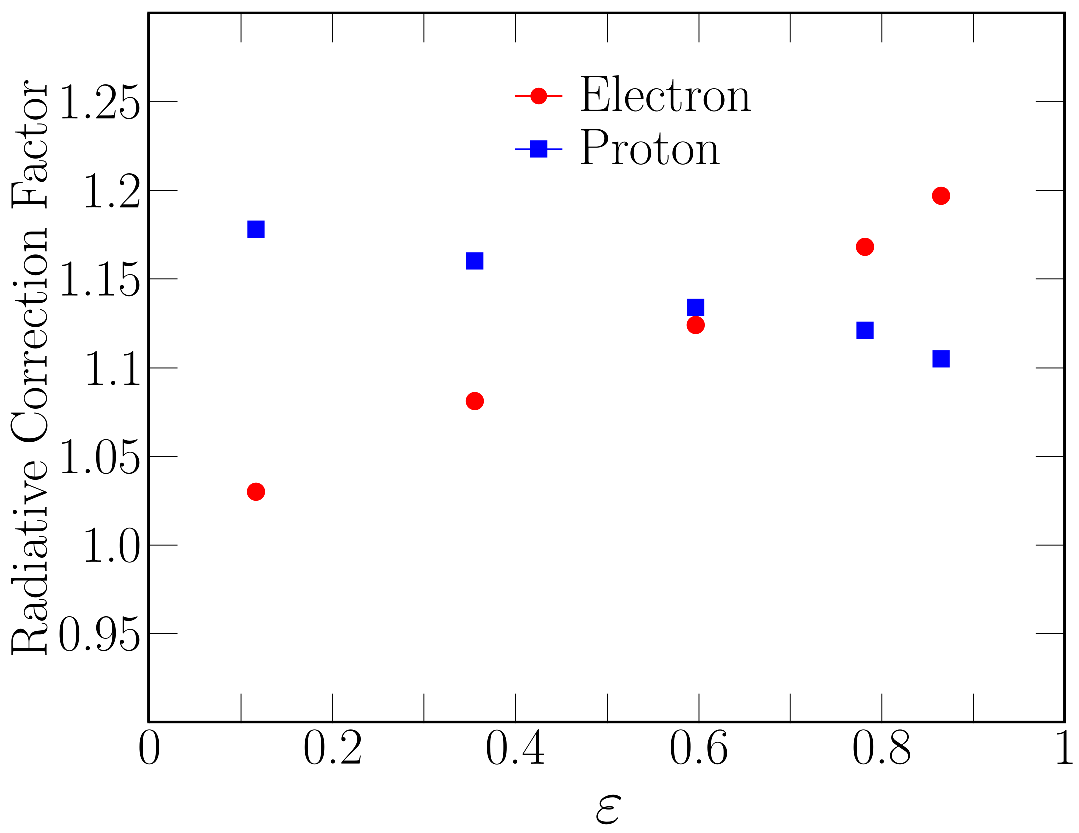}
\end{center}
\caption{(Color online) The proton and electron radiative correction factor (internal corrections only) as a function of $\varepsilon$ at $Q^2$ = 2.64~GeV$^2$.}
\label{fig:rad_corr_epsilon}
\end{figure}

Fig.~\ref{fig:rad_corr_epsilon} shows the internal radiative correction factor at $Q^2$ = 2.64~GeV$^2$ for protons and electrons, from the calculation of Afanasev \etal~\cite{afanasev01}. While the overall magnitude of the corrections is similar, the proton correction factor changes by 8\% over the $\varepsilon$ range, while the electron correction varies by $\approx$17\%.  Thus, the impact of the large, $\varepsilon$-dependent radiative corrections should be smaller when detecting the proton.  In addition, because the $\varepsilon$ dependence has the opposite sign, a comparison of electron and proton detection yields a test of the $\varepsilon$ dependence of these corrections.

We note that proton detection also has disadvantages compared to electron scattering. Because the collimator used in the spectrometer is not as effective in stopping protons, we use a tight software cut to restrict the solid angle to the center of the acceptance. This reduces the cross section, although it is not a significant limitation as there is a significant enhancement in the proton-detection cross section, compared to electron detection, for kinematics where the electron cross section limits Rosenbluth separations. In addition, there is a larger normalization uncertainty from knowing the exact solid angle associated with the software cut. This increases the uncertainties on $\gep$ and $\gmp$, but because these are scale uncertainties, this contribution cancels out in the ratio of interest for this experiment, $\mugegmp$.

In addition, proton detection requires rejecting positive pions or deuterons with kinematics overlapping the proton peak. This is done through a combination of time-of-flight information and aerogel Cerenkov detectors, as discussed in detail in the analysis section. For the kinematics of this experiment, high proton  efficiency with good pion/deuteron rejection was achieved and this was not a significant issue.

Most importantly, proton detection requires dealing with larger backgrounds. The subtraction of protons coming from the target endcaps is larger. In addition, there are backgrounds that are not relevant for electron detection: protons coming from Compton scattering of neutral pion production. The subtraction of these backgrounds is described in detail in Sec.~\ref{sec:analysis}, and this subtraction has an important contribution to the systematic uncertainties of the measurement, especially at the largest $Q^2$ value.

\section{Beamline Instrumentation} \label{expt_equip}

For experiment E01-001, the Continuous Electron Beam Accelerator Facility of the Thomas Jefferson National Accelerator Facility provided an unpolarized electron beam in the range of $1.912< E_b < 4.702$~GeV with beam currents up to 70 $\micro$A.  Details of the beamline instrumentation are provided in the following sections.

\subsection{Beam Energy Measurements} \label{sec:beam_energy}

A detailed description of the accelerator is provided in Ref.~\cite{leemann01}, and the equipment in Hall A in Ref.~\cite{alcorn04}. Precise measurements of the beam energy are required to extract the proton form factors from elastic e-p cross sections. There are two different measurements that can be performed to determine the energy of the incident electron beam with precision as high as $\delta E_b/E_b$ = 2$\times 10^{-4}$. These measurements are known as the arc \cite{marchandthesis, alcorn04} and ep~\cite{alcorn04, ravelthesis} measurements.
Table~\ref{beam_energy} lists the beam energy measurements for the E01-001 experiment. The Tiefenback energy $E_{\mbox{Tief}}$ measurement is similar to the arc measurement but is based on the Hall A arc beam position monitors (BPMs) measurements rather than the superharps. The Tiefenback value for the beam energy, which is consistent with the arc measurements, is used in the E01-001 analysis.

\begin{table}[!htbp]
\begin{center}
\caption{The Tiefenback energy $E_{\mbox{Tief}}$, Tiefenback quoted uncertainty $\Delta E$, arc, and ep beam energy measurements of the E01-001 experiment. The Tiefenback energy was used for the analysis with final uncertainty of 0.03\% offset combined with 0.02\% point-to-point uncertainty. See text for details.}
\begin{tabular}{c c c c c c c} \hline \hline
Pass & $E_{\mbox{Tief}}$ & $\Delta E$ &$E_{\mbox{Arc}}$ & $E_{\mbox{ep}}$& Arc/Tief. & ep/Arc \\
     & (MeV)           & (MeV)      &(MeV)     & (MeV)   &           &          \\
\hline
2    & 1912.94         & 0.69       & --       & --      & --        & --        \\
2    & 2260.00         & 0.81       & 2260.20  & 2260.83 & 1.000088  & 1.000279   \\
3    & 2844.71         & 1.03       & --       & --      & --        & --          \\
4    & 3772.80         & 1.36       & 3773.10  & 3775.23 & 1.000080  & 1.000565     \\
5    & 4702.52         & 1.70       & --       & --      & --        & --            \\
5    & 5554.60         & 2.00       & 5555.17  & --      & 1.000103  & --             \\
\hline \hline
\end{tabular}
\label{beam_energy}
\end{center}
\end{table}

Random (point-to-point) and scale uncertainties of 0.01\% and 0.05\%, respectively, have been reported on the non-invasive arc measurements (``Tiefenback energy'' measurements), and a 0.02\% random uncertainty in the full invasive arc measurement or the ep measurement~\cite{alcorn04}. Since the Tiefenback results were consistent with the full arc and ep measurements where they were taken, we assume that the absolute uncertainty in the Tiefenback is closer to the 0.02\%. Therefore, we apply scale and random uncertainties of 0.03\% and 0.02\%, respectively, on the Tiefenback energy.

As noted in Sec.~\ref{sec:systematics}, we separate the impact of the kinematic offsets into an overall normalization uncertainty, a correction that changes the slope of the reduced cross section versus $\varepsilon$, and $\varepsilon$-uncorrelated contribution. An overall offset in the beam energy will impact the cross-section extraction at all kinematics, with a larger impact at small $\varepsilon$ values. To separate this into normalization, slope, and random uncertainties, we evaluate the correction corresponding to a 0.03\% energy offset and take the average cross section change as the normalization uncertainty, the slope of the change as the slope uncertainty, and the residual scatter of the correction around the linear fit as a point-to-point contribution. The right arm results show an average scale, random, and slope uncertainties of 0.034\%, 0.01\%, and 0.29\%, respectively. The left arm results show an average scale, random, and slope uncertainties of 0.13\%, 0.02\%, and 0.073\%, respectively. The impact of the 0.02\% point-to-point uncertainty in the beam energy is evaluated for each kinematic setting and treated as an uncorrelated error in the cross sections. This yields an additional random uncertainty of (0.01--0.03)\% for the right arm and (0.04--0.08)\% for the left arm.


\subsection{Beam Position Measurements} \label{sec:beam_position}

Because the endcap of the cryogenic target is not flat, the effective target length changes slightly if the beam position is offset from the center of the cell.  In addition, a vertical beam offset yields a small shift in the reconstructed momentum of the detected proton. To determine the beam position and direction at the target, two beam position monitors (BPMs)~\cite{alcorn04, xzhengthesis} located 7.516 m and 2.378 m upstream from the target were used. The BPM is a cavity with a 4-wire antenna in one plane with frequency tuned to match the RF frequency of the beam (499 MHz). The absolute position of the beam is determined by the BPMs by calibrating the BPMs with respect to wire scanners (superharps) located adjacent to each BPM at 7.345 and 2.214 m upstream of the target.

\begin{figure}[!htbp]
\begin{center}
\includegraphics*[width=0.98\columnwidth]{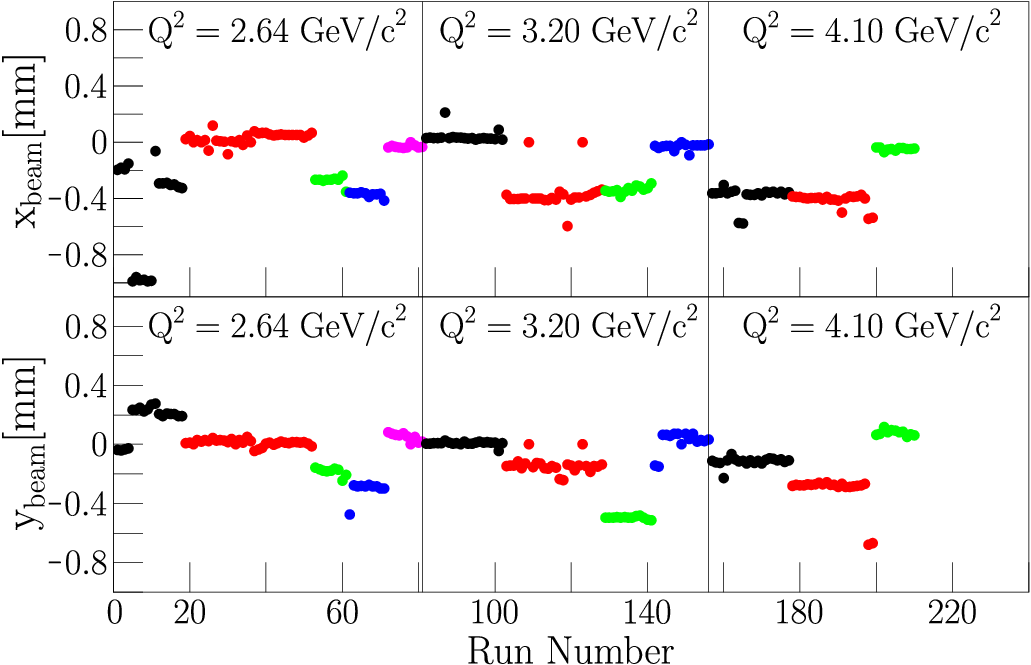}
\end{center}
\caption{(Color online) The top (bottom) plot shows the $x$ ($y$) coordinate of the beam at the target.  The different colors represent the different beam energy settings in order of increasing $\varepsilon$.}
\label{fig:beam_position}
\end{figure}

From these BPMs, the position and angle of the beam on target were reconstructed. Figure~\ref{fig:beam_position} shows the $x$ and $y$ coordinates of the beam at the target for all runs. The beam was well focused on the target with an average ($x$,$y$) position of (--0.2,--0.1)~mm, and an average beam drift of $\pm$0.30~mm. The position uncertainties of 0.30~mm in two BPMs roughly 5~m apart yield a 0.07~mrad angle uncertainty in the beam angle.  The horizontal beam position offset is accounted for in the target thickness and the vertical offset used to correct the reconstructed momentum as discussed in Secs.~\ref{target_leng_corr} and~\ref{recon}.


\subsection{Beam Current Measurements} \label{sec:beam_current}

Two identical beam current monitors (BCMs)~\cite{alcorn04}  located approximately 25~m upstream from the target are used to determine the total charge on target. The two BCMs are calibrated at several beam currents relative to a parametric current transformer or Unser monitor. The Unser monitor is located halfway between the two BCMs and is calibrated by passing a precisely known current through it. The BCMs are designed to provide a continuous, non-invasive measurement of the beam current.

Based on the analysis of all of the calibration runs, the gains and offsets were stable to within $\pm$0.1\% and $\pm$0.01\%, respectively. The effect of the offset drift is negligible, so a random uncertainty in the charge measurement of 0.1\% is assigned for both arms. Between experiments, the BCMs and Unser are calibrated using a calibrated current source and the results are stable to within $\pm$0.5\%~\cite{alcorn04}. Therefore, a scale uncertainty in the charge measurement of 0.5\% is assigned for both arms.


\section{Hall A Target System} \label{sec:HallAtarg}

The cryogenic target system of Hall A is mounted on a ladder inside the scattering vacuum chamber. The ladder contains sub-systems for cooling, gas handling, temperature and pressure monitoring, and target control and motion. The ladder also contains a selection of solid targets such as dummy target for background measurements and BeO, $^{12}$C, and optics targets for beam viewing and calibration. The cryogenic target has three independent loops. Two loops were configured to hold liquid hydrogen (LH2) or deuterium (LD2), and one for gaseous helium. Fans are used to circulate the liquid or the gas through each loop. Each of the LH2 and LD2 loops has two aluminum cylindrical target cells of either 4 or 15~cm in length and 6.35~cm in diameter. The 4-cm long liquid hydrogen target was used in this experiment. The side walls of the cells are 0.178~mm thick with entrance (upstream) and exit (downstream) windows of 0.071 and 0.102~mm thick, respectively.

During E01-001, the LH2 target was operated at a constant temperature of 19~K and pressure of 25~psi with a density of 0.0723~g/cm$^3$. The temperature of the target was stabilized using a high-power heater to compensate for the effect of any beam intensity variation on the target's temperature. The target was cooled using 15~K helium coolant supplied by the End Station Refrigerator. Because of the small spot size of the beam, the beam was rastered using a rectangular raster~\cite{alcorn04} producing a 2~mm $\times$ 2~mm spot size at the target to prevent any damage to the target windows and to reduce density fluctuations due to local heating.


\section{Hall A High-Resolution Spectrometers} \label{sec:HallA_spect}

Protons were detected simultaneously using the two identical high-resolution spectrometers. The left arm (HRS-L) was used to measure three $Q^2$ points of 2.64, 3.20, and 4.10~GeV$^2$. Simultaneous measurements at $Q^2 = 0.5$~GeV$^2$ were carried out using the right arm (HRS-R) which served as a check on uncertainties due to beam charge, current, and target density fluctuations. The two spectrometers are identical in design and can provide a maximum central momentum of approximately 4~GeV with momentum resolution better than 2$\times$10$^{-4}$, horizontal (vertical) scattering angle resolution better than 0.6 (2.0) mrad, solid angle acceptance of $\Delta \Omega$ = 6.7~msr, and transverse target length acceptance of $\pm$5~cm.  A detailed description of the spectrometer design can be found in~\cite{alcorn04} and references therein.


\subsection{Detector Package}

The detector package of each HRS is located in a large steel and concrete detector hut behind the magnet system. During E01-001, the two spectrometers included similar detector packages with only slight differences. A detailed description of the detector packages can be found in Refs.~\cite{alcorn04, qattanphd}. The following detectors were used in each arm: two Vertical Drift Chambers (VDCs) for tracking, scintillator planes for trigger and time-of-flight measurements, aerogel Cerenkov detectors for particle identification, and a gas Cerenkov for $e$ identification during coincidence measurements.  These detectors are described in detail in the following sections.

While the gas Cerenkov detectors and the HRS-R mirror aerogel (A$_M$) were not used in the production running, they do contribute to proton absorption in the detector stack, as they are located before the last scintillator plane, S$_2$, which is used in the trigger.  Additional detectors (electromagnetic calorimeters) positioned after S$_2$ were not used and have no impact on the measurement.


\subsubsection{Vertical Drift Chambers} \label{VDCs}

The vertical drift chambers (VDCs) are used for tracking (reconstruction of the position and slope of the particle trajectory) of the scattered particles. The first VDC is located at the focal plane of the spectrometer, and the second VDC is positioned parallel to and 23~cm away from the first VDC. Both VDCs intersect the spectrometer central ray at an angle of 45$^o$, which is the approximate nominal angle at which the particle trajectory crosses the wire planes of each VDC.  A detailed description of the hardware and operation of these detectors can be found in Refs.~\cite{alcorn04, fissum01}


\subsubsection{Scintillators and Triggers} \label{sec:scin}

Event triggers were formed using the hodoscope scintillators. Each spectrometer has two scintillator planes, S$_1$ and S$_2$, which are used for triggering and time-of-flight determination. Both scintillator planes are perpendicular to the nominal central ray of the spectrometer and separated by approximately 2~m. Each scintillator plane has 6 identical overlapping scintillator paddles made of thin plastic (0.5~cm thick BICRON 408) to minimize hadron absorption. A photo-multiplier tube (PMT) is attached to each end of each scintillator paddle to collect the photons produced by particles passing through the scintillator. We refer to these as the left and right PMTs of the scintillator paddle.

The active area is approximately 170~cm~$\times$~36~cm for S$_1$ and 220~cm~$\times$~60~cm for S$_2$. The time resolution for each scintillator plane is approximately 0.30~ns. An additional scintillator counter, S$_0$, was added to the left arm trigger system for measurement of the efficiency. S$_0$ was installed before the S$_1$ scintillator plane and it is a 1~cm thick scintillator paddle with an active area of 190~cm~$\times$~40~cm and PMTs on both ends.

The ADC signal measures the energy deposited in the scintillator. For the lower momentum settings, heavier particles deposit more energy in the scintillator, and this can be used to separate pions, protons, and deuterons. The TDC signal provides timing information for the different types of triggers used.  Given the time-of-flight of the particle between the scintillator planes and the particle momentum determined from the track reconstruction, we can reconstruct the particle velocity, $\beta$ = $v/c$.

\begin{table}[!htbp]
\begin{center}
\caption {Trigger type definitions.}
\begin{tabular}{c c c} \hline \hline
Trigger	& Definition & Purpose\\ \hline
T$_1$	& S$_1$ AND S$_2$ & HRS-L physics \\
T$_2$	& S$_1$ XOR S$_2$ & HRS-L efficiency \\
T$_3$	& ~S$_0$ AND S$_1$ AND S$_2$~   & HRS-R physics	\\
T$_4$	& ~S$_0$ AND (S$_1$ XOR S$_2$)~ & HRS-R efficiency	\\
T$_5$	& T$_1$ AND T$_3$ & e-p coincidence \\ \hline\hline
\end{tabular}
\label{trig_def}
\end{center}
\end{table}

There are five basic triggers (event types) generated from the scintillator timing information. They are classified as primary physics triggers (T$_1$, T$_3$, and T$_5$) and efficiency triggers (T$_2$ and T$_4$). A sample of the efficiency triggers, which require only one of the primary hodoscope planes, is taken to monitor the hodoscope efficiency. Table~\ref{trig_def} gives the definitions of the trigger types used in this experiment. The trigger signal for a single plane requires that both of the PMTs fire on any one of the elements in that plane.


\subsubsection{Gas and aerogel Cerenkov Detectors}

Two aerogel Cerenkov detectors, A$_1$ and A$_2$~\cite{alcorn04, comanthesis}, were used to separate protons from pions. The A$_2$ aerogel detector was used with the right arm, while the A$_1$ aerogel was used with the left arm.  A mirror aerogel, A$_M$, was installed on the right arm between the S$_1$ and S$_2$ scintillators but was not used in the analysis. The two aerogel detectors A$_1$ and A$_2$ have similar designs. The A$_1$ aerogel has a 9~cm aerogel radiator with index of refraction $n_{A_1}$=1.015 while A$_2$ has a 5~cm aerogel radiator with index of refraction $n_{A_2}$ = 1.055.  This corresponds to threshold momentum of 5.4~GeV for protons (0.8~GeV for pions) for A$_1$ and 2.8~GeV for protons (0.4~GeV for pions) for A$_2$, such that protons are always below the threshold and pions are well above the threshold for all settings.

The gas Cerenkov detector used in Hall A~\cite{alcorn04,iodice98} was installed between the scintillator S$_1$ and S$_2$ planes.  The detector was filled with CO$_2$ to give an index of refraction of $n$ = 1.00041 such that electrons(hadrons) are always above(below) the Cerenkov threshold. In this experiment, the Cerenkov was only used for electron identification during the coincidence runs, where a cut corresponding to approximately two photoelectrons was applied to reject pions.


\subsection{Data-Acquisition System}~\label{sec:DAQ}

The data-acquisition (DAQ) system used in Hall A is the CEBAF on-line data-acquisition (CODA) system~\cite{alcorn04, coda95, watson93} which is developed by the JLab data-acquisition group and designed for nuclear physics experiments. CODA is a toolkit that is composed of a set of software and hardware packages from which DAQ can be built to manage the acquisition, monitoring, and storage of data.

During the running, the ADC and TDC signals are read out whenever one of the physics triggers (Sec.~\ref{sec:scin}) is generated.  In addition, the scalers and control and beam status information are read out and recorded every few seconds by the EPICS~\cite{epics} system and included in the data stream.  The data are written to a local disk and then transferred to the mass storage system.


\section{Analysis} \label{sec:analysis}

In this section we discuss the corrections and cuts applied to the data and the procedure used to estimate the detector efficiencies, corrections, as well as the systematic uncertainties applied to the e-p elastic scattering cross sections. The philosophy for separating systematic uncertainties into random, scale, and slope was discussed in Sec.~\ref{sec:systematics}, and illustrated in detail in Sec.~\ref{sec:beam_energy}.

The raw data collected in the experiment are saved to disk by the data-acquisition system, as summarized in Sec.~\ref{sec:DAQ}. These files are read in and analyzed using the standard Hall A event processing software of the time: ESPACE (Event Scanning Program for Hall A Collaboration Experiments)~\cite{alcorn04}. ESPACE reconstructs the physical variables of interest needed for the analysis on an event-by-event basis, such as detector hits, tracks, and particle identification (PID) signals, and the coordinates of the reaction vertex in the target. 


\subsection{The Effective Charge} \label{qeff_intro}

To extract the cross section, we determine the number of detected protons, the accumulated charge, and the various detector efficiencies, livetimes, and other correction factors.  For each run, we calculate an effective charge, $Q_{\mbox{eff}}$, which is the measured beam charge, $Q$, reduced to account for any loss of events due to deadtime, inefficiency, and prescaling of the triggers:
\begin{eqnarray} \label{eq:qeff}
Q_{\mbox{eff}} = \frac{1}{ps} \Big(Q \times \mbox{ELT} \times \mbox{CLT} \times \epsilon_{\mbox{VDC}} \times \epsilon_{\mbox{VDCH}}\times \epsilon_{S_1} \times\nonumber\\
\epsilon_{S_2} \times \epsilon_{\mbox{PID}} \times C_{\mbox{Absorption}} \times C_{\mbox{TB}} \times C_{\mbox{TL}}\Big),~~~~~
\end{eqnarray}
where the dimensionless corrections applied above are the computer and electronic livetimes (CLT and ELT), VDC tracking efficiency ($\epsilon_{\mbox{VDC}}$), VDC multiplicity cut efficiency ($\epsilon_{\mbox{VDCH}}$), scintillator efficiency (product of $\epsilon_{S_1}$ and $\epsilon_{S_2}$), particle identification efficiency ($\epsilon_{\mbox{PID}}$), proton absorption correction ($C_{\mbox{Absorption}}$), target boiling correction ($C_{\mbox{TB}}$), and target length correction ($C_{\mbox{TL}}$). Here $ps$ is the prescale factor and is defined as: the prescale factor $n$ for the trigger type T$_i$ ($i = 1,\cdots,5$) means that the Trigger Supervisor will read out every $n$th event of type T$_i$. Having applied these corrections and efficiencies to the beam charge, we refer to the corrected charge as the effective charge or $Q_{\mbox{eff}}$. Because the prescale differs for each trigger type, $Q_{\mbox{eff}}$ also differs.

By applying the efficiency and other related corrections to the charge, we can simply sum the counts from different runs at the same kinematic settings and sum the combined $Q_{\mbox{eff}}$. In this way, pure counting statistics can be used for our individual LH2 and dummy runs, with weights applied only when making dummy and background subtractions. The target thickness is taken into account in the simulation, as it is used to normalize the number of scattering events and also used in calculating the radiative corrections. As a general data quality/consistency check, we also compare the run-by-run normalized yield for all runs at a given kinematic setting to ensure that there are not statistically significant jumps or outliers. An example of this, used to compare the normalized yield for runs at different beam currents, is shown in Sec.~\ref{tgt_boiling}.


\subsection{HRS Optics}  \label{e01001_optics}

The position and angle at the focal plane are measured in the VDCs and the target kinematics are reconstructed using the optics database - the transformation matrix between the focal plane and target variables. These reconstructed target variables are the in-plane ($\phi$ or $\phi_{\mbox{tg}}$) and out-of-plane ($\theta$ or $\theta_{\mbox{tg}}$) scattering angles, the $y$-coordinate of the target position $y_{\mbox{tg}}$, and the deviation of the particle's momentum from the central momentum setting of the spectrometer, $\delta = (p-p_o)/p_o$.

The optical properties of the spectrometers were studied using sieve slit collimators~\cite{alcorn04}. The sieve slit is a 5~mm thick stainless steel sheet with a pattern of 49 holes (7 $\times$ 7), spaced 12.5~mm apart horizontally and 25~mm apart vertically. Two of the holes, one in the center and one displaced two rows vertically and one horizontally, are 4~mm in diameter, while the rest are 2~mm in diameter to verify the orientation of the sieve slit. The slit is positioned (1.184$\pm$0.005)~m and (1.176$\pm$0.005)~m from the target on the left and right arms, respectively.  

A software cut was applied to limit the solid angle to 1.6~msr, with a cut on the out-of-plane angle of $ |\theta_{\mbox{tg}}|<$40~mrad and in-plane angle of $ |\phi_{\mbox{tg}}|<$10~mrad. Based on the accuracy of the reconstructed sieve slit hole pattern, we estimate that the size of the region accepted by the cuts is accurate to 0.20~mrad. This translates into a 2.0\% scale uncertainty in the in-plane angle and 0.5\% scale uncertainty in the out-of-plane angle. The sum in quadrature of the two angle scale uncertainties was used to determine the final estimated scale uncertainty in the 1.6~msr solid angle cut of 2.06\%. Because the 1.6~msr solid angle cut is identical for all kinematics, the uncertainty in the solid angle contributes fully to the scale uncertainty, but yields a negligible contribution to the uncertainty in the extraction of the ratio $\gegm$.


\subsection{Spectrometer Pointing} \label{spect_mispoint}

Extraction of the e-p cross sections requires knowledge of the scattering angle which depends on the spectrometer optics and offsets, target position, and beam position. Ideally, the spectrometer points directly at the center of the target. However, due to translational movements of the spectrometer around the hall center, the central ray of the spectrometer can miss the hall center in both the horizontal and vertical directions. For the inclusive measurements we use to extract $\gegm$, the vertical offset corresponds mainly to rotation in the azimuthal angle and has minimal impact on the cross-section extraction. The horizontal offset or the horizontal distance between the hall center and the central ray of the spectrometer is referred to as the spectrometer mispointing. There are two different and reliable methods by which the horizontal offsets can be measured: the survey method and the carbon-pointing method, discussed below.

The two spectrometers were surveyed at several kinematic settings. For the HRS-R, the spectrometer angle was surveyed and determined at all 5 $\varepsilon$ points, and the survey angles were used in the analysis. For the left arm, there were several spectrometer settings where a survey was not performed, and so the carbon-pointing method was used. In the carbon pointing method, the spectrometer mispointing, $\Delta h$, and spectrometer angle, $\theta_{s}$, are determined using ``pointing'' runs where electrons are scattered by a thin carbon foil at a known position. From the foil position and the spectrometer central angle, $\theta_{o}$ as determined from the hall floor marks, the target position as reconstructed by the spectrometer, $y_{\mbox{tg}}$, and the target offset along the beam direction, $z_{\mbox{off}}$, as measured by the target survey group, allow for an extraction of the mispointing $\Delta h$:
\begin{equation} \label{eq:offset}
\Delta h = \pm y_{\mbox{tg}} + z_{\mbox{off}} \sin(\theta_o),
\end{equation}
and hence the spectrometer scattering angle (angle setting) can be calculated as
\begin{equation} \label{eq:spec_angle}
\theta_{s} = \theta_{o} +  \Delta h/L,
\end{equation}
where the plus(minus) sign in front of $y_{\mbox{tg}}$ in Eq. (\ref{eq:offset}) is used with the right(left) arm and $L$ is the distance between the hall center and the floor marks where the angles are scribed and has a value of 8.458 m~\cite{qattanphd}. Note that $\frac{\Delta h}{L} = \Delta\theta_{o}$ in Eq. (\ref{eq:offset}) represents the correction to the central scattering angle of the spectrometer, $\theta_{s} = \theta_{o} \pm \Delta\theta_{o}$. The spectrometer is said to be mispointed downstream(upstream) if $\Delta h$ is positive(negative).

\begin{figure}[!htbp]
\begin{center}
\includegraphics*[width=0.95\columnwidth]{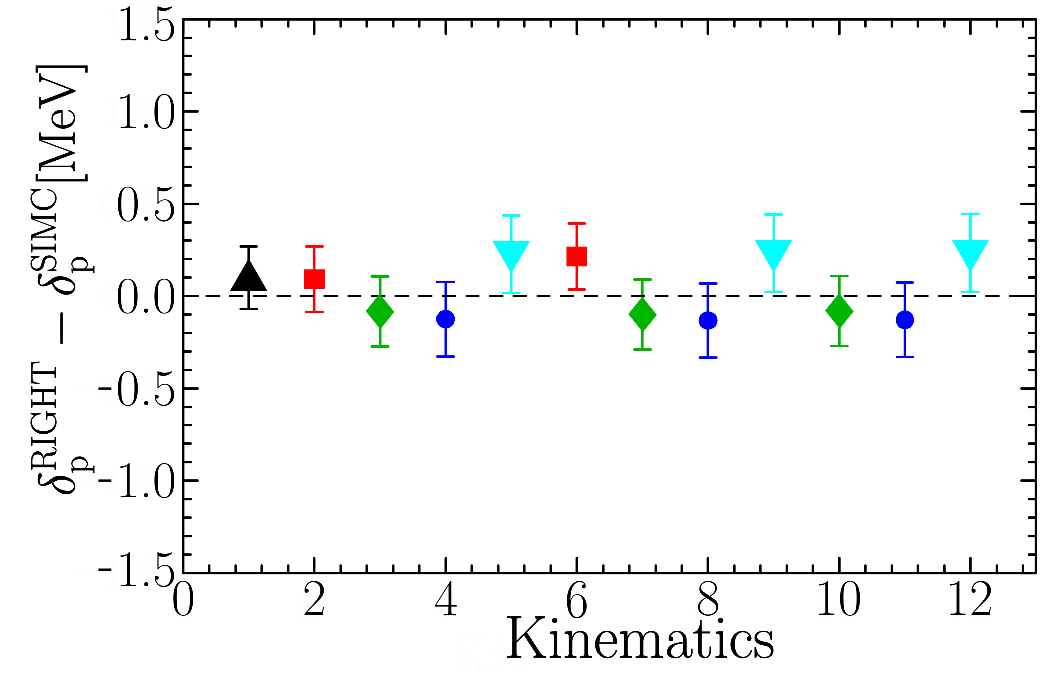} \\
\includegraphics*[width=0.95\columnwidth]{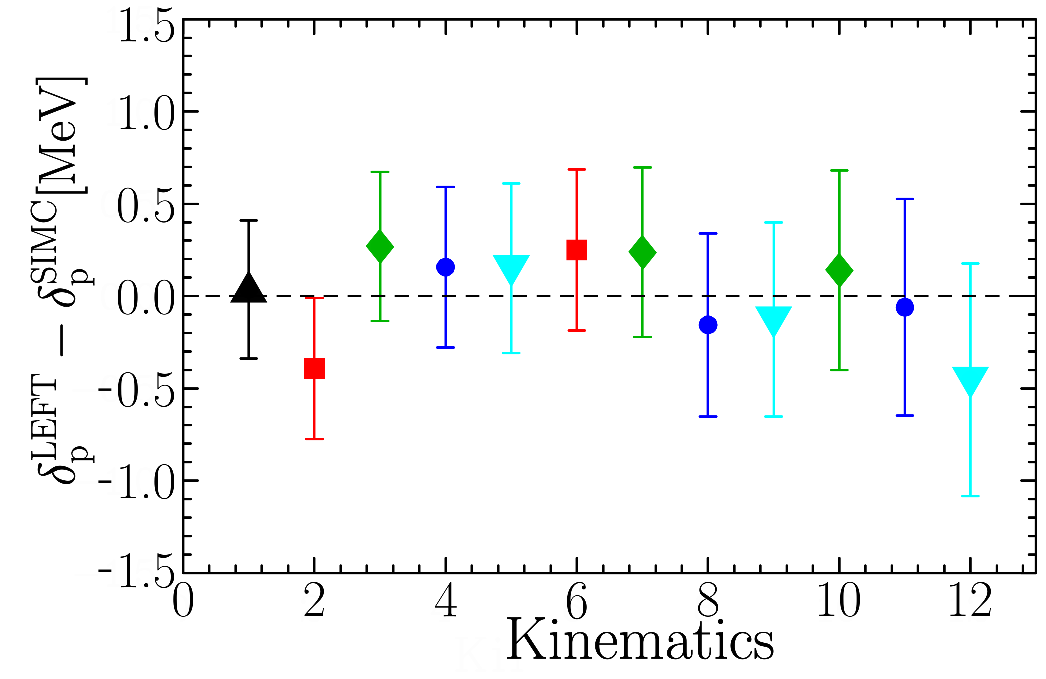}
\end{center}
\caption{(Color online) The elastic peak offset between data and simulations for the right (top) and left (bottom) spectrometers after applying a 0.19~mrad offset to the left arm. The points are sorted according to the left arm kinematics; 1--5 corresponds to $Q^2$ = 2.64~GeV$^2$, and 6--9 (10--12) correspond to 3.20 (4.10)~GeV$^2$. For each $Q^2$ value, the points are sorted by $\varepsilon$ (low to high).  Note that for the right arm only the first five measurements are truly independent; later points are additional measurements at the same kinematics and color coded by $\varepsilon$ to identify points at the same kinematics.}
\label{fig:deltap_angle_offset}
\end{figure}

We performed an additional check using the overconstrained kinematics for elastic scattering by checking the reconstructed $\Delta P$ spectrum for elastic events(Eq. (\ref{eq:delta_p})). The elastic peak position should be near $\Delta P=0$~MeV, with small corrections due to energy loss and radiative corrections, which are modeled in the Monte Carlo simulation program SIMC~\cite{makinsphd, oneillphd, mohring02}, described in Sec.~\ref{ep_simc}. We compare the elastic peak position in the measured $\Delta P$ spectrum to the simulated elastic peak. For the left arm, an overall angular offset of 0.19~mrad was applied to the pointing angles to best center the elastic peak position from data to that of simulations at each kinematics. Note that an offset of 0.28~mrad is needed to match the carbon-pointing angles to the survey ones. The two offsets are consistent within the 0.18~mrad scale uncertainty assigned in the scattering angle (discussed below). For the right arm, the survey angles are used and yielded a good $\Delta P$ peak position and no additional offset was needed. Figure~\ref{fig:deltap_angle_offset} shows the difference in the elastic peak position from data and that of simulations after applying the angular offset and for the left arm. The error bars assume random uncertainties of 0.10~mrad for the angle (see discussion below) and 0.02\% for the beam energy (see Sec.~\ref{expt_equip}). With these uncertainties, the values of $\delta_P$ from data are in good agreement with the simulations.

Table~\ref{tab:kinematics} shows the final beam energy and the spectrometer angle values used for the analysis. The kinematic settings go from $a$ to $r$, based on the chronological ordering of the runs; note that some letters are not used as only settings used for extraction of the elastic cross section are included and kinematics used for calibration purposes are not included.
A 0.10~mrad random uncertainty in the angular offset is assigned, with 0.07~mrad coming from beam angle drifts as determined from the BPMs, 0.05~mrad from the 0.1 MeV uncertainties in determining the peak position, and 0.05~mrad from the pointing determination due to run-to-run scatter and uncertainty in determining the target position. The scale uncertainty is taken to be 0.18~mrad, with 0.13~mrad due to energy loss and radiative effects on the elastic peak, 0.07~mrad from possible beam angle offset, and 0.10~mrad coming from the change in the best angle for a 0.03\% shift in the beam energy.

\begin{table}[!htbp]
\begin{center}
\caption{The beam energy, $E_b$, and spectrometer angles used in the analysis.
The right-arm settings $\theta_{R}$ are the survey angles, while
the left-arm settings $\theta_{L}$ are the angles as determined by
carbon-pointing runs measurements corrected by 0.19 mrad to center the elastic
peak. See text for details.}
\begin{tabular}{c c c c c c} \hline\hline
kin & $E_{\mbox{Tief}}$  & $\theta_{L}$ & $\theta_{R}$ & $Q_L^2$ & $\varepsilon_L$\\
           & (MeV)             & (deg)   & (deg)     & (GeV$^2$) &   \\
\hline
~$o$~      & ~1912.94~         & ~~12.631~~ & ~58.309~  & ~2.64~& ~0.117~  \\
$a$        & 2260.00           & 22.159     & 60.070    & 2.64  &  0.356  \\
$i$        & 2844.71           & 29.459     & 62.038    & 2.64  &  0.597  \\
$q$        & 3772.80           & 35.151     & 63.871    & 2.64  &  0.782  \\
$l$        & 4702.52           & 38.251     & 64.981    & 2.64  &  0.865  \\
\hline
$b$        & 2260.00           & 12.523     & 60.070    & 3.20  &  0.131  \\
$j$        & 2844.71           & ~23.390~   & 62.038    & 3.20  &  0.443  \\
$p$        & 3772.80           & 30.480     & 63.871    & 3.20  &  0.696  \\
$m$        & 4702.52           & 34.123     & 64.981    & 3.20  &  0.813  \\
\hline
$k$        & 2844.71           & 12.681     & 62.038    & 4.10  &  0.160  \\
$r$        & 3772.80           & 23.659     & 63.871    & 4.10  &  0.528  \\
$n$        & 4702.52           & 28.374     & 64.981    & 4.10  &  0.709  \\

\hline\hline
\end{tabular}
\label{tab:kinematics}
\end{center}
\end{table}

We estimated the impact of a 0.18~mrad shift and 0.10~mrad random variations on the extracted cross sections.  A 0.10 mrad angle fluctuation changes the cross section by (0.10--0.12)\% for the right arm and (0.02--0.10)\% for the left arm, applied as random uncertainties for each $\varepsilon$ point.  A 0.18~mrad shift to all angles for the right arm yields average scale, random, and slope variations of 0.20\%, 0.02\%, and 0.67\%, respectively, while for the left arm they yield average scale, random, and slope uncertainties of 0.13\%, 0.01\%, and 0.18\%, respectively.


\subsection{VDC Multiplicity Cuts} \label{vdcs_multiplicity_eff}

A typical single-track event making an angle of 45$^o$ with the VDC surface is expected to have a multiplicity (number of hit wires in the cluster of wires with a signal) of 4--6~\cite{fissum01, alcorn04}. Some of the good single-track events have fewer hits in the VDC cluster or have noise hits not associated with the true particles. This can lead to tracks that do not reproduce the true particle trajectory, introducing long tails in the distribution of the reconstructed physical quantities. If included in the analysis, such events will be lost in acceptance and elastic kinematic cuts, even though they may correspond to elastic events. To make the inefficiency caused by such events less sensitive to kinematic cuts, we choose to remove the events that yield these long tails by requiring one cluster of hits per plane with 3--6 hits per cluster.

\begin{figure}[!htbp]
\begin{center}
\includegraphics*[width=0.9\columnwidth]{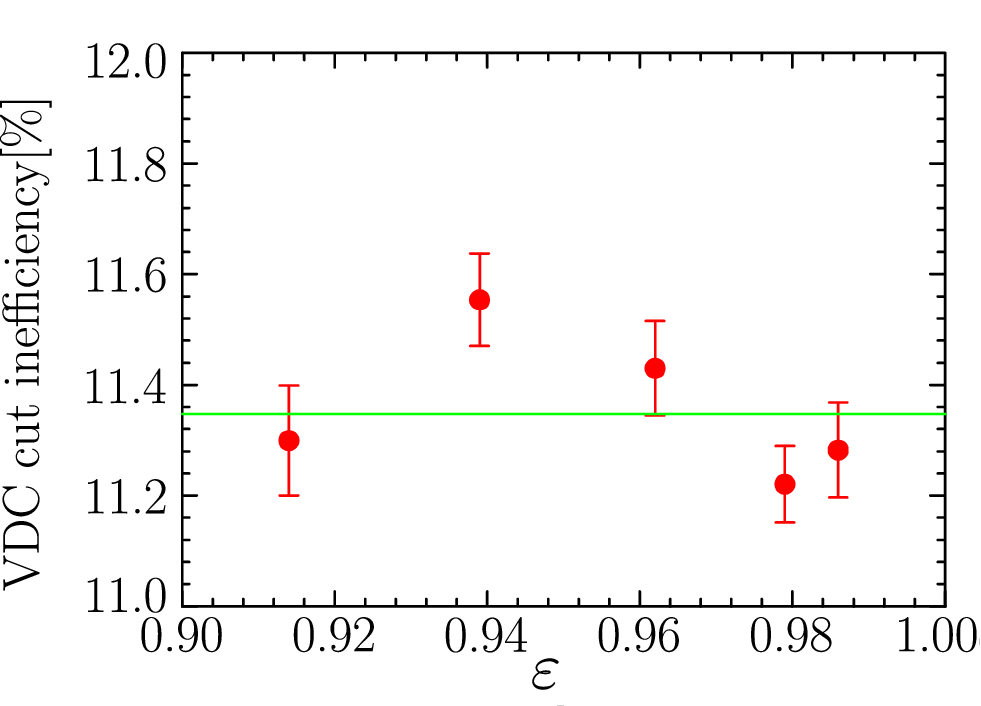}
\includegraphics*[width=0.9\columnwidth]{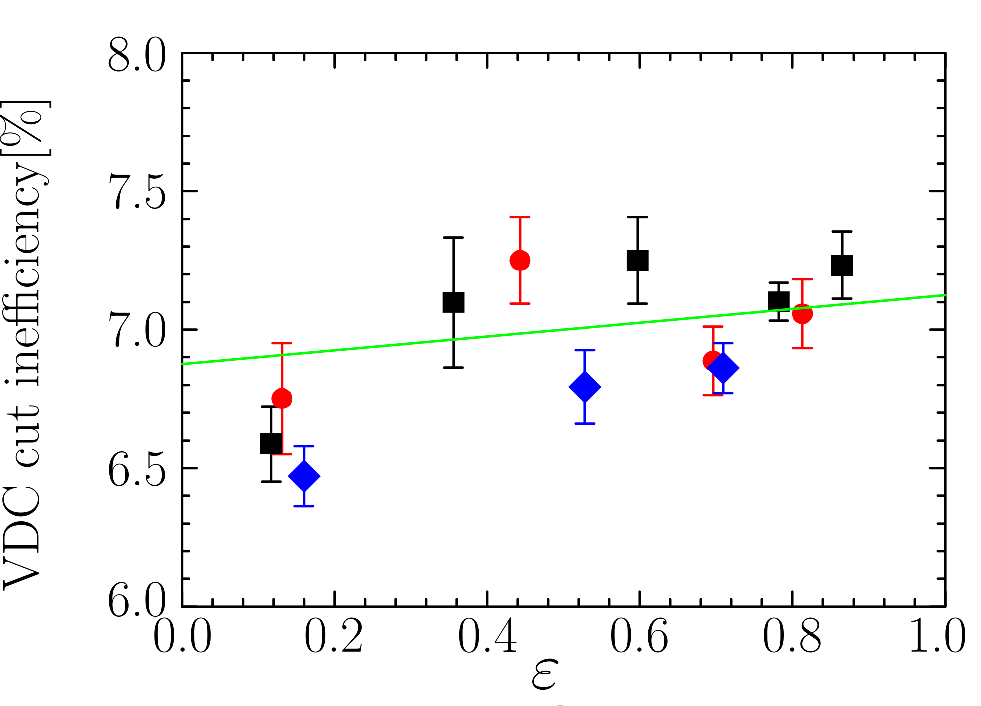}
\end{center}
\caption{(Color online) VDC cut inefficiency ($\epsilon_{\mbox{VDCH}}$) vs. $\varepsilon$ for all kinematics for the right (top) and left (bottom) HRS. For the left arm, the black, red, and blue points correspond to $Q^2$=2.64, 3.20, and 4.10~GeV$^2$, respectively.  The solid green line is the inefficiency used in the analysis: 11.35\% for HRS-R, 7.0\%+0.25\%($\varepsilon$$-$0.50) for HRS-L.}
\label{fig:left_vdc_multiplicity}
\end{figure}

Figure~\ref{fig:left_vdc_multiplicity} shows the inefficiency of the VDC multiplicity cuts applied for the right arm (top) and left arm (bottom) for all of the settings. This inefficiency represents the fraction of good events that were lost after applying the VDC multiplicity cuts to remove the long tails seen in the reconstructed kinematics. In determining the inefficiency, we applied a set of physics/data quality cuts to exclude junk events (see Sec.~\ref{recon} for a full list of cuts applied) and the results have some cut dependence. The right arm has an average efficiency of $\epsilon_{\mbox{VDCH}}$ = 0.8865 with no $\varepsilon$ dependence, and we apply scale (random) uncertainties of 0.5\% (0.1\%). The scale uncertainty accounts for the variation of the efficiency with different VDC cuts.

For the left arm, there is some apparent $\varepsilon$ dependence, yielding an average slope of (0.60$\pm$0.14)\%. For a fixed $Q^2$ value, the proton central momentum is the same at each $\varepsilon$ point, while the event rate, proton momentum range, and spatial distribution of the elastic events are very similar. As such, we do not expect to see a significant $\varepsilon$ dependence. While these small changes could introduce a $\varepsilon$ dependence, it could also be that the efficiency for good elastic events is constant, and the extracted $\varepsilon$ dependence changes due to the somewhat larger change between the elastic event rate, mainly the center of the detectors, and the rate of inelastic protons which populate the full acceptance. We take a conservative approach that should allow for either of these explanations. We apply an efficiency correction with a slope of (0.25$\pm$0.25)\% to allow for the possibility of some $\varepsilon$ dependence while also being only one standard deviation for the expected $\varepsilon$ independence. By applying this as a slope uncertainty, we allow for an $\varepsilon$ dependence that yields a larger impact on $\gegm$ than a simple uncorrelated error. In addition to the slope uncertainty, we apply scale and random uncertainties of 0.5\% and 0.1\%, respectively.


\subsection{VDC Tracking Efficiency} \label{vdcs_track_eff}

To estimate the tracking efficiency, we first define a sample of candidate events for which we believe a good track should be found. The candidate events must have hits in both VDC planes, 30 hits or less per VDC plane, and be in the central region of the detector, i.e. hit either paddle 3 or 4 in both the $S_1$ and $S_2$ scintillator planes. Events with zero or many ($>$30) wire hits in either VDC are excluded, as they are not expected to yield good tracks: having no hits in a chamber indicates that the event missed the chamber, while events with $>$30 hits in one VDC are almost always caused by the incoming high-energy particle hitting an aperture near the end of the magnets and generating a shower of low-energy particles in the chambers. We also apply particle identification cuts using A$_1$ and A$_2$ aerogels to exclude $\pi^{+}$ from the proton tracking efficiency measurement. See Sec.~\ref{pid_cuts} for a more detailed description of the PID cuts. 

We then determine the fraction of these that have a single track, as we reject events with zero tracks or with multiple tracks. The tracking efficiency is $\epsilon_{\mbox{VDC}} = N_{\mbox{one}} / (N_{\mbox{zero}} + N_{\mbox{one}} + N_{\mbox{mult}})$ where $N_{\mbox{zero}}$, $N_{\mbox{one}}$, and $N_{\mbox{mult}}$ are the number of candidate events with zero, one, or multiple tracks, respectively. The efficiency is calculated and applied to each run separately. 

The zero-track inefficiency for the left arm is 0.05--0.20\%, varying with $Q^2$ but not $\varepsilon$, and a 0.1\% scale uncertainty is applied. For the right arm, the inefficiency is smaller, and the uncertainty is taken to be negligible. The multiple-track inefficiency for the left arm is 0.15--0.30\% with a small $Q^2$ dependence and no $\varepsilon$ dependence. We apply a 0.1\% scale uncertainty and no slope uncertainty. The multiple-track inefficiency for the right arm, for which the event rate is much higher, is 0.60--1.30\%, varying mainly with the event rate in the spectrometer, but also showing a 0.05\% variation over the $\varepsilon$ range. We assume a 0.1\% scale uncertainty and a 0.7\% slope uncertainty, based on the 0.05\% variation over a $\Delta \varepsilon$ range of 0.07. Run-to-run variations were used to estimate a random uncertainty of 0.02\% for the left arm, and below 0.01\% for the right arm.


\subsection{Scintillator Efficiency} \label{scint_effic}

Scintillator inefficiency can cause loss of good events if a physics trigger, requiring both scintillator planes, is not formed. The inefficiency is estimated by examining events that did not yield a primary main physics trigger T$_1$ or T$_3$, but still yielded a signal in the corresponding efficiency trigger (T$_2$ or T$_4$). We examine single-track events that fell inside the scintillator boundaries by projecting the track to the scintillator plane and excluding events that missed the detectors. We also apply the usual VDC multiplicity cuts, 1.6~msr solid angle cut, and particle identification cuts to reject pions. Each scintillator plane efficiency $\epsilon_{S1,S2}$ was calculated using
\begin{equation} \label{eq:scinteff}
\epsilon_{S1(S2)} = \frac{N_{1(3)} + N_5}{N_{1(3)} + N_5 + N_{2(4)}},
\end{equation}
where $N_i$ ($i = 1,\cdots,5$) is the number of events of trigger type $i$, corrected for prescaling factor and electronic and computer deadtimes, where a track fell inside the scintillator boundaries as defined by Table~\ref{focal_plane_cuts}.  Note that the efficiency trigger requires a hit in $S_1$ or $S_2$ but not both, and the primary and efficiency trigger will never fire for the same event.

\begin{table}[!htbp]
\begin{center}
\caption{Cuts applied to define the scintillator fiducial volume. The x and y values at the VDC1 and VDC2 are determined by projecting the track from the focal plane to z=1.381~m and 3.314~m for the HRS-R and z=1.287 and 3.141~m for the HRS-L.}
\begin{tabular}{c c c}
\hline\hline
Arm~~            & S$_1$ Plane Boundary       &~~ S$_2$ Plane Boundary \\
\hline
Right~~          & $-1.05< x_{S1} <0.90$ &~~ $-1.30< x_{S2} <1.00$ \\
                 & $-0.18< y_{S1} <0.18$ &~~ $-0.32< y_{S2} <0.32$ \\
\hline
Left~~           & $-1.05< x_{S1} <0.90$ &~~ $-1.30< x_{S2} <1.00$ \\
                 & $-0.18< y_{S1} <0.18$ &~~ $-0.32< y_{S2} <0.32$ \\
\hline\hline
\end{tabular}
\label{focal_plane_cuts}
\end{center}
\end{table}

The total scintillator efficiency for any run is the product of the two scintillator efficiencies. The total scintillator efficiency, typically $\geq$ 99.5\%, was calculated and applied for each run. The efficiency is nearly constant, except for the HRS-L at the initial kinematics, where a tilted scintillator paddle led to a small gap and reduced the average efficiency by 0.15\%. However, this gap was far from the position of the elastic peak on the detectors, and so the efficiency was increased from the measured value by 0.15\%. For both arms, we estimate the random and scale uncertainties to be 0.05\% and 0.10\%, respectively, with no slope uncertainty.


\subsection{Particle Identification Efficiency} \label{pid_cuts}

Particle identification (PID) cuts are needed to obtain a clean proton sample and the efficiency of these cuts as well as any misidentification of other particles as protons must be determined. The PID cuts and their efficiencies for both spectrometers are discussed below.


\subsubsection{The efficiency of the right arm $\beta$ cut} \label{beta_eff}

\begin{figure}[!htbp]
\begin{center}
\includegraphics*[width=0.95\columnwidth]{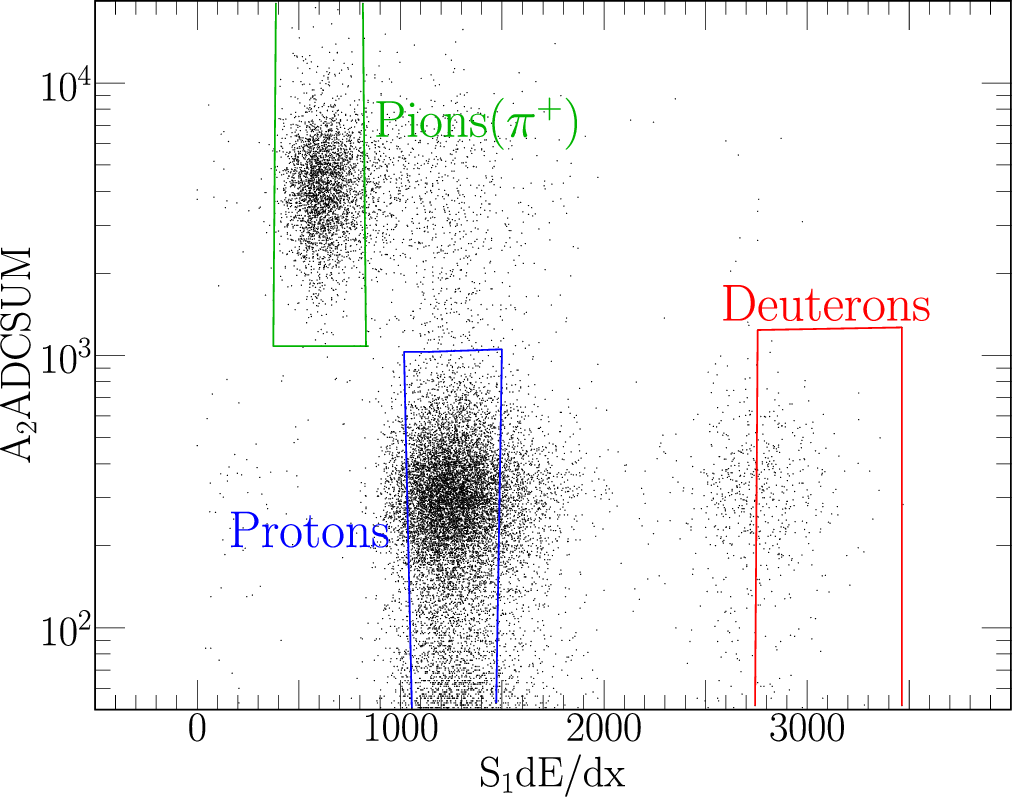} 
\end{center}
\caption{(Color online) Aerogel signal (A$_2$ ADCSUM) vs energy loss in S1 for run 1730 from kinematics $n$ (Tab.~\ref{tab:kinematics}). Clean samples of pions, protons, and deuterons are selected by applying the cuts shown. The deuteron cut is offset to reduce proton contamination since the efficiency of this cut is not important as long as a clean and unbiassed sample of deuterons is obtained.}
\label{fig:a2_s1dedx}
\end{figure}

\begin{figure}[!htbp]
\begin{center}
\includegraphics*[width=0.95\columnwidth]{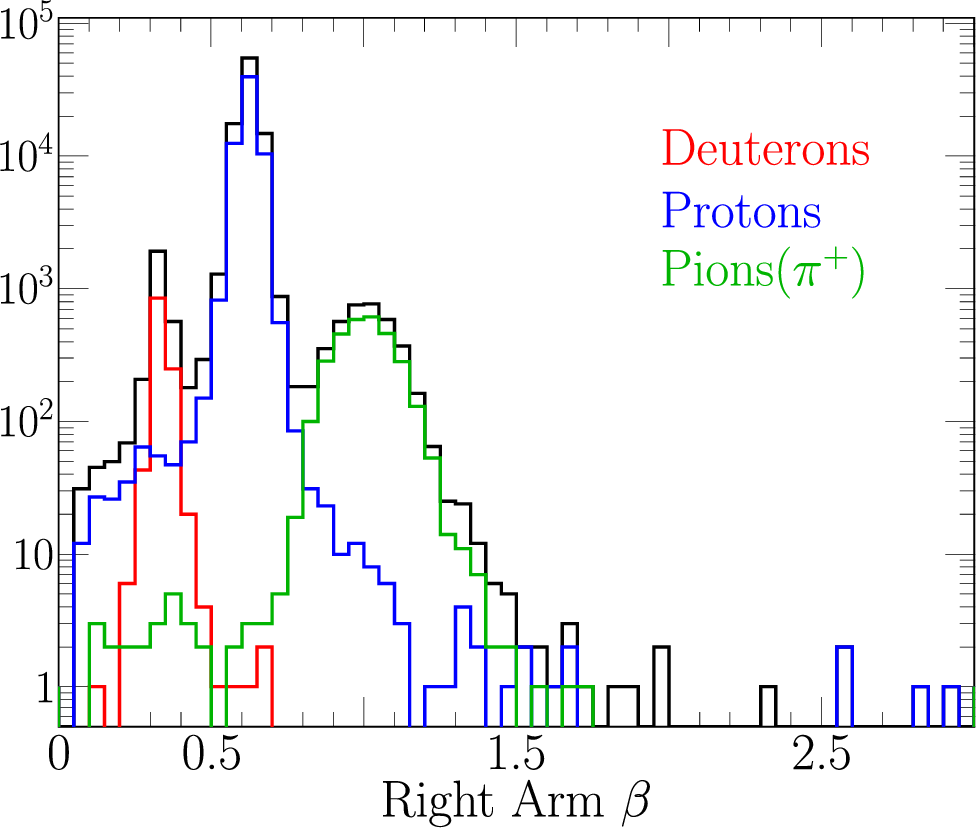} 
\end{center}
\caption{(Color online) The contribution of pions (green), deuterons (red), and protons (blue) to the full right arm $\beta$ spectrum (solid black line).}
\label{fig:right_beta}
\end{figure}

To determine the efficiency of the $\beta$ cut that we are applying, we first select clean samples of pions, protons, and deuterons using the A$_2$ aerogel and energy deposition ($dE/dx$) in the first scintillator plane, as shown in Fig.~\ref{fig:a2_s1dedx}. Figure~\ref{fig:right_beta} shows the $\beta$ spectrum for each particle type. From this, we can determine the fraction of protons lost to the $\beta$ cut applied to identify protons along with any contamination of pions or deuterons.

\begin{table}[!htbp]
\begin{center}
\caption{The incident energy, deuteron contamination, pion contamination, and proton efficiency for the $\beta$ cut.}
\begin{tabular}{c c c c}
\hline \hline
Beam Energy &  Max. Deuteron    & Max. Pion      & Proton   \\
(MeV)           &  Contam.(\%)  & Contam.(\%)    & Effic. ($\epsilon_{\beta}$)\\
\hline
1912.94         & 0.000            & 0.229        & 0.993     \\
2260.00         & 0.010            & 0.189        & 0.994     \\
2844.71         & 0.009            & 0.182        & 0.993     \\
3772.80         & 0.012            & 0.194        & 0.993     \\
4702.52         & 0.014            & 0.206        & 0.992     \\
\hline\hline
\end{tabular}
\label{beta_effic}
\end{center}
\end{table}

For each kinematic setting, the $\beta$ spectrum for each particle type is generated and a $\beta$ cut to select protons is chosen. The proton efficiency, $\epsilon_{\beta}$, is taken as the fraction of the proton $\beta$ spectrum within the $\beta$ cut. For the pions and deuterons, the fraction of the $\beta$ spectrum within the proton window is taken as the fractional misidentification, and used to estimate the contamination of the final proton sample, after correcting for the total number of deuterons and pions in the sample. This analysis was done for multiple runs at all 5 beam energies, and the results are summarized in Tab.~\ref{beta_effic}. The deuteron contamination is negligible while the pion contamination is small after the $\beta$ cut, and is negligible after applying the aerogel cut.

Note that $\epsilon_{\beta}$ was determined assuming that the proton $\beta$ spectrum represents a pure proton sample. However, any contamination from pions or deuterons to the proton sample in Fig.~\ref{fig:a2_s1dedx} will decrease the extracted efficiency by adding events to the tails of the reference proton beta spectrum. We estimate the possible pion contamination in the proton sample of Fig.~\ref{fig:right_beta} by scaling down the pion peak until it is consistent with the proton spectrum in the high-$\beta$ region. This assumes that all of the events in the proton tail at the position of the pion peak come from pions, giving an upper limit to the error made in extracting the proton efficiency. The same procedure is used for deuterons. We use this upper limit to estimate the uncertainty in the $\beta$ cut efficiency. The same procedure is used to determine the uncertainty in $\epsilon_{A_2}$, extracted in the next section.

\subsubsection{The efficiency of the right arm A$_2$ aerogel cut} \label{A2_eff}

\begin{figure}[!htbp]
\begin{center}
\includegraphics*[width=0.95\columnwidth]{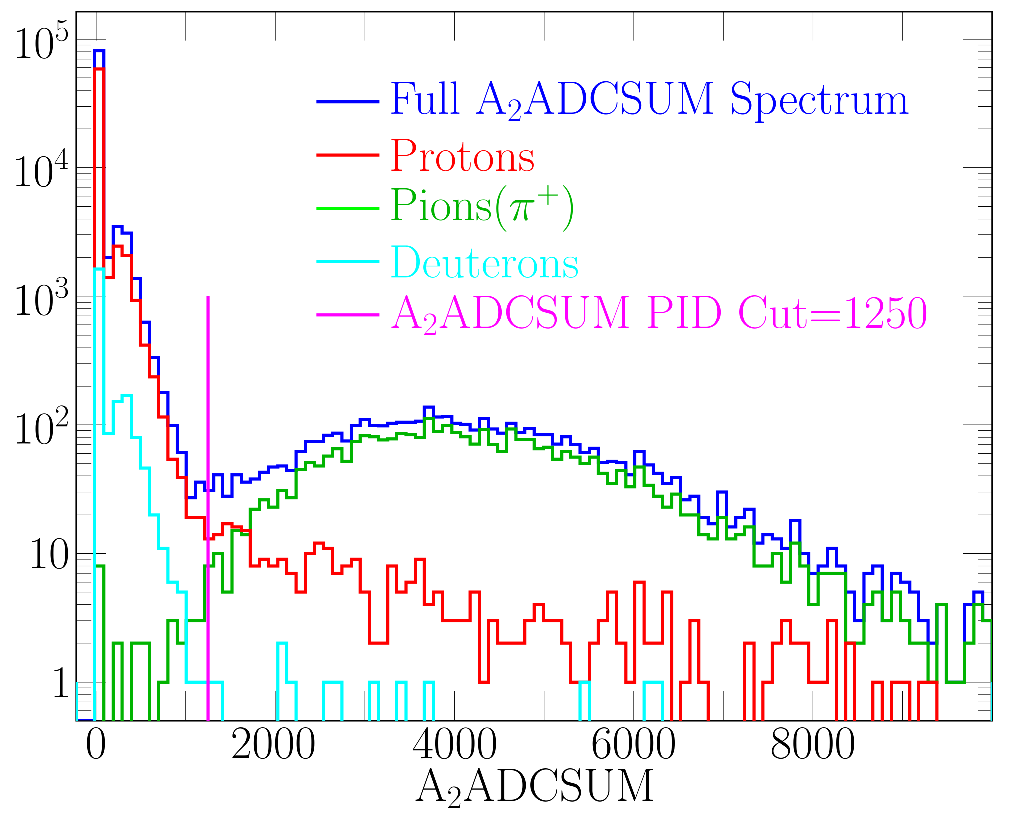} 
\end{center}
\caption{(Color online) The pedestal-subtracted A$_2$ ADCSUM spectrum (blue). Also shown the deuterons (cyan), protons (red), and pions ($\pi^+$) (green) contribution to the full A$_2$ ADCSUM signal for run 1730 from kinematics $n$  (Tab.~\ref{tab:kinematics}). The magenta line at A$_2$ ADCSUM = 1250 is the PID cut used.}
\label{fig:ra2_adcsum}
\end{figure}

For all of the settings shown in Table~\ref{beta_effic}, we generate the A$_2$ ADCSUM spectra for pions, protons, and deuterons using tight cuts on $\beta$ and $S_1~dE/dx$ as shown in Table~\ref{beta_s1dedx}. Figure~\ref{fig:ra2_adcsum} shows the full A$_2$ ADCSUM spectrum and its constituents.

\begin{table}[!htbp]
\begin{center}
\caption{The $\beta$ and $S_1~dE/dx$ cuts used for the A$_2$ efficiency analysis.}
\begin{tabular}{c c c}
\hline\hline
Particle~~    & ~~$\beta$ Range      & ~~$S_1~dE/dx$ Range\\
\hline
Deuterons     & ~~0.20$<\beta<$0.40  & ~~2500$<S_1~dE/dx<$3500\\
Protons       & ~~0.60$<\beta<$0.70  & ~~1000$<S_1~dE/dx<$1800 \\
Pions         & ~~0.80$<\beta<$1.20  & ~~400$<S_1~dE/dx<$900   \\
\hline\hline
\end{tabular}
\label{beta_s1dedx}
\end{center}
\end{table}

\begin{table}[!htbp]
\begin{center}
\caption{The pion contamination and proton efficiency as determined using the A$_2$ ADCSUM cut.}
\begin{tabular}{c c c c}
\hline\hline
Run         & Incident Energy& Pion          & Proton   \\
Number      & (MeV)          & Contamination & Efficiency \\
            &                & (\%)          & ($\epsilon_{A_2}$) \\
\hline
1772  & 1912.94        & 0.040        & 0.996        \\
1252  & 2260.00        & 0.041        & 0.996     \\
1653  & 2844.71        & 0.035        & 0.995      \\
1823  & 3772.80        & 0.040        & 0.995         \\
1730  & 4702.52        & 0.050        & 0.995       \\
\hline\hline
\end{tabular}
\label{ra2_effic}
\end{center}
\end{table}

We exclude events with A$_2$ ADCSUM $>$ 1250 to provide further suppression of the pions. The efficiency of this cut for protons is given in Tab.~\ref{ra2_effic}, along with the fractional contamination of pions after applying only the aerogel cut. The pion rejection from the aerogel, combined with the $\beta$ cut, yields negligible pion contamination in the final proton sample. This can be seen in Fig.~\ref{fig:ra2_adcsum}, where the green spectrum represents the ADCSUM spectrum for pions, which has approximately 30 events below the cut at 1250, compared to the $>$60k protons. While this is only an estimate, as the tight cuts used to identify the clean pion and proton samples have 20-30\% inefficiencies, it is well below 0.1\% and will be further suppressed by the $\Delta P$ cuts.

The final proton efficiency is the product of the proton efficiency from the $\beta$ cut and the A$_2$ ADCSUM cut. This efficiency is fairly constant, ranging from (98.75--99.03)\%, as expected since all data are taken at a fixed proton momentum. The efficiency of the aerogel cut is typically $\geq$ 99.4\% with random fluctuations of 0.05\%. We estimate the error in the extracted efficiency due to contamination of our clean proton sample following the same approach as for the $\beta$ cut efficiency. The possible impact of contamination is essentially independent of beam energy, and an overall 0.25\% scale uncertainty was assigned.


\subsubsection{The efficiency of the left arm A$_1$ aerogel cut} \label{A1_eff}

The PID efficiencies are more difficult to determine in the left arm, as the higher proton momenta make the $\beta$ and $dE/dx$ cuts less effective. In this case, we use the coincidence runs with protons in the left arm to generate a pure proton spectrum for A$_1$ ADCSUM. The total number of events in the A$_1$ ADCSUM spectrum as well as the number below and above A$_1$ ADCSUM = 350 (the pion rejection cut used for the left arm) was determined for the pure proton sample for each coincidence setting. The two $Q^2$ = 2.64~GeV$^2$ settings yielded consistent proton inefficiencies, (1.037$\pm$0.029)\% and (1.030$\pm$0.013)\%, while the $Q^2$ = 4.10~GeV$^2$ setting had an inefficiency of (1.940$\pm$0.054)\%. For $Q^2 =$ 3.20~GeV$^2$, we did not have coincidence data. Instead, we generate the proton A$_1$ spectrum by applying kinematics cuts on the elastic peak ($\Delta P$, $-$15.5$<\Delta P<$30.5 MeV) and the target position ($-$0.0044$<y_{\mbox{tg}}<$$-$0.001~m) to eliminate the endcap scattering. This yielded a proton inefficiency of (1.550$\pm$0.049)\%.

\begin{figure}[!htbp]
\begin{center}
\includegraphics*[width=0.95\columnwidth]{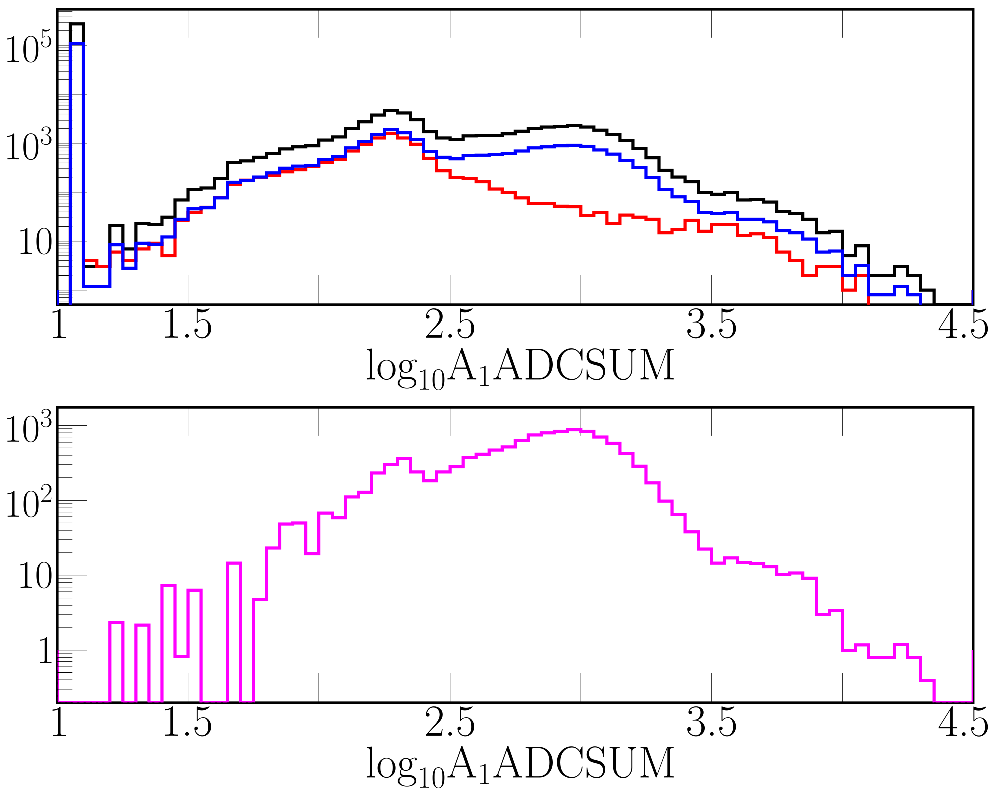} 
\end{center}
\caption{(Color online) The A$_1$ ADCSUM spectra from singles kinematics $n$ from Tab.~\ref{tab:kinematics} (top black), coincidence kinematics coin1 (top red), and scaled kinematics $n$ (top blue). The difference between the A$_1$ ADCSUM spectrum from the scaled kinematics $n$ and the coin1 spectrum (blue minus red) is the pions spectrum (bottom magenta).  Note that the cut at 350 channels corresponds to log$_{10}$(ADCSUM)=2.544.}
\label{fig:pions_shape}
\end{figure}

We also need to determine the $\pi^+$ contamination. To do that, we need the number of $\pi^+$ and fraction of $\pi^+$ that are identified as protons ($\pi^+$ efficiency). We compared the A$_1$ ADCSUM spectrum from trigger type T$_3$ events (LH2 singles events, which include both protons and $\pi^+$) with that of the trigger type T$_5$ events (LH2 coincidence events, which have only protons). The difference in the spectra should represent the A$_1$ ADCSUM spectrum for the background $\pi^+$. When we compare the two A$_1$ ADCSUM spectra, we scale the A$_1$ ADCSUM spectrum for the singles events so that the two peaks match. Figure~\ref{fig:pions_shape} illustrates this procedure where the A$_1$ ADCSUM spectrum with trigger type T$_3$ events is compared with that of trigger type T$_5$ events. The peak of the A$_1$ ADCSUM spectrum from singles was scaled to match that of coincidence. The difference between the A$_1$ ADCSUM spectrum from the scaled singles and the coincidence spectrum should represent the A$_1$ ADCSUM spectrum for pions. Having determined the pion efficiency and proton inefficiency, the pion contamination is then calculated for each kinematics based on the A$_1$ ADCSUM spectrum, i.e. the number of pions below the A$_1$ ADCSUM cut in the bottom panel of Fig.~\ref{fig:pions_shape} relative to the number of protons below the cut in the upper panel.

For the proton inefficiency, we take the average values: 1.03\% at $Q^2$ = 2.64~GeV$^2$ and 1.88\% at $Q^2$ = 4.10~GeV$^2$. For $Q^2$ = 3.20~GeV$^2$, the average of these two values (1.45\%) was taken, which was consistent with the attempt to estimate the inefficiency from the proton singles data. Based on the scatter of the extracted efficiencies for different runs and different methods, we assign a 0.2\% scale uncertainty, 0.1\% random uncertainty, and no slope uncertainty for the A$_1$ PID efficiency.

The fraction of pions misidentified as protons for $Q^2$ = 2.64 (4.10)~GeV$^2$ was found to be below 5.5\% (11\%). We use these upper limits (and assume 11\% for 3.20~GeV$^2$) to estimate the maximum pion contamination.  With these misidentification probabilities, the pion contamination after the aerogel cut is always below 1.0\%. Applying kinematics cuts and subtracting the target endcaps further reduces the pion contamination as most of the pions are generated by Bremsstrahlung scattering in the endcaps and have a much broader kinematic distribution. Thus, pion contamination in the final results, after kinematic cuts and the dummy subtraction, is always below 0.1\%.

\subsection{Proton Absorption} \label{proton_absorp}

The struck protons have to pass through material in the target, spectrometer, and the detector stack before they can generate a physics trigger. Some protons will undergo nuclear interactions and be absorbed or scattered such that they do not yield triggers. We account for absorption in all materials the protons pass through on their way to the scintillators~\cite{alcorn04, schultethesis, lingyanzhuthesis}. We estimate the absorption using $\bar{\lambda}$ (effective absorption length), which depends on the mean free path between nuclear collisions (total interaction length, $\lambda_T$), and the mean free path between inelastic interactions (inelastic interaction length, $\lambda_I$)~\cite{hagiwara02, eidelman04}. In this analysis, $\bar{\lambda}$ was determined using two different definitions making different assumptions about the impact of elastic scattering: $\bar{\lambda}=(\lambda_T+\lambda_I$)/2, averaging the two contributions, and $\bar{\lambda}=2\lambda_T\lambda_I/(\lambda_T+\lambda_I)$ assuming that half of the elastic and inelastic scattering contribute to the absorption. 

The ratio of $X/\bar{\lambda}$ is calculated for each absorber using both estimates of $\lambda$, where $X$ is the product of the material density and thickness. The ratios are then added, i.e, $\sum_{i=1}^n (X_i/\bar{\lambda_i})$ where $i$ runs over all absorbers $n$. The proton absorption is given by
\begin{equation}
\mbox{proton absorption} = 1.0 - e^{-\sum_{i=1}^n (X_i/\bar{\lambda_i})},
\end{equation}
where the full list of materials is given in Ref.~\cite{qattanphd} The final proton absorption used is the average value of the calculated proton absorption from the two definitions, yielding 5.19\% for the right arm and 4.91\% for the left arm, giving a proton absorption correction of $C_{\mbox{Absorption}}$ = 0.948 and 0.951 for the right arm and left arm, respectively, with a scale uncertainty of 1\%. Because the scattered proton will travel different distances through the LH2 target at each angle, we include a slope uncertainty in this correction of 0.03\% and 0.10\% for the left and right arm, respectively, and apply no random uncertainty.

\subsection{Target Length Correction} \label{target_leng_corr}

\begin{figure}[!htbp]
\begin{center}
\includegraphics*[width=0.98\columnwidth,trim={3mm 32mm 4mm 22mm},clip]{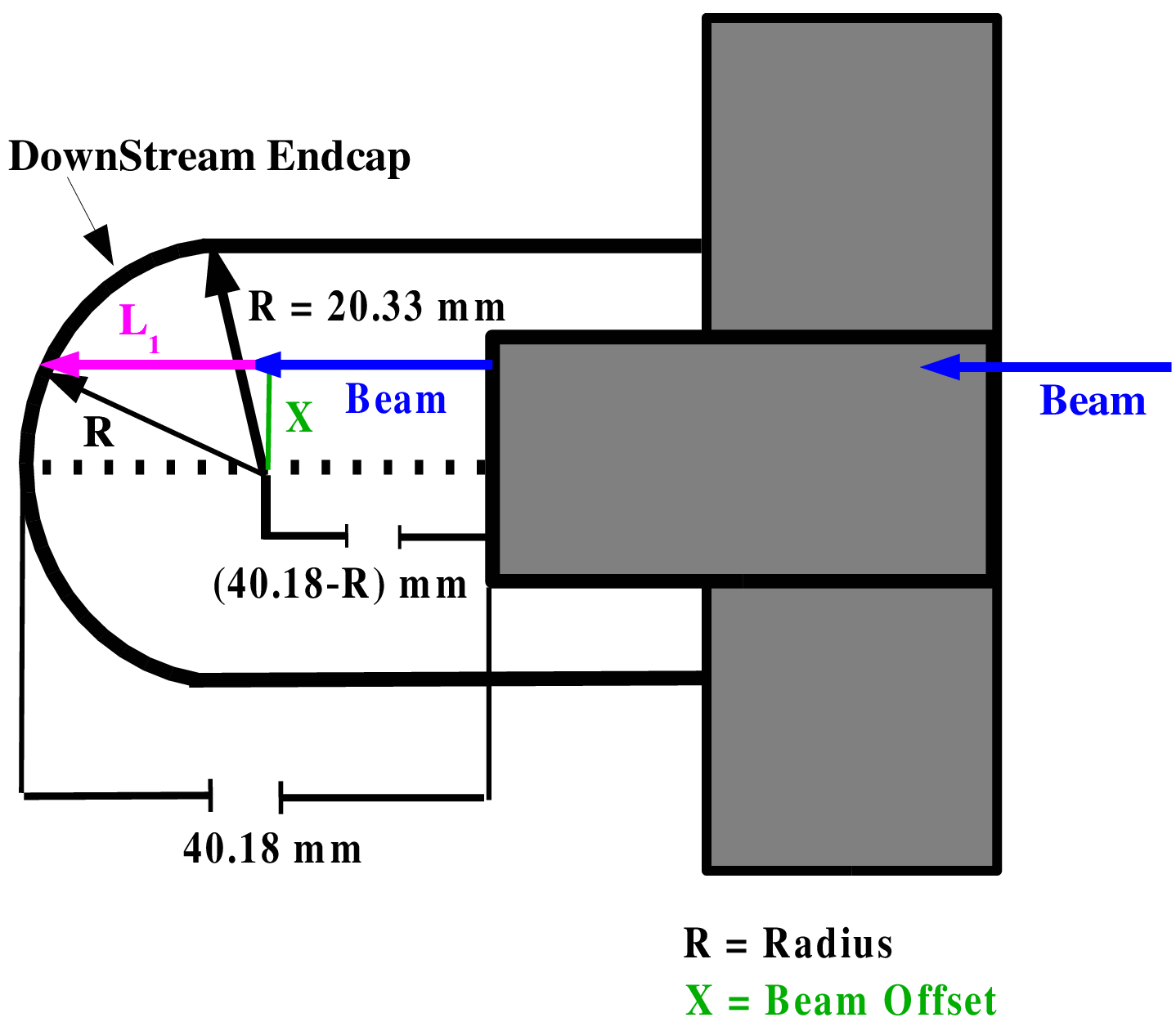}
\end{center}
\caption{(Color online) Geometry of the 4-cm LH2 cell. The electron beam is going from the right to the left side with an offset X from the central axis of the cell (black dashed line).}
\label{fig:target_offset}
\end{figure}

Figure~\ref{fig:target_offset} shows the geometry of the 4-cm LH2 cell used in the experiment. The cell wall is made of 0.14~mm Al. The upstream endcap (not shown) is the beam entrance window and is made of Al 7075 T6 with 0.142~mm thickness, while the downstream endcap window is uniformly machined Al in the shape of a hemisphere and has a thickness of 0.15~mm and radius R = 20.33~mm. The length of the central axis of the cell is 40.18~mm (black dashed line). The 4~cm dummy target used is made of Al 6061 T6 with density 2.85~g/cm$^3$. The thickness is 0.2052~g/cm$^2$ for the upstream foil and 0.2062~g/cm$^2$ for the downstream one. The distance between the two foils is (40$\pm$0.13)~mm.

If the beam (blue arrow) is displaced a distance $\mbox{X}$ (green line) from the central axis, then the length that the beam will travel in the target is L = L$_1$ + (40.18 - R)~mm where L$_1<$ R and is shown as a magenta arrow. This length is effectively the new target length. A survey showed that the target was displaced down (2.0$\pm$0.5)~mm relative to the center of the beam. Furthermore, the beam was rastered to produce a 2~mm $\times$ 2~mm spot. Therefore, the effective target length is the average beam path length in the target, after applying the beam position offset and averaging over the raster spot size.  The 2~mm offset yields a 0.266\% reduction in the target length which is applied as a correction to the target thickness, and the change in target length for a 0.5~mm shift from the central position is 0.12\%, which we take as the scale uncertainty. The average beam position drift is approximately 0.30~mm (Sec.~\ref{sec:beam_position}), contributing a 0.07\% point-to-point uncertainty.


\subsection{Computer and Electronics Deadtime} \label{comp_elec_deadtime}

Electronic deadtime in the trigger logic yields loss of events if the trigger electronics are still processing a signal for one trigger when the next event occurs. The trigger modules are busy for a latency time $\tau \approx ($80--120)~ns processing hits from one event, yielding an electronic livetime of $e^{-r\tau}$ where $r$ is the event rate. This is calculated based on the measured rates and is applied as a correction on a run-by-run basis. For the HRS-L, the rates are below 1~kHz, yielding a correction well below 0.1\% and a negligible uncertainty. For the HRS-R, the deadtime is slightly larger, up to 0.3\% at 30~kHz. The typical deadtime is 0.2\%, and we use the 20\% variation in trigger latency time in different detector elements and estimate an uncertainty of 0.04\% for each setting, and a possible 0.04\% variation over the $\varepsilon$ range due mainly to small rate variations, yielding a slope uncertainty of 0.5\%.

Computer deadtime occurs when the DAQ system is unable to record an event because it is busy recording another event. The combination of this deadtime and (known) prescaling factor is directly measured by counting the number of triggers generated in a scaler and comparing this to the number of events recorded to disk. The deadtimes are 10--20\% for the HRS-R and 2--16\% for the HRS-L and are applied on a run-by-run basis. The computer deadtime effect is well measured and known to better than 1\%. We apply a scale uncertainty of 0.10\% for both arms, and a slope uncertainty of 0.10\% for the left arm, and 1.0\% for the right arm (taking the $<$0.10\% shift over a $\Delta\varepsilon$ window of 0.07).


\subsection{Target Boiling Correction} \label{tgt_boiling}

The LH2 target density $\rho_0$ can decrease due to localized density variations caused by the energy deposition of the beam. Runs were taken at several beam current values using LH2 and carbon targets. We take the normalized yield, $Y$, to be the total number of events $N$ normalized to the effective beam charge $Q_{\mbox{eff}}$ from Eq.~\ref{eq:qeff}. The normalized yield for the LH2 target decreased linearly with increasing beam current. The target density $\rho$ was parameterized as
\begin{equation} \label{yield_boiling2}
\rho(I) = \rho_0 \cdot (1 - B I),
\end{equation}
where $\rho_0$ = $\rho(I = 0)$ and $B$ is the current dependence.

We extract $B$ by looking at the charge-normalized scaler rates, corrected for current-independent rate (cosmic ray triggers), versus beam current. For carbon, the slope is 0.32$\pm$0.32\%/100$\micro$A, consistent with zero as it should be. For LH2, the slope is (1.38$\pm$0.15)\%/100$\micro$A, but there is a slight nonlinearity at lower current. This can be caused by a small offset in the beam current measurement, but can also be caused by nonlinearity in the BCM calibration at very low current or uncertainty in the correction for cosmic trigger rate, so this may be a slight overestimate. However, the total rate will give a slight underestimate, as the normalized yield from endcap scattering does not depend on current. A similar analysis, using the full elastic cross section analysis, yields a somewhat larger current dependence but is still consistent with the scaler-based analysis within its larger uncertainties. While the elastic analysis avoids some of the systematics associated with the scaler analysis, it is not sufficient to serve as a primary measurement due to lack of statistics. 

\begin{figure}[!htbp]
\begin{center}
\includegraphics*[width=0.95\columnwidth]{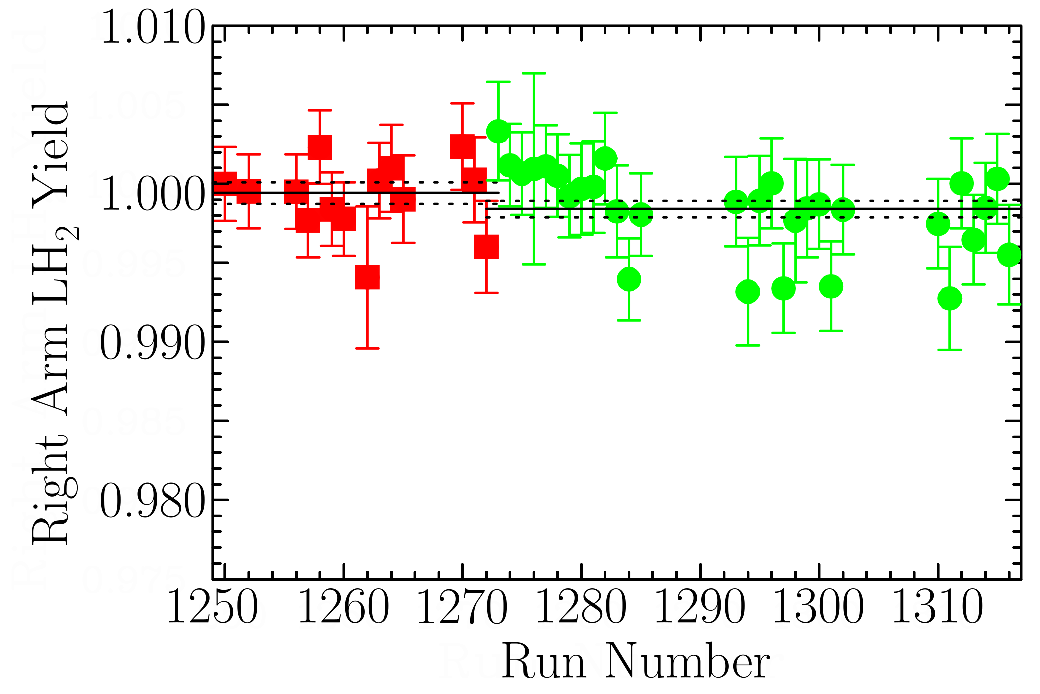}
\end{center}
\caption{(Color online) The right arm normalized yield for elastic events after applying the target boiling correction for all the runs at 30$\micro$A (1250--1271, shown in red) and 50$\micro$A (1272--1316, shown in green). The 0.5\% difference in the boiling corrections brings the low and high current running into good agreement. The solid and dashed black lines indicate the mean and uncertainty of the yields at 30 and 50$\micro$A.}
\label{fig:rightkinab_boil}
\end{figure}

As a final, high-statistics check of this correction, we can examine runs from kinematics $a$ (Tab.~\ref{tab:kinematics}) where some of the data were taken at 30$\micro$A and some at 50$\micro$A. Looking at the high-statistics data in the right arm, we find a correction of (3.0$\pm$0.5)\% per 100$\micro$A, once again larger than the scaler-based analysis. Therefore, we apply a final correction of (2.5$\pm$1.5)\%/100$\micro$A, with the large uncertainty accounting for the somewhat different results from different extractions. Figure~\ref{fig:rightkinab_boil} shows the consistency of the data at 30 and 50$\micro$A before and after applying this correction.  We apply an overall normalization uncertainty of 1.05\%, corresponding to the uncertainty in the correction at 70$\micro$A, and apply an additional correction to the runs taken at lower current. The corrections and uncertainties are summarized in Tab.~\ref{tgt_boiling_corr}. We also assign a 0.1\% random uncertainty to account for the fact that the raster size can vary by (5--10)\% from the nominal value at different beam energies. Note that while rate-dependent corrections can introduce a current-dependent effect if they are not correctly determined, the final test in Fig.~\ref{fig:rightkinab_boil} includes the full set of corrections used in the analysis. In addition, data were also taken on carbon at a range of beam currents during the boiling studies, and these did not show any current (rate) dependence.

\begin{table}[!htbp]
\begin{center}
\caption{The target boiling correction $C_{\mbox{TB}}=\rho(I)/\rho_0$ and its uncertainty for different beam currents. Note that $\delta C_{\mbox{TB}}$(relative) is the uncertainty relative to data at 70$\micro$A.}
\begin{tabular}{c c c c}
\hline \hline
Kinematics & $I(\micro$A) &$C_{\mbox{TB}}$ $\pm$ $\delta C_{\mbox{TB}}$&$\delta C_{\mbox{TB}}$(relative) \%\\
\hline
a (run $\leq$ 1269)  & 30  &  0.9925$\pm$0.0045 &$\pm$ 0.60\\
a (run $>$ 1269)     & 50  &  0.9875$\pm$0.0075 &$\pm$ 0.30\\
b                    & 50  &  0.9875$\pm$0.0075 &$\pm$ 0.30\\
i-r                  & 70  &  0.9825$\pm$0.0105 &$\pm$ 0.00\\
\hline \hline
\end{tabular}
\label{tgt_boiling_corr}
\end{center}
\end{table}

\subsection{Event Reconstruction and Cuts} \label{recon}

For each run, Q$_{\mbox{eff}}$ is calculated from the measured beam current, target thickness, efficiency, and livetime. The detectors are decoded for each event, tracking and particle identification information are reconstructed, and cuts are applied to select protons associated with elastic scattering. Events that satisfy these cuts are accepted for the final data analysis.

For the main analysis, we select only events with triggers T$_1$ and T$_3$, the main physics triggers for the HRS-R and HRS-L.  We require clean events in the chamber; one cluster of 3--6 wires per VDC plane, with only one track found.  We also apply PID cuts using the aerogel detectors and time-of-flight information to select protons.

Before track-based cuts are applied, we apply a correction to the reconstructed momentum, $\delta$. A small out-of-plane angle dependence was observed for both arms, appearing as a dependence of the position of the elastic peak as a function of $\theta_{\mbox{tg}}$. To remove this dependence, arising from imperfect reconstruction of the kinematics, the following correction was applied:
\begin{eqnarray}
& \delta_1^{HRS-L} = \delta_0 + 0.025\theta_{\mbox{tg}} + 0.250\theta_{\mbox{tg}}^2, \\
& \delta_1^{HRS-R} = \delta_0 + 0.0274\theta_{\mbox{tg}}.
\end{eqnarray}
In addition, the vertical position offset at the target yields a small shift in the reconstructed momentum.  From the current in the raster magnet, we can determine the offset of the beam on target and apply an event-by-event correction to the reconstructed momentum:
\begin{equation} \label{eq:dp_vs_raster}
\delta_{\mbox{final}} = \delta_1  + \alpha \Delta y_{\mbox{rast}},
\end{equation}
where the parameter $\alpha$ depends on the beam energy. Note that the parameter $\alpha$ would not be necessary if the beamline optics between the raster and the target remained fixed with respect to the electron beam energy when the energy was changed. In that case, the $y$ offset due to the raster would scale with the raster current divided by the beam energy. However, there are quadrupoles between the raster and target that were adjusted to give an optimized beam tune when the beam energy was changed. Because of this, the mapping between the raster current and $y$ offset at the target has an additional scaling factor for each energy, which is accounted for by applying the factor $\alpha$. A similar correction for the overall vertical offset of the beam is also applied. The value of $\alpha$ at each energy is determined by examining the position of the elastic peak as a function of $\Delta y_{\mbox{rast}}$ for each energy setting, and varying $alpha$ until the elastic peak position is independent of $\Delta y_{\mbox{rast}}$.

\begin{table}[!htbp]
\begin{center}
\caption{Cuts used in the E01-001 analysis.  $\phi_{\mbox{tg}}$ ($\theta_{\mbox{tg}}$) is the in-plane (out-of-plane) angle measured relative to the spectrometer central ray.  The upper section of the table list cuts on raw detector quantities while the second half requires reconstructed tracks. The `hourglass' cut defines an hourglass-shaped region in focal plane coordinates that is slightly larger than the nominal acceptance for good events; the cut listed in the table is applied to the quantity defined in Eq.~\ref{eq:hourglass}. See text for more details.}
\begin{tabular}{c c c}
\hline \hline
Cut Type Applied & Left Arm Cut   & Right Arm Cut                                  \\
\hline
Event Type              & 3                             & 1              \\
aerogel PID             &A$_1$ADCSUM$<$350              & A$_2$ADCSUM$<$1250   \\
PID $\beta$             & --                            & 0.45$<$$\beta$$<$0.85    \\
Hits/cluster (VDC)      & 3--6                          & 3--6                         \\
Clusters per plane      & 1                             & 1                         \\
Number of Tracks        & 1                             & 1               \\
\hline
$Q_3$ radius (m)        & $<$0.29                       & --                 \\
hourglass (m)           &$<$$0.02\sin\theta_{L}$+0.01   & $<$$0.02\sin\theta_{R}$+0.02\\
$\delta$ Momentum (\%)  &$-$5.0$<$$\delta$$<$5.0        &$-$5.0$<$$\delta$$<$5.0 \\
$y_{\mbox{tg}}$ (m)     & $-$0.05$<$$y_{\mbox{tg}}$$<$0.05          & $-$0.05$<$$y_{\mbox{tg}}$$<$0.05      \\
$\theta_{\mbox{tg}}$ (mrad) & $-$40.0$<$$\theta_{\mbox{tg}}$$<$40.0 & $-$40.0$<$$\theta_{\mbox{tg}}$$<$40.0 \\
$\phi_{\mbox{tg}}$ (mrad)   & $-$10.0$<$$\phi_{\mbox{tg}}$$<$10.0   & $-$10.0$<$$\phi_{\mbox{tg}}$$<$10.0    \\
\hline \hline
\end{tabular}
\label{recon_cuts}
\end{center}
\end{table}

Once we have fully reconstructed tracks, we apply acceptance and background-rejection cuts, summarized in Table~\ref{recon_cuts}. The cuts on $\theta_{\mbox{tg}}$ and $\phi_{\mbox{tg}}$ limit events to the central 1.6~msr acceptance, roughly one-quarter of the full acceptance, to maintain full acceptance of the 4~cm target at all scattering angles. The cut on $y_{\mbox{tg}}$ is longer than the target, and there are no real events coming from upstream or downstream of the target. The cut is applied to reject background events, e.g. cosmic ray triggers, and, more importantly, events where poor track reconstruction yields unphysical target quantities.  We apply two other cuts to remove background events that are not within the acceptance of the spectrometer but which yield tracks at the focal plane.

The $Q_3$ cut is applied to reject events that scrape the exit pipe of the $Q_3$ quadrupole. We project the track to the exit of the $Q_3$ dipole, 2.64~m before the focal plane, and take events only within 29~cm of the center of the vacuum pipe. These background events generally give poor reconstruction to the target and would largely be removed in any event, but the cut at the $Q_3$ exit allows us to clearly identify these as true background events rather than simply poor reconstruction.  This is an important background at large $Q^2$ for kinematics where the elastic rate is extremely small and the rate just below the nominal momentum acceptance can be very large. In this case, even a small fraction of these high-rate inelastic events that hit an aperture in a field-free region may provide a non-negligible background in the low-rate elastic peak region. There is no evidence that events were scraping the exit pipe of the $Q_3$ quadrupole in the right arm. Note that elastic events are far from the $Q_3$ edges and the loss of events due to the $Q_3$ cut is negligible.

The other background rejection cut is referred to as the ``hourglass'' cut (as shown in Fig. 5.3 of Ref.~\cite{qattanphd}). For a short target, such as the 4-cm LH2 target used here, all events within the spectrometer acceptance fall within an hourglass-shaped distribution at a plane 69~cm in front of the first VDC.  We apply a cut to eliminate events which are well outside of this, and which correspond to events that did not fall inside the HRS acceptance: cosmic ray triggers, events which hit an aperture somewhere in the spectrometer and are scattered back into the acceptance, etc.  We apply a cut, listed in Tab.~\ref{recon_cuts} on the following quantity:
\begin{equation}
\Big|\mbox{y}_{fp}-0.69\mbox{y}^\prime_{fp}+0.005 \Big| - C_1\Big|\mbox{x}_{fp}-0.69\mbox{x}^\prime_{fp}\Big|,
\end{equation}\label{eq:hourglass}
where x$_{fp}$ and x$^\prime_{fp}$ are the vertical offset and slope of the track at the focal plane,  y$_{fp}$ and y$^\prime_{fp}$ are the horizontal offset and slope, and $C_1$=0.045(0.017) for the right (left) arm.


\section{Extraction of the cross section} \label{sigma_R}

The measured proton yield is corrected for experimental inefficiencies, e.g. detector and tracking inefficiency, as well as any dead time associated with the data-acquisition system. We then subtract the contribution from the target endcaps and compare the scattering yields from the liquid hydrogen to simulated elastic scattering and background processes.

The LH2 spectrum is dominated by the elastic e-p peak. In addition, there are backgrounds due to quasi-elastic and inelastic scattering from the aluminum target windows and protons generated from photoreactions $\gamma p \to \pi^0 p$ and $\gamma p \to \gamma p $. We isolate the elastic scattering signal from the data by subtracting the measured endcap contribution, using data taken on a 'dummy' target consisting of two aluminum foils at the target window positions. The protons associated with $\pi^0$ photoproduction and Compton scattering are simulated, normalized to the inelastic region of the data, and subtracted from the full yield to isolate the elastic scattering events. The background processes, as well as the elastic scattering, are simulated using a detailed simulation of the spectrometer to account for spectrometer acceptance and resolution, radiative corrections, multiple scattering, etc. 

Once the proton elastic events are isolated, we compare the distribution of measured elastic protons to the simulated spectrum. We use the simulation of elastic scattering events to determine the conversion of the measured yield into the underlying Born cross section. If the simulation accounts for all experimental corrections, then it provides a translation from the Born cross section at the central kinematics to the observed yield integrated over the spectrometer acceptance, with identical cuts applied to the data and simulation. Potential shortcomings of the simulation are evaluated to estimate the systematic uncertainties of our extraction procedure. This also corrects for bin-centering effects, e.g. the difference between the cross section at the quoted kinematics and the cross section averaged over the spectrometer acceptance. These corrections are small, due to the small acceptance of the spectrometers after applying the tight solid angle cut, and yield negligible uncertainty due to the well-understood elastic cross section.

\subsection{The elastic e-p Simulations (SIMC)} \label{ep_simc}

The simulation program SIMC is used to simulate elastic scattering for all kinematics and for both arms. The elastic e-p simulations are a crucial component of the analysis as they are used to extract the reduced cross section. SIMC was adapted from the (e,e$'$p) code SIMULATE that was written for SLAC experiment NE18~\cite{makins94, oneill95, arrington01} and used in several JLab experiments (described in Refs.~\cite{gaskell01, mohring02, dohrmann04, ambrozewicz04}). The two main components of SIMC are the event generator, which includes the cross-section weighting and radiative corrections~\cite{makinsphd, ent01}, and the spectrometer models. First, SIMC randomly generates the energy and position of the incident proton at the target to match the energy and spatial spread of the beam, accounting for the target length and the beam raster. The beam energy is then corrected for event-by-event ionization losses in the target. SIMC then randomly generates the angle of the scattered electron over a large angular acceptance, such that all proton angles that fall within the HRS acceptance are fully sampled.  Having generated a basic event at the scattering vertex, SIMC applies corrections to the Born cross section (taken from the form factor parameterization of Ref.~\cite{bosted95}) to account for internal radiative correction and simulates the emission of real photons for all generated particles~\cite{ent01}. The struck proton is transported through the target where ionization energy loss and multiple scattering in the target material, cells, and vacuum chamber windows are applied. Finally, the scattered protons are transported through the spectrometers.

Transporting the protons through the spectrometer was done using the spectrometer optics models included in the Monte Carlo simulation program COSY~\cite{cosy95}. COSY generates both the forward and backward sets of matrix elements to simulate the optical resolution of the magnetic systems in the spectrometer. Note that we apply energy loss and multiple scattering for all of the materials the particles traverse, but do not account for proton absorption in the simulations since this correction is applied to the data. The forward matrix elements transport the particle vectors from the entrance window of the spectrometer to its focal plane going through every major aperture in the spectrometer. SIMC applies an aperture cut for each of these steps, including the initial rectangular collimator, apertures at the front, middle, and back of each magnetic element, and the vacuum chamber after the $Q_3$ quadrupole.  Events are also required to be within the active area of any detectors required in the trigger or analysis. The positions at the VDCs are recorded, including a randomly-generated offset to simulate VDC resolution, and the focal plane track is then fitted and reconstructed to the target position using the simulation's HRS reverse matrix elements. In reconstructing the target quantities, we apply the same average corrections for energy loss that are applied to the data. The simulated $\Delta P$ spectrum is then normalized to the accumulated effective charge for the data at that setting.

In the initial simulations, the width of the measured elastic peak was slightly broader than the simulation and had significantly larger non-Gaussian tails. We used the coincidence data, where we can cleanly isolate the tails of the elastic peak, to determine the shape of the non-Gaussian tails, and apply an additional smearing in the angular resolution of the spectrometer in the simulations.

\subsection{The $\gamma p \to \pi^0 p$ and $\gamma p \to \gamma p$ Simulations} \label{pi0} 

When the electron beam passes through the target, electrons lose energy by the emission of real photons. These real photons which impinge upon the target have a maximum energy just below the beam energy and can generate high-momentum protons in kinematics similar to elastic scattering through Compton scattering or pion photoproduction. To model the $\Delta P$ spectrum for the pion photoproduction, we first calculate the Bremsstrahlung cross section and reconstruct the $E_{\gamma}$ spectrum for these photons~\cite{schultethesis, schulte02, meekinsthesis, meekins99}, and then randomly generate photons according to the $E_{\gamma}$ spectrum. The next step is to uniformly and randomly generate protons over the spectrometer acceptance using the same event generation procedure as presented earlier. The generated event is then weighted by an $s^{-7}$ cross-section dependence, as predicted by the high-energy approximation and based on the constituent counting rules or $s^{-n}$, forming the shape of the $\Delta P$ spectrum used for the $\gamma p \to \pi^0 p$ contribution. The generated protons are transported through the same spectrometer model as the elastic events and protons that make it through the spectrometer are used to generate a reconstructed $\Delta P$ spectrum.  The absolute normalization of this background is determined in a fit to the data below the elastic peak.

Previous experiments~\cite{shupe79} indicated that the ratio of the $\gamma p \to \gamma p$ to the $\gamma p \to \pi^0 p$ cross sections was (1--5)\%.  The ratio has a clear dependence on energy, and was well parameterized as a function of energy as $(0.92 E - 1.2)$\% with $E$ in~GeV for our kinematics. We use this fit to scale the Compton spectrum to the simulated yield. The simulated spectra are compared to data in Sec.~\ref{extract_sigma}.


\subsection{Subtraction of endcap contributions} \label{add_histo}

\begin{figure}[!htbp]
\begin{center}
\includegraphics*[width=0.95\columnwidth]{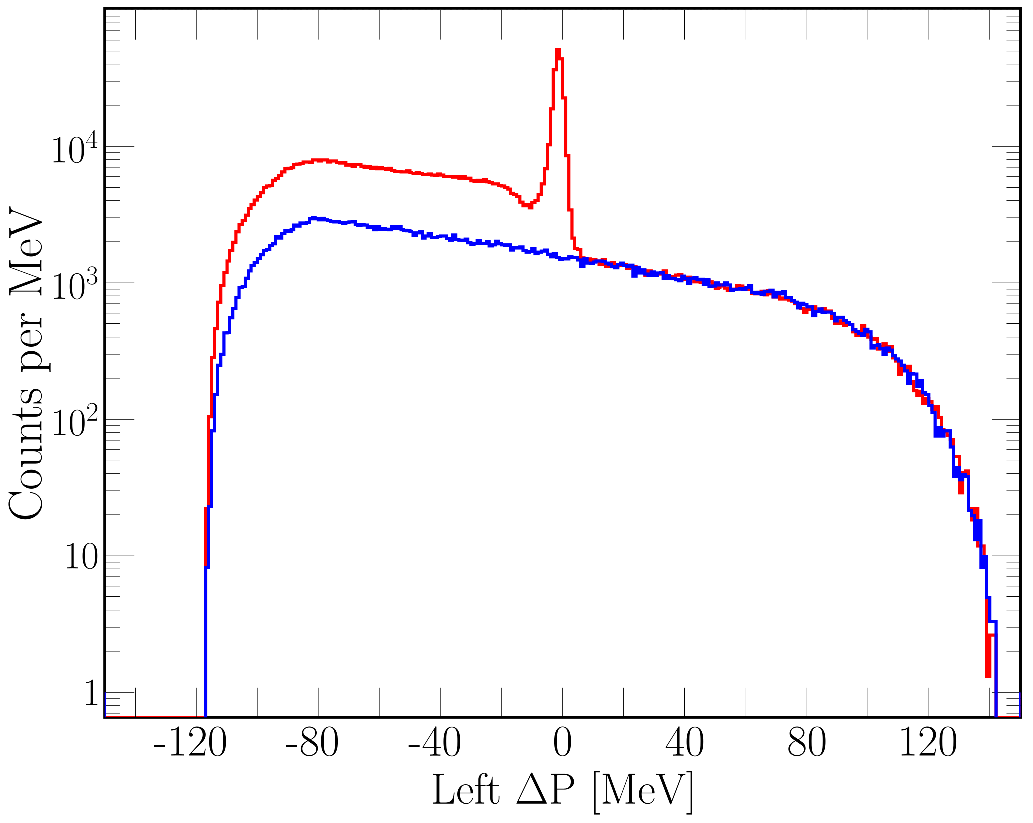} 
\end{center}
\caption{(Color online) Left arm $\Delta P$ spectrum for LH2 (red) and dummy (blue) targets from kinematics $m$ (Tab.~\ref{tab:kinematics}). The dummy spectrum is normalized to match the LH2 spectrum at large $\Delta P$. The spectra are similar for all settings, with the elastic peak becoming broader relative to the $\Delta P$ acceptance, for lower $Q^2$ settings and for larger $\varepsilon$ kinematics.}
\label{fig:corr_ldelta_p}
\end{figure}

For each run, the effective charge and a histogram of events versus $\Delta P$ is generated.  For all runs at a given kinematic setting, these are combined to yield a total effective charge and distribution of events. From this, we extract normalized $\Delta P$ distributions for LH2 and dummy target runs at each setting. Only scattering from the aluminum endcaps can contribute to the super-elastic region, $\Delta P > 0$, and so the positive side of the $\Delta P$ spectrum can be used to normalize the dummy target contribution to yield the endcap spectrum for the LH2 target. This is done, as shown in Fig.~\ref{fig:corr_ldelta_p}, and the normalized dummy target data is used to subtract out the endcap contributions.  The dummy target normalization factor obtained by normalizing the dummy spectrum to the LH2 data is close to 4.1, the relative thickness of the dummy targets and the aluminum endcaps on the LH2 target. However, there is some kinematic dependence in this scaling factor arising from the fact that the radiative correction factor is different for the thick Al dummy windows than in the LH2 target. The observed deviations are qualitatively consistent with simulations used to estimate the effect, but the evaluation could only be done using pure elastic scattering, while the endcap is a combination of quasielastic and inelastic scattering. Note that the thicker Al foils in the dummy target lead to a change in the ratio of scattering from the upstream and downstream windows compared to the LH2 target.  However, the distributions are very smooth, and the $\Delta P$ spectrum was identical in shape for the upstream and downstream windows, and so knowing the combined normalization is sufficient to reliably subtract the contributions. 

The dummy subtraction is approximately a 10\% correction for both arms, with a 2\% $\varepsilon$ dependence in the left arm and a 0.5\% dependence in the right arm. We assign a conservative 5\% systematic uncertainty in the dummy subtraction which yields a 0.5\% scale uncertainty and slope uncertainties of 0.10\% for the left arm and 0.36\% (0.025\%/0.07) for the right arm when we account for the $\Delta \varepsilon$ range of 0.07. The random uncertainty is taken from the statistical uncertainty in the determination of the effective dummy thickness (See Sec.~\ref{extract_sigma}).

\begin{figure}[!htbp]
\begin{center}
\includegraphics*[width=0.95\columnwidth]{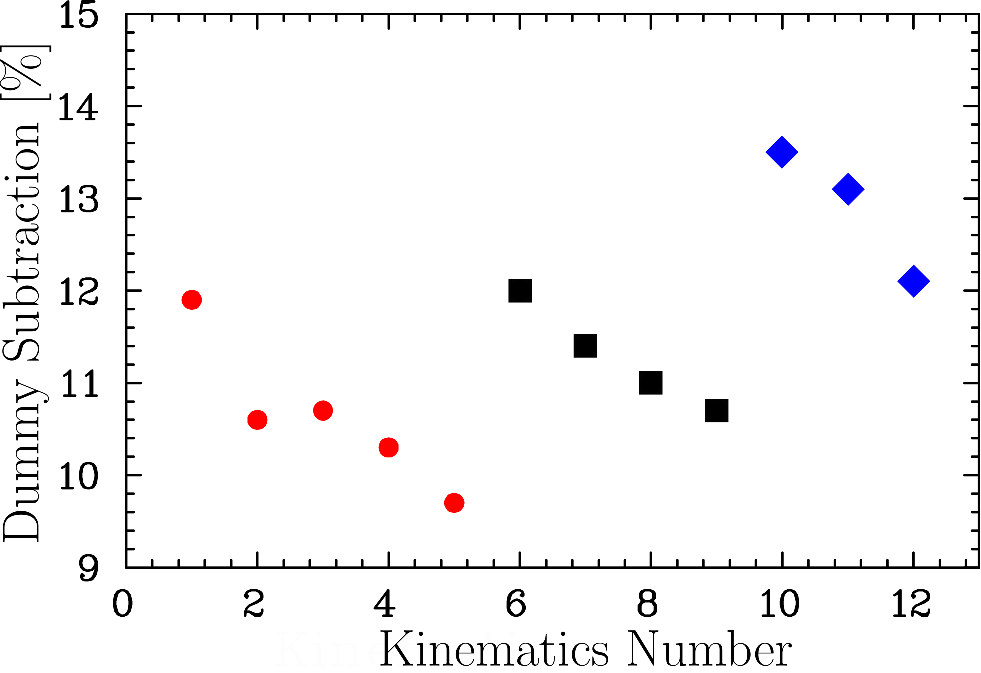}
\end{center}
\caption{(Color online) The ratio of $N_{\mbox{dummy}}$ to $N_{\mbox{LH}_2}$ events or ``Dummy Subtraction (\%)'' for the left arm in the elastic $\Delta P$ window used in the analysis. The points are color coded according to the left arm kinematics: $Q^2$ = 2.64~GeV$^2$ as red circles, 3.20~GeV$^2$ as black squares, and 4.10~GeV$^2$ as blue diamonds. For each $Q^2$ value, the points are sorted by $\varepsilon$ (low to high).  The right arm yields a correction of 9.0-9.5\%, with no systematic $\varepsilon$ dependence.}
\label{fig:show_dummysub_both}
\end{figure}

\subsection{$\gamma p \to \pi^0 p$ and $\gamma p \to \gamma p$ Subtraction} \label{normalized_pi0_simul}

The next step is to subtract the protons generated from photoreactions $\gamma p \to \pi^0 p$ and $\gamma p \to \gamma p$ from the endcap-subtracted LH2 $\Delta P$ spectrum. The elastic simulation is normalized to match the elastic peak in the dummy-subtracted LH2 spectrum, and then the simulated elastic peak is subtracted, yielding an estimate of the total photoproduction backgrounds. This spectrum is used to normalize the simulated pion production and Compton scattering spectra, matching the counts for $-50<\Delta P<-20$~MeV (away from the elastic peak contributions, but avoiding more-inelastic kinematics where two-pion production may become important). We then obtain the pure hydrogen elastic spectrum by subtracting this normalized background contribution from the dummy-subtracted LH2 spectrum.

\begin{figure}[!htbp]
\begin{center}
\includegraphics*[width=0.95\columnwidth]{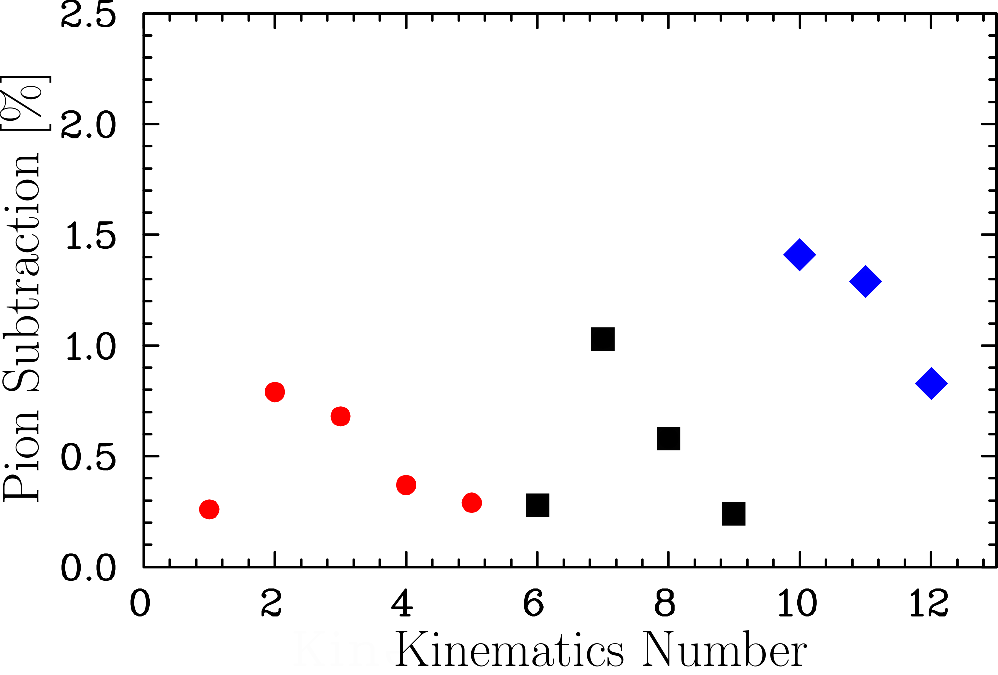}
\end{center}
\caption{(Color online) The ratio of ($N_{\gamma p \to \pi^0 p} + N_{\gamma p \to \gamma p}$) to $N_{\mbox{LH}_2}$ events or ``Pion Subtraction (\%)''for the left arm in the $\Delta P$ window cut used as a function of kinematics number. The points are color coded as in Fig.~\ref{fig:show_dummysub_both}.}
\label{fig:show_pionsub_left}
\end{figure}

Figure~\ref{fig:show_pionsub_left} shows the ratio of background events to the elastic events for the left arm in the $\Delta P$ window used in the analysis. The subtraction is typically (0.5-1.5)\% in the left arm with a large scatter and $\varepsilon$ dependence. It is generally smaller for the highest and lowest $\varepsilon$ values, as the background becomes small at large $\varepsilon$, while it is better separated from the elastic peak at small $\varepsilon$.  Because the subtraction is sensitive to the details of the simulations, in particular the resolution in $\Delta P$, we take the total uncertainty to be half of the correction, separated into a slope uncertainty of 40\% of the correction and a random uncertainty of 30\% of the correction. Note that when the background contribution is subtracted, the statistical uncertainty in the background normalization factor is applied as an additional contribution to the uncertainty, as with the dummy subtraction. For the right arm, this background is essentially negligible, with a maximum contribution well below 0.1\%. We do not apply any correction to the right arm results, and assign a 0.05\% random uncertainty.


\subsection{Extracting the Reduced Cross Section $\sigma_{R}$} \label{extract_sigma}

\begin{figure}[!htbp]
\begin{center}
\includegraphics*[width=0.95\columnwidth]{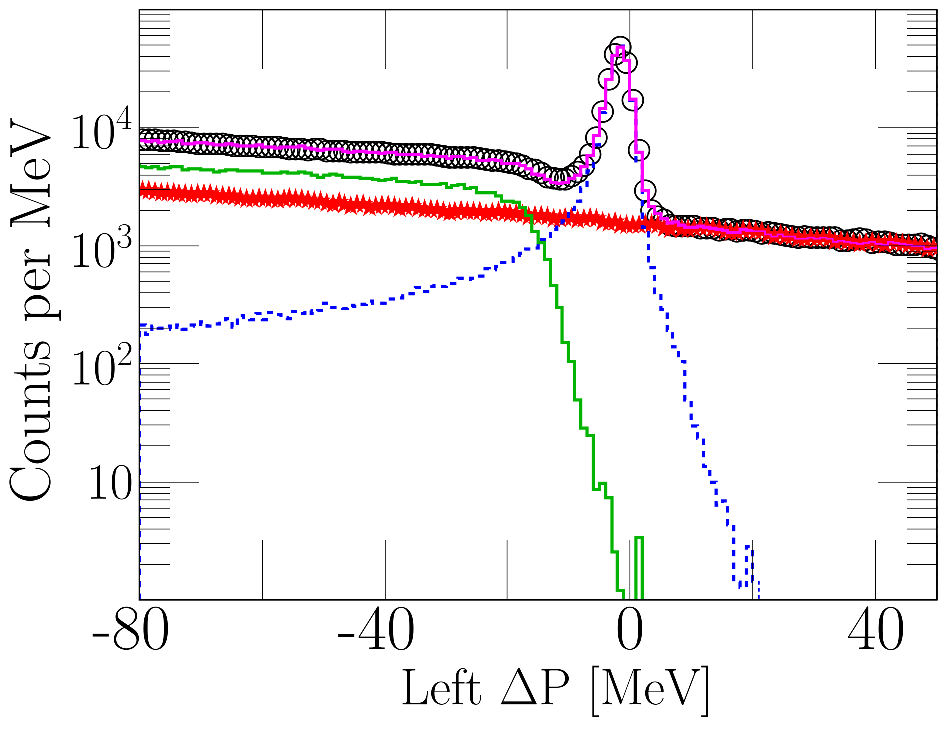} \\
\includegraphics*[width=0.95\columnwidth]{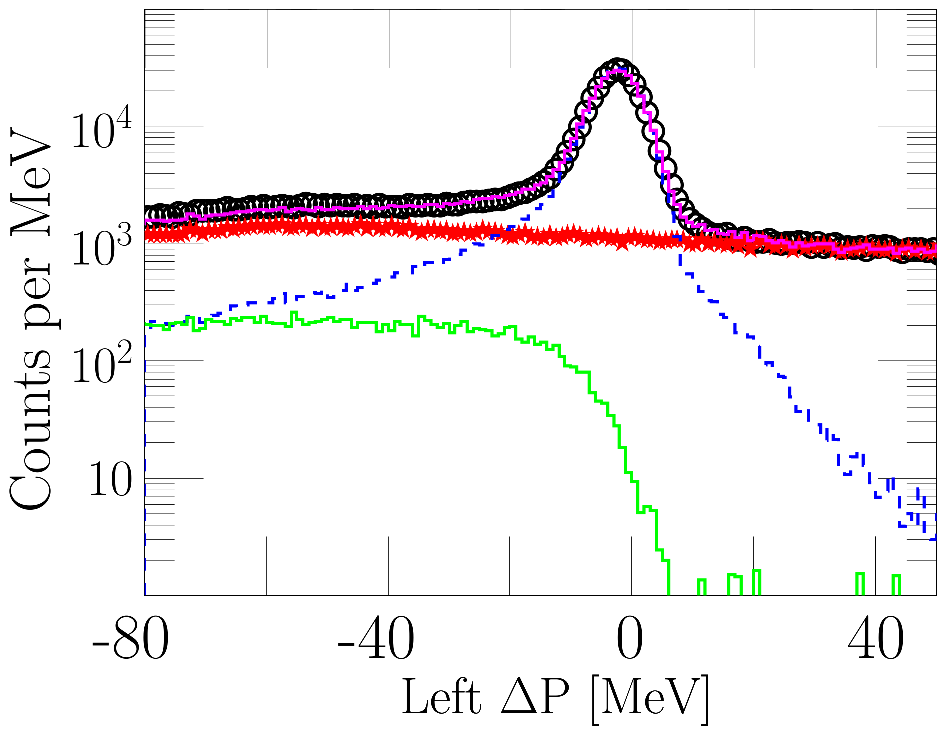} 
\includegraphics*[width=0.95\columnwidth]{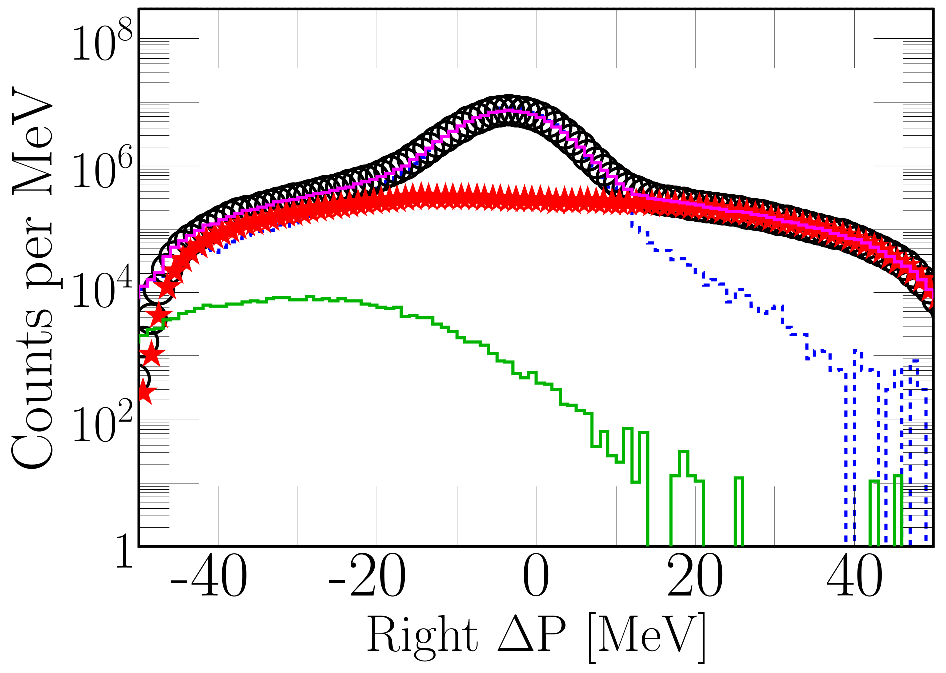} 
\end{center}
\caption{(Color online) Contributions to the $\Delta$P spectrum for the HRS-L at kinematics $b$ (top - low $\varepsilon$) and kinematics $m$ (middle - high $\varepsilon$), and for HRS-R for $b$ (bottom). The curves indicate the normalized yield for LH2 (black circles) and dummy (red stars) data, scaled elastic e-p simulation (blue dotted line), normalized $\gamma p \to \pi^0 p$ and $\gamma p \to \gamma p$ backgrounds (solid green), and sum of dummy plus simulated elastic and backgrounds (solid magenta). At higher $\varepsilon$, the elastic peak broadens but the inelastic contributions become much smaller.}
\label{fig:kinb_all_ldp}
\end{figure}

Figure~\ref{fig:kinb_all_ldp} shows the individual contributions to the $\Delta P$ spectra for the left and right arms. After subtracting the endcap and background contributions, we integrate the counts in a narrow $\Delta P$ window around the elastic peak. The $\Delta P$ window was also chosen to minimize size and $\varepsilon$ dependence of the dummy and pions subtraction while maintaining a high acceptance of elastic protons. These cuts remove much of the elastic tail, which is accounted for in the simulation of the experiment, but if they are too restrictive, they will begin to remove events in the gaussian-like part of the peak, making the result more sensitive to the agreement between the resolution in the experiment and in the simulation. The $\Delta P$ windows used in the analysis yield a high acceptance ($>$95\%) for the main peak at all left arm settings, but roughly 85\% for the right arm.

Table~\ref{kinematics} lists the $\Delta P$ window used for all settings. Table~\ref{sigma_R_data} shows the extracted reduced cross sections and their uncertainties for all kinematic settings. The statistical uncertainty in $\sigma_{R}$ includes the statistical uncertainty from all of the yields in $N_{\mbox{Elastic-Data}} = (N_{\mbox{LH}_2} - N_{\mbox{dummy}} - N_{\gamma p \to \pi^{0} p} - N_{\gamma p \to \gamma p})$, as well as the statistical uncertainty in the extracted dummy normalization factor. In addition to removing non-elastic scattering events, the $\Delta P$ cut yields a loss of elastic events in the radiative tail. For all kinematics, about 1\% of the total elastic events are in the non-Gaussian tails of the elastic peak but outside of the $\Delta P$ window. While we match the non-Gaussian tails to a clean proton sample using the coincidence runs, this may not be a perfect reproduction of the resolution for all settings, and we assign a 50\% uncertainty to these events, yielding a 0.5\% scale uncertainty for both spectrometers.

\begin{table}[!htbp]
\begin{center}
\caption{The elastic e-p reduced cross section, $\sigma_{R}$ in Eq.~\ref{eq:reduced}, along with the statistical ($\delta_{\mbox{stat}}$), random ($\delta_{\mbox{random}}$), and total ($\delta_{\mbox{tot}}$) point-to-point uncertainties.  Note that for the right arm data, there are multiple measurements of $\sigma_R$ at the same kinematics, corresponding to different left arm kinematics.}
\begin{tabular}{c c c c c}
\hline \hline
$\varepsilon$ & $\sigma_{R}$ & $\delta_{\mbox{stat}}$ & $\delta_{\mbox{random}}$ & $\delta_{\mbox{tot}}$ \\
\hline \hline
\multicolumn{5}{c}{$Q^2=0.5$~GeV$^2$} \\
\hline
 0.914 & 0.2286 & 0.54$\times$$10^{-3}$ & 1.04$\times$$10^{-3}$ & 1.17$\times$$10^{-3}$ \\
 0.939 & 0.2325 & 0.46$\times$$10^{-3}$ & 1.49$\times$$10^{-3}$ & 1.56$\times$$10^{-3}$ \\
 0.962 & 0.2319 & 0.51$\times$$10^{-3}$ & 1.06$\times$$10^{-3}$ & 1.18$\times$$10^{-3}$ \\
 0.979 & 0.2347 & 0.51$\times$$10^{-3}$ & 1.07$\times$$10^{-3}$ & 1.19$\times$$10^{-3}$ \\
 0.986 & 0.2349 & 0.66$\times$$10^{-3}$ & 1.08$\times$$10^{-3}$ & 1.27$\times$$10^{-3}$ \\
 0.939 & 0.2323 & 0.48$\times$$10^{-3}$ & 1.48$\times$$10^{-3}$ & 1.56$\times$$10^{-3}$ \\
 0.962 & 0.2319 & 0.45$\times$$10^{-3}$ & 1.06$\times$$10^{-3}$ & 1.15$\times$$10^{-3}$ \\
 0.979 & 0.2349 & 0.52$\times$$10^{-3}$ & 1.07$\times$$10^{-3}$ & 1.19$\times$$10^{-3}$ \\
 0.986 & 0.2346 & 0.63$\times$$10^{-3}$ & 1.08$\times$$10^{-3}$ & 1.25$\times$$10^{-3}$ \\
 0.962 & 0.2315 & 0.47$\times$$10^{-3}$ & 1.05$\times$$10^{-3}$ & 1.15$\times$$10^{-3}$ \\
 0.979 & 0.2348 & 0.47$\times$$10^{-3}$ & 1.07$\times$$10^{-3}$ & 1.17$\times$$10^{-3}$ \\
 0.986 & 0.2335 & 0.70$\times$$10^{-3}$ & 1.07$\times$$10^{-3}$ & 1.27$\times$$10^{-3}$ \\
\hline \hline
\multicolumn{5}{c}{$Q^2=02.64$~GeV$^2$} \\
\hline
 0.117 & 1.340$\times$$10^{-2}$ &~0.40$\times$$10^{-4}$~& 0.51$\times$$10^{-4}$ &~0.65$\times$$10^{-4}$~ \\
 0.356 & 1.382$\times$$10^{-2}$ & 0.35$\times$$10^{-4}$ & 0.82$\times$$10^{-4}$ & 0.89$\times$$10^{-4}$ \\
 0.597 & 1.430$\times$$10^{-2}$ & 0.47$\times$$10^{-4}$ & 0.55$\times$$10^{-4}$ & 0.72$\times$$10^{-4}$ \\
 0.782 & 1.468$\times$$10^{-2}$ & 0.44$\times$$10^{-4}$ & 0.57$\times$$10^{-4}$ & 0.72$\times$$10^{-4}$ \\
 0.865 & 1.483$\times$$10^{-2}$ & 0.38$\times$$10^{-4}$ & 0.58$\times$$10^{-4}$ & 0.69$\times$$10^{-4}$ \\
\hline \hline
\multicolumn{5}{c}{$Q^2=3.2$~GeV$^2$} \\
\hline
 0.131 & 0.869$\times$$10^{-2}$ & 0.30$\times$$10^{-4}$ & 0.42$\times$$10^{-4}$ & 0.52$\times$$10^{-4}$ \\
 0.443 & 0.908$\times$$10^{-2}$ & 0.29$\times$$10^{-4}$ & 0.35$\times$$10^{-4}$ & 0.45$\times$$10^{-4}$ \\
 0.696 & 0.931$\times$$10^{-2}$ & 0.30$\times$$10^{-4}$ & 0.36$\times$$10^{-4}$ & 0.47$\times$$10^{-4}$ \\
 0.813 & 0.953$\times$$10^{-2}$ & 0.27$\times$$10^{-4}$ & 0.37$\times$$10^{-4}$ & 0.46$\times$$10^{-4}$ \\
\hline \hline
\multicolumn{5}{c}{$Q^2=4.1$~GeV$^2$} \\
\hline 0.160 & 0.473$\times$$10^{-2}$ & 0.23$\times$$10^{-4}$ & 0.22$\times$$10^{-4}$ & 0.31$\times$$10^{-4}$ \\
 0.528 & 0.497$\times$$10^{-2}$ & 0.17$\times$$10^{-4}$ & 0.23$\times$$10^{-4}$ & 0.29$\times$$10^{-4}$ \\
 0.709 & 0.512$\times$$10^{-2}$ & 0.19$\times$$10^{-4}$ & 0.24$\times$$10^{-4}$ & 0.31$\times$$10^{-4}$ \\
\hline \hline
\end{tabular}
\label{sigma_R_data}
\end{center}
\end{table}

We vary the $\Delta P$ window cut used on the left arm by $\pm$2~MeV to estimate the cut dependence, and find that the ratio of the data to simulated yield varies by about 0.20\%. Since this could potentially have some correlation with $\varepsilon$, the 0.20\% uncertainty is broken down equally as 0.14\% random uncertainty and 0.14\% slope uncertainty. Accounting for the average $\Delta \varepsilon$ range of 0.7 in the left arm, this yields a slope uncertainty of 0.20\%. For the right arm, the elastic acceptance shows a 1.5\% $\varepsilon$ dependence.  Varying the $\delta$ cut changes the correction by about 30\%, so we take 30\% of the 1.5\% correction as the acceptance uncertainty. We divide this evenly into scale, random, and slope uncertainties and assign a 0.30\% uncertainty to each. Accounting for the small $\Delta \varepsilon$ range of the right arm, this yields a slope uncertainty of 4.3\%.

\subsection{Summary of Systematic Uncertainties}

\begin{table*}[!htbp]
\begin{center}
\caption{Summary of the systematic uncertainties; numbers in bold indicate the dominant contributions. For target boiling, the random uncertainty is 0.45\% and 0.30\% for kinematics $a$ and $b$ (taken below 70$\micro$A), and zero for all others. For the pion subtraction, the left arm $Q^2$ = 4.10~GeV$^2$ uncertainty is 0.30\% while other kinematics have 0.15\%. For the uncertainties associated with angle and energy fluctuations, the values applied on kinematics-by-kinematics basis. Since $\Delta \varepsilon \approx 0.07$ for the right arm, the slope uncertainty is much larger for the right arm.}
\begin{tabular}{|c c | c c c | c c c |}
\hline \hline
				&		 &\multicolumn{3}{c|}{Right Arm Systematics}&\multicolumn{3}{c|}{Left Arm Systematics} \\
Source                          &Section                 &~Scale~     &Random       &~Slope~  &~Scale~     &Random       &Slope\\
                                &                        &(\%)        &(\%)         &(\%)        &(\%)        &(\%)         &(\%) \\\hline
BCM Calibration                 &\ref{sec:beam_current}  &0.50        &0.10         & -  	 &0.50        &0.10         & -  \\
Target Boiling                  &\ref{tgt_boiling}       &{\bf{1.05}} &{\bf{0.30-0.45}}& -  	 &{\bf{1.05}} &{\bf{0.30--0.45}}& -  \\
Raster Size                     &\ref{tgt_boiling}       & -          &0.10         & -  	 & -          &0.10         & -  \\\hline
Target Length                   &\ref{target_leng_corr}  &0.12        &0.07         & -   	 &0.12        &0.07         & -  \\
Electronic Deadtime             &\ref{comp_elec_deadtime}& -          &0.04         &0.50        & -          & -           & -  \\
Computer Deadtime               &\ref{comp_elec_deadtime}&0.10        & -           &1.00        &0.10        & -           &0.10\\\hline
VDC Zero-Track Inefficiency     &\ref{vdcs_track_eff}    & -          & -           & -  	 &0.10        & -           & -  \\
VDC Multiple-Track Inefficiency &\ref{vdcs_track_eff}    &0.10        & -           &0.70        &0.10        &0.02         & -  \\
VDC Multiplicity Cuts               &\ref{vdcs_multiplicity_eff} &0.50        &0.10         & -  	 &0.50        &0.10         &{\bf{0.25}}\\\hline
Scintillator Efficiency         &\ref{scint_effic}       &0.10        &0.05         & -  	 &0.10        &0.05         & -  \\
PID Efficiency ($\beta$)        &\ref{beta_eff}~~        &0.15        &0.05         & -  	 & -          & -           & -  \\
PID Efficiency (A$_2$)          &\ref{A2_eff}            &0.25        &0.05         & -  	 & -          & -           & -  \\\hline
PID Efficiency (A$_1$)          &\ref{A1_eff}            & -          & -           & -  	 &0.20        &0.10         & -  \\
Pion Contamination              &\ref{pid_cuts}          & -          & -           & -  	 & -          &0.10         & -  \\
Proton Absorption               &\ref{proton_absorp}     &{\bf{1.00}} & -           &0.10	 &{\bf{1.00}} & -           &0.03\\\hline
Solid Angle Cut                 &\ref{e01001_optics}     &{\bf{2.06}} & -           & -  	 &{\bf{2.06}} & -           & -  \\
Pion Subtraction                &\ref{extract_sigma}     & -          &0.05         & -  	 & -          &{\bf{0.15--0.30}} &{\bf{0.20--0.40}}\\
Dummy Subtraction               &\ref{extract_sigma}     &0.50        & -           &0.40	 &0.50        & -           &0.10\\\hline
$\Delta P$ Cut-Dependence       &\ref{extract_sigma}     &0.50        &{\bf{0.14}}  &{\bf{2.00}} &0.50        &{\bf{0.14}}  &{\bf{0.20}}\\
$\delta$ Cut-Dependence         &\ref{extract_sigma}     &0.30        &{\bf{0.30}}  &{\bf{4.30}} & -          & -           & -  \\
0.18 mrad Angle Offset          &\ref{spect_mispoint}    &0.20        &0.02         &{\bf{0.67}} &0.13        &0.01         &{\bf{0.18}}\\\hline
0.10 mrad Angle Fluctuations    &\ref{spect_mispoint}& -    &0.097--0.125 & -  	 & -          &0.03--0.10   & -  \\
0.03\% Beam Energy              &\ref{expt_equip}        &0.03        &0.01         &0.29	 &0.13        &0.02         &0.07\\
0.02\% Beam Energy Fluctuations &\ref{expt_equip}& -       &0.015--0.029 & -          & -          &0.043--0.081 & -  \\\hline
Radiative Corrections           &\ref{rad_corrections}   &{\bf{1.00}} &{\bf{0.20}}  &{\bf{2.00}} &{\bf{1.00}} &{\bf{0.20}}  &{\bf{0.30}}\\
\hline
 & & & & & & & \\
& & $\delta_{\mbox{scale}}$ & $\delta_{\mbox{random}}$ & $\delta_{\mbox{slope}}$ & $\delta_{\mbox{scale}}$ & $\delta_{\mbox{random}}$&$\delta_{\mbox{slope}}$ \\
\hline
Total (\%)                               & &2.93  &0.454--0.640 &5.38    & 2.91  &0.384--0.593  &0.539--0.801 \\
\hline \hline
\end{tabular}
\label{r_systematic_summary}
\end{center}
\end{table*}

Table~\ref{r_systematic_summary} summarizes the systematic uncertainties for the right and left arms. The systematic uncertainty for each source is broken down into scale, random, and slope contributions. The contribution of each uncertainty type from all sources is then added in quadrature to form the total uncertainty in $\sigma_R$ for that uncertainty type. Only the range on $\delta_{\mbox{random}}$ and $\delta_{\mbox{slope}}$ is given in these tables, while the actual value of the total uncertainty $\delta_{\mbox{tot}}$ in $\sigma_R$ at each $\varepsilon$ point is listed in Table~\ref{sigma_R_data}. It must be mentioned that the scale uncertainty in the pion subtraction for the left arm is 0.20\% for all kinematics except those of $Q^2$ = 4.10 GeV$^2$ which is 0.40\%. This results in a $\delta_{\mbox{slope}} = 0.539$\% for all kinematics at $Q^2$ = 2.64 and 3.20~GeV$^2$, and $\delta_{\mbox{slope}} = $ 0.641\% for all kinematics at $Q^2$ = 4.10 GeV$^2$. In addition, $\delta_{\mbox{slope}}$ for $Q^2$ = 4.10~GeV$^2$ is then scaled up by 25\% to account for the $\Delta \varepsilon$ range difference among the three $Q^2$ points, which results in a $\delta_{\mbox{slope}} = $ 0.801\%. 

In the initial proposal for the experiment~\cite{e01001}, the plan was to use the right arm as a luminosity monitor. While the kinematics of the right arm change for the different left-arm $\varepsilon$ values, the extremely small $\varepsilon$ range means that the reduced cross section has minimal variation with angle, and that variation is well known, allowing for its use as a luminosity monitor. However, because the data were taken at fixed beam current for each $\varepsilon$ setting at a given $Q^2$, with the exception of kinematic $a$, the luminosity-related uncertainties were smaller than anticipated and the use of the right arm as a luminosity monitor was not beneficial, as the systematic uncertainties associated with the low-$Q^2$ measurements were larger than the uncertainties associated with the luminosity measurements. Thus, the right arm is analyzed independently and serves mainly as a check on the assumed systematic uncertainties, many of which are common to both spectrometers.


\section{Results and Discussion}

In this section, we describe the general procedure used to extract the proton form factors and their ratio from the reduced cross sections given in Table~\ref{sigma_R_data}. While the extraction of $\gegm$ for the high-$Q^2$ points was previously published~\cite{qattan05}, this work includes the right arm data as well as updated systematic uncertainties for the left arm. We then compare the results to previous Rosenbluth and polarization transfer results and discuss their impact on constraining two-photon-exchange (TPE) contributions. This includes evaluating the discrepancy between these precise Rosenbluth results and polarization data, and examining the data for indications of non-linearity which would indicate a deviation from the Born approximation. Finally, the results of this experiment raised the interest in the physics of the TPE and Coulomb distortion corrections and laid down the foundations for new measurements~\cite{meziane11, e05017, moteabbed13, CLAS:2014xso, CLAS:2016fvy, vepp_proposal, Rachek:2014fam, Nikolenko:2015xsa, OLYMPUS:2016gso, e12+23-008, e12+23-012} aimed at measuring the size of the TPE corrections. We conclude with a brief discussion of recent calculations and experiments aimed at estimating/measuring the size of the TPE effect in e-p elastic scattering.


\subsection{$\gep$ and $\gmp$ Extraction}\label{results_conclusion}

\begin{figure}[!htbp]
\begin{center}
\includegraphics*[width=0.95\columnwidth]{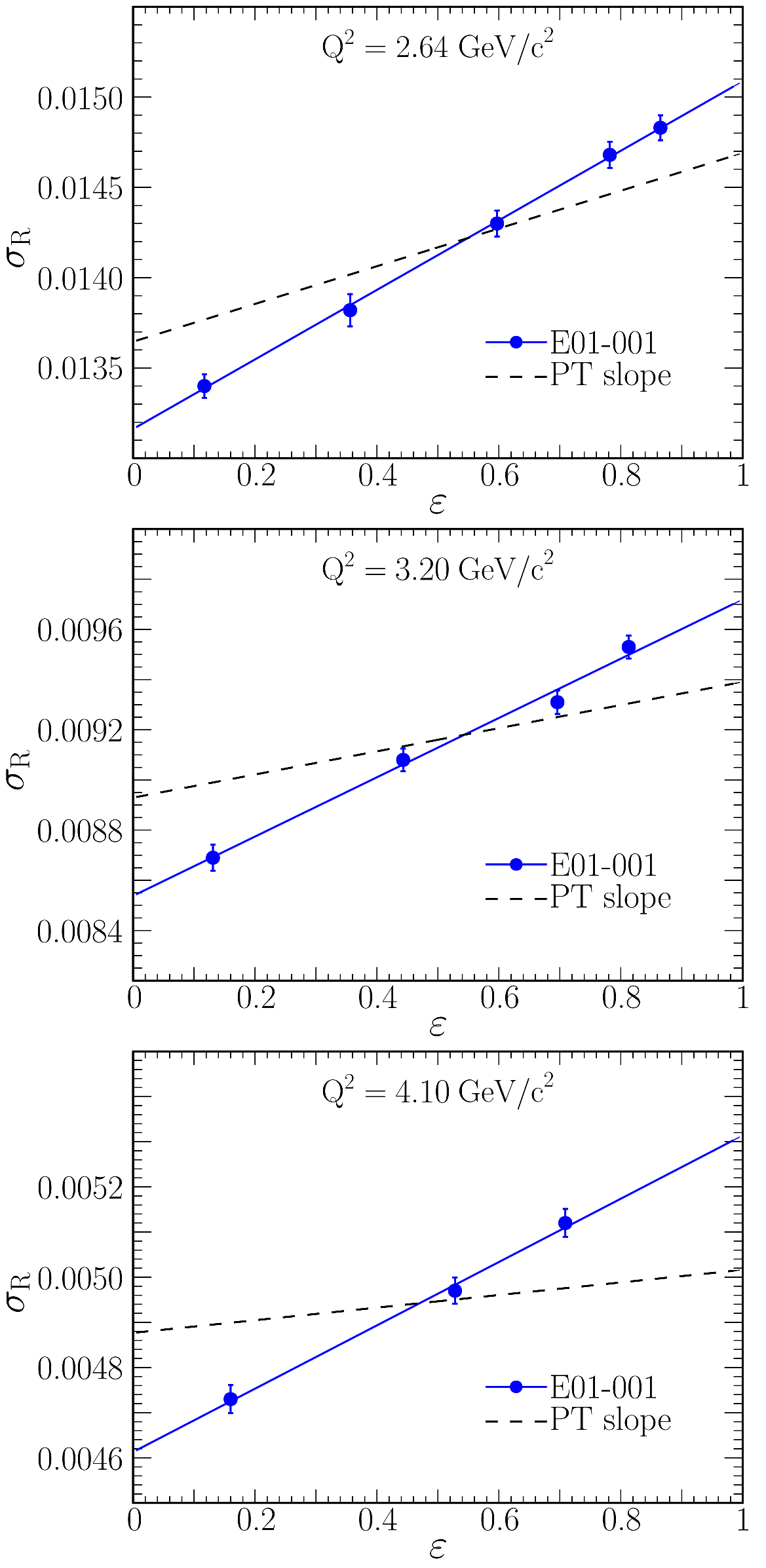}
\end{center}
\caption{(Color online) Linear fit to $\sigma_{R}$ vs $\varepsilon$ for $Q^2$ = 2.64 (top), 3.20 (middle), and 4.20~GeV$^2$ (bottom). The uncertainties shown are the combined statistical and random systematic uncertainties. The dashed black line is the fit to the E01-001 data and the solid blue line is the slope from a fit to earlier polarization measurements~\cite{Christy:2021snt}.} 
\label{fig:2.64_lt_sigma_r}
\end{figure}

The classification of the systematic uncertainties into scale, random, and slope uncertainties is explained in Sec.~\ref{expt_equip}. We describe here how these are treated in our extraction procedure. Taking the data and total point-to-point uncertainties from Tab.~\ref{sigma_R_data}, a linear fit of $\sigma_{R}$ to $\varepsilon$ at a fixed $Q^2$ is performed to obtain $\gep$ and $\gmp$ and their ''random'' uncertainty contributions. Figure~\ref{fig:2.64_lt_sigma_r} shows the fits done for the left arm measurements using only the random uncertainties.

We evaluate the contribution of the slope uncertainty in $\sigma_{R}$ to the uncertainty in the form factors by varying the slope in the $\sigma_{R}$ results according to the total slope uncertainty,
\begin{equation} \label{eq:sigma_R_stat_random_slope_correc}
\sigma_{R}^{\mbox{slope}} = \sigma_{R}\Big(1.0 + \varepsilon \delta_{\mbox{\mbox{slope}}}\Big), 
\end{equation}
and then repeating the fits described above. We then take the change in the form factors as the slope contribution
to the uncertainty. Note that because this is applied as a linear correction proportional to $\varepsilon$, there is no impact on the extracted value of $\gmp$, even for the right arm where the linear fit is extrapolated from $\varepsilon > 0.9$ down to $\varepsilon=0$. The scale uncertainty, $\delta_{\mbox{scale}}$, yields a simple renormalization of all of the cross sections, shifting $\gep$ and $\gmp$ but having no impact on their ratio. Tables~\ref{formfactor_results} gives the values and uncertainties for $\gep$, $\gmp$, and their ratio.

\begin{table}[!htbp]
\begin{center}
\caption{The extracted values of $\gep$, $\gmp$, and $R=\gegm$, and their associated uncertainties. No TPE corrections have been applied.}
\begin{tabular}{c c c c c c}
\hline \hline
$Q^2$ & $\gep$ & $\delta \gep^{\mbox{rand.}}$ & $\delta \gep^{\mbox{slope}}$ & $\delta \gep^{\mbox{scale}}$ & $\gep / \gd$ \\
\hline
0.50  & 0.29060 & 0.03589 & 0.02680 & 0.003182 & 0.844$\pm$0.1304 \\
2.64  & 0.04385 & 0.00128 & 0.00091 & 0.000639 & 0.976$\pm$0.0377 \\
3.20  & 0.03436 & 0.00138 & 0.00075 & 0.000496 & 1.042$\pm$0.0499 \\
4.10  & 0.02666 & 0.00149 & 0.00077 & 0.000378 & 1.224$\pm$0.0789 \\
\hline \hline
      &$\gmp$&$\delta \gmp^{\mbox{rand.}}$&$\delta \gmp^{\mbox{slope}}$&$\delta \gmp^{\mbox{scale}}$ & $\gmp / (\mu_p \gd)$\\
\hline
0.50 &1.0334 &0.06726  & \textbf{0.0125} &0.01702 & 1.0747$\pm$0.0733 \\
2.64 &0.1325 &0.000347 & 0	&0.00191 & 1.0563$\pm$0.0155 \\
3.20 &0.0969 &0.000320 & 0	&0.00140 & 1.0523$\pm$0.0156 \\
4.10 &0.0629 &0.000282 & 0	&0.00091 & 1.0339$\pm$0.0157 \\
\hline \hline
      & $R$ & $\delta R^{\mbox{rand.}}$ & $\delta R^{\mbox{slope}}$ & $\delta R^{\mbox{scale}}$ & $\mugegm$ \\
\hline
0.50 &0.2803 &0.0530  &0.0304 & 0 & 0.7828$\pm$0.1707 \\
2.64 &0.3308 &0.0104  &0.0069 & 0 & 0.9240$\pm$0.0349 \\
3.20 &0.3545 &0.0153  &0.0078 & 0 & 0.9900$\pm$0.0479 \\
4.10 &0.4238 &0.0254  &0.0122 & 0 & 1.1837$\pm$0.0787 \\
\hline \hline
\end{tabular}
\label{formfactor_results}
\end{center}
\end{table}

\begin{figure}[!htbp]
\begin{center}
\includegraphics*[width=0.95\columnwidth, trim={0mm 5mm 0 3mm}, clip]{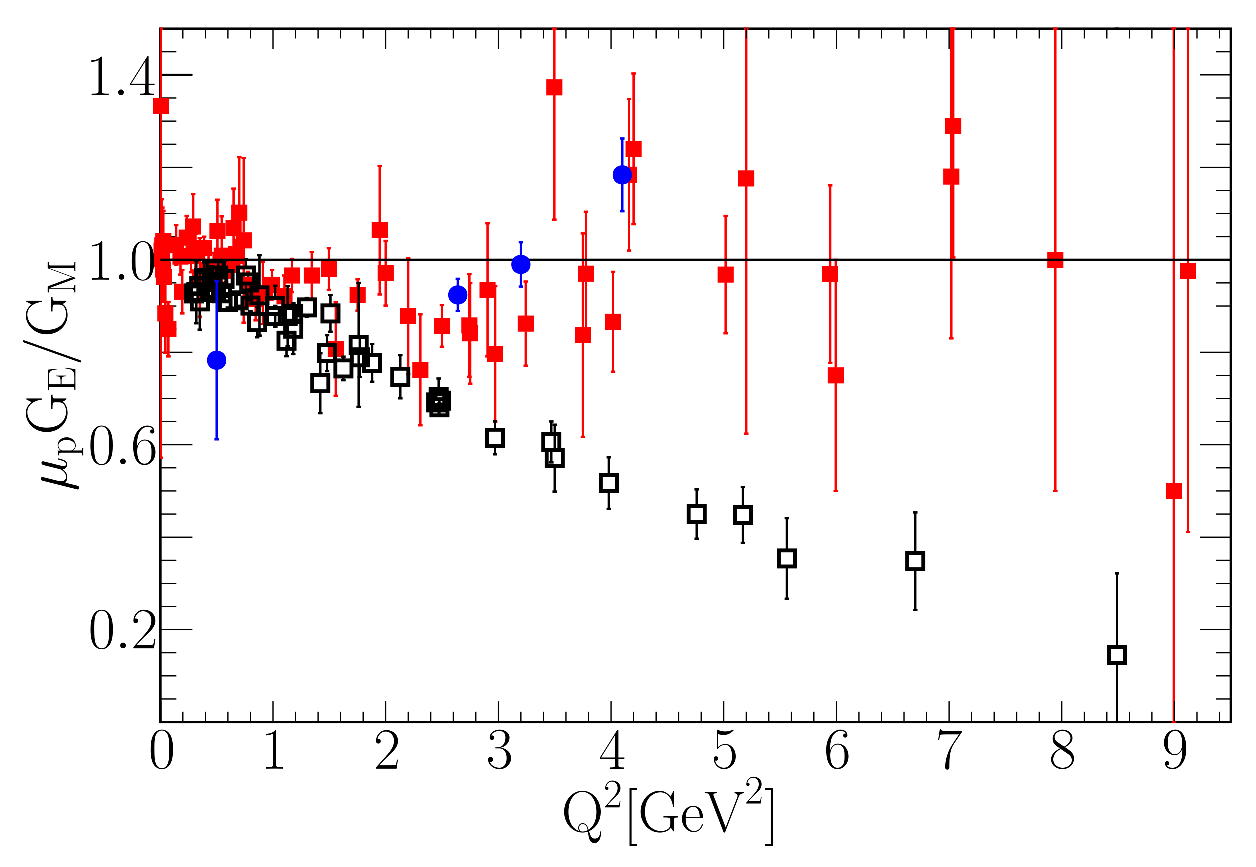}
\end{center}
\caption{Extractions of $\mugegm$ from Rosenbluth separations~\cite{arrington04a, christy04, Christy:2021snt} (solid red squares) and polarization measurements~\cite{gayou01, punjabi05, puckett12, strauch02, jones06, maclachlan06, zhan11, paolone10, Puckett:2017flj} (hollow black squares) at large $Q^2$ compared to the updated E01-001 extraction (solid blue circles).}
\label{fig:gep_gmp_all}
\end{figure}

Figure~\ref{fig:gep_gmp_all} shows extractions of $\mugegm$ from previous Rosenbluth extractions~\cite{arrington03a, christy04} (solid red squares), high-$Q^2$ polarization extractions~\cite{punjabi05, puckett12, puckett10} (hollow squares) and the E01-001 extractions (solid blue circles). We see that the E01-001 results are in good agreement with previous Rosenbluth extractions, with uncertainties that are comparable to the polarization measurements. These data clearly confirm the discrepancy between high-$Q^2$ Rosenbluth and recoil polarization measurements and provide an improved quantitative measure of this discrepancy. This supports the hypothesis that the discrepancy is not simply an experimental error in the Rosenbluth measurements or technique, as our measurement has different and significantly smaller experimental systematics as well as very different radiative corrections. This is consistent with the hypothesis that TPE corrections explain the discrepancy, as TPE corrections modify the e-p cross section without modifying the kinematics, and so yield identical corrections for electron-detection and proton-detection measurements.

The following sections examine the size of TPE contributions needed to resolve this discrepancy, assuming that TPE explains the entire discrepancy, use the data to set limits on the component of TPE that is nonlinear in $\varepsilon$, and examine the consistency of electron and proton detection measurements as a test of the conventional radiative correction procedures applied.

\subsection{Two-Photon-Exchange (TPE) and Possible Sources for the Discrepancy} \label{two_photon_exchange_coulomb}

The results of the E01-001 experiment confirmed the discrepancy between the Rosenbluth separations and recoil polarization results and provided a much improved precision in the comparison of the two techniques. Important questions to address are why the two techniques disagree and which form factors are correct? As mentioned previously, the significantly reduced $\varepsilon$-dependent corrections of this measurement allow us to rule out several possible experimental errors as the source of the discrepancy. We find that our high-precision Super-Rosenbluth extractions are consistent with conventional electron detection measurements, and an earlier reanalysis of form factor data~\cite{arrington03a} confirmed that previous Rosenbluth separations were consistent when excluding results which combined data from multiple experiments. That reanalysis, combined with the results from E01-001, strongly argues against the hypothesis that the discrepancy is associated with one or two bad data sets or incorrect normalization factors in the combined Rosenbluth analyses. Similarly, the recoil polarization measurements have small systematic uncertainties and different measurements taken under different conditions yield consistent results. This suggests that the explanation for the discrepancy is most likely a common correction, such as TPE contributions.

If TPE contributions are small (of order $\alpha$), then they would be expected to impact the observables at the few percent level. This would have a small impact on the polarization extractions of $\gegm$, as well as a small impact on the extraction of $\gmp$ from cross-section measurements. But a few percent correction with a linear $\varepsilon$ dependence could modify the slope significantly at large $Q^2$, where the slope due to $\gep^2$ is very small. A combination of polarization extractions of $\gegm$ and Rosenbluth extractions of $\gmp$ would have smaller corrections, but a precise extraction of the form factors requires an improved understanding of the TPE contribution to ensure that TPE contributions do not limit the extraction of the form factors~\cite{arrington07c}.

Finally, while the form factors extracted by Rosenbluth separations may not represent the true form factors for the proton due to TPE contributions, they do indeed provide a reliable parameterization of the elastic e-p cross sections, where the impact of TPE is absorbed into the effective $\gep$ and $\gmp$ values. Thus, the use of TPE-uncorrected form factors in evaluating the elastic e-p cross section should not introduce significant errors to measurements which rely on the e-p cross section, e.g. extraction of neutron form factors from light nuclei~\cite{CLAS:2014xso,Santiesteban:2023rsh} or in accounting for the e-p cross section in proton knockout measurements~\cite{dutta00, Garrow:2001di, dutta03}, as discussed in Ref.~\cite{arrington04a}. If the Born form factors are used, it is important to apply consistent TPE contributions that match those removed in the analysis of the form factors from elastic scattering data.

\subsubsection{Recent Theoretical and Phenomenological Two-Photon-Exchange Studies} \label{sec:TPE_theory_Phenomen}

Before addressing what we learn from the E01-001 results, we will provide a brief summary of theory and phenomenology associated with two-photon exchange effects. In some cases, the comparison of Rosenbluth and Super-Rosenbluth data to polarization measurements can provide model-independent constraints on TPE contributions. In other cases, more detailed information can be extracted using a specific model or formalism for the $\varepsilon$ dependence.

The interference of the OPE and TPE amplitudes represents the leading TPE correction to the elastic e-p scattering cross section. A simple way to account for the TPE contribution to $\sigma_{R}$ is to add the real function $F(Q^2,\varepsilon)$ to the Born reduced cross section:
\begin{equation} \label{eq:reduced3}
\sigma_{R} = \gmp^2\Big(1 + \frac{\varepsilon}{\tau} R^2\Big) + F(Q^2,\varepsilon),
\end{equation}
where a fit to the polarization transfer data, assumed to be insensitive to TPE, is used for $R = \gegm$. This is a commonly used approach, although the full picture is more complicated, with a full modeling of the electron-proton scattering including three complex form factors, rather than two real form factors~\cite{guichon03}. We provide a basic overview of some of the formalisms for estimating TPE contributions below. Additional details can be found in reviews of the form factor extractions and TPE corrections~\cite{punjabi05, arrington07a, carlson07, arrington11b, arrington11a, yang13, Afanasev:2017gsk}.

Several calculations in the 1950s and 1960s tried to estimate the size of the TPE contributions to the unpolarized elastic e-p cross sections~\cite{drell57, drell59, werthamer61, greenhut69, lewis56}. While some calculations used only the proton intermediate state~\cite{lewis56}, others included the excited intermediate states of the proton~\cite{drell57,drell59,werthamer61,greenhut69}. The TPE corrections estimated from these calculations were extremely small ($\leq$ 1\%) and were not included in the standard radiative correction procedures. However, many of these focused on the low $Q^2$ and large $\varepsilon$ regions where the cross sections were larger, but where TPE contributions are generally smaller.

Section~\ref{sec:earlytpe} summarized two works that examined TPE in the context of the initial observations of the Rosenbluth-Polarization discrepancy~\cite{jones00, gayou01, gayou02}. The first provided a more complete formalism for scattering beyond the OPE approximation~\cite{guichon03}, and the other presented calculations of the TPE contribution within a hadronic framework~\cite{blunden03}. We provide additional information on these and other approaches below, as these can be useful in interpreting the Super-Rosenbluth results. 

The initial hadronic calculation~\cite{blunden03} included only the unexcited proton in the intermediate state, yielding a small (2\%) linear $\varepsilon$ dependence for $Q^2 \gtorder 2$~GeV$^2$ and minimal non-linear contributions.  Inclusion of an improved form factor parameterization enlarged the corrections~\cite{blunden05a}, and the discrepancy was largely resolved up to $Q^2$ = (2--3)~GeV$^2$ but not for $Q^2 >$ 3~GeV$^2$. The effect of including the $\Delta$ resonance as an intermediate excited state to the elastic box and crossed-box calculations was also investigated~\cite{kondratyuk05} and found to be significantly smaller in magnitude than the proton intermediate state. Related studies examined the model dependence and/or inclusion of additional intermediate states~\cite{borisyuk06a, kondratyuk07, Borisyuk:2013hja, Ahmed:2020uso}. In general, these works find that the proton intermediate state dominates up to $Q^2=5$-6~GeV$^2$, and that the deviations from linearity in $\varepsilon$ are also small except at larger $Q^2$ values~\cite{arrington11b}. 

Chen \etal~\cite{chen04} evaluated a high energy model at the quark-parton level which calculates the TPE correction using generalized parton distributions (GPDs) for the quark distribution to describe the emission and re-absorption of the partons by the nucleon. At large $Q^2$, their calculations showed a significant $\varepsilon$ dependence with nonlinearity to the TPE correction and weak $Q^2$ dependence. Their TPE correction resolves roughly half of the observed discrepancy at large $Q^2$ values. However, the calculation is not expected to work at low $Q^2$ values, and the TPE contribution is very sensitive to the choice of GPD parameterization~\cite{afanasev05a}.

Other authors used different approaches to make predictions for TPE contributions. Calculations based on dispersion relations~\cite{borisyuk08, Tomalak:2014sva, Tomalak:2016vbf, Tomalak:2017shs} were used to estimate TPE contributions including a range of different intermediate states. Predictions based on perturbative QCD were also made to examine the behavior of the corrections at large $Q^2$ values~\cite{borisyuk09}. Discussions of the various approaches, as well as comparisons of the predictions, can be found in some of the more recent TPE review articles~\cite{arrington11b, Afanasev:2017gsk, Afanasev:2023gev}.

\subsection{Extracting TPE-corrected form factors}

As noted in the previous section, we can use the difference between the Rosenbluth (or Super-Rosenbluth) data and the polarization measurements to constrain the TPE contributions. Earlier analyses following this approach~\cite{guichon03, arrington03a, arrington04a} suggested that the difference in the $\mugegm$ ratio from Rosenbluth separations and recoil polarization results could be explained by a common (5--8)\% $\varepsilon$-dependent correction to the cross section, allowing for an extraction of the TPE-corrected form factors. Given the limited precision of the Rosenbluth data and the limited $Q^2$ coverage of polarization measurements, it was difficult to make a precision extraction of the size or the $Q^2$ dependence of the TPE corrections.

\begin{figure}[!htbp]
\begin{center}
\includegraphics*[width=0.95\columnwidth,trim={0mm 2mm 0mm 1mm},clip]{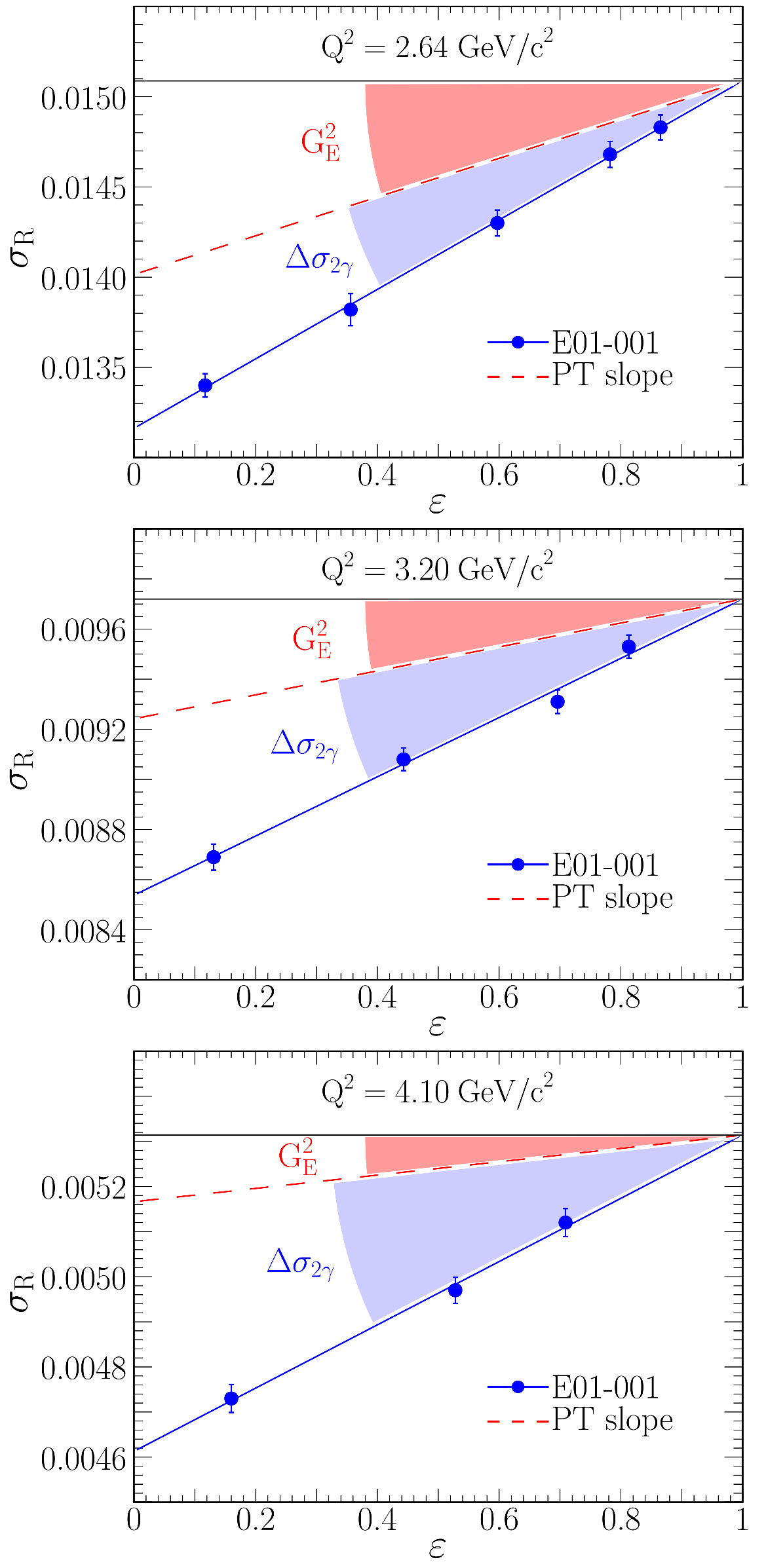}
\end{center}
\caption{(Color online) The $\varepsilon$ dependence of the reduced cross section at $Q^2$ = 2.64 (top), 3.20 (middle) and 4.10~GeV$^2$ (bottom) as measured in the E01-001 experiment (blue solid circles), a linear fit to the cross sections (blue line), and the slope predicted by recoil polarization data (red dashed line). The recoil polarization  measurement constrains only the slope and has been normalized to match the Rosenbluth extraction at $\varepsilon=1$. Assuming the recoil polarization results represent the true contribution of $\gep$ (red shaded region), then TPE contribution yields the additional slope (blue shaded region).}
\label{fig:2.64_sigma_lt_pol_2gamma}
\end{figure}

We can perform a more detailed analysis if we make additional assumptions about the nature of the TPE corrections. Most such analyses assumed that the polarization extraction of $\mugegm$ is unaffected by TPE and that the TPE contributions to the unpolarized cross section are linear in $\varepsilon$ and constrained to be zero at $\varepsilon=1$ due to crossing symmetry~\cite{chen07, arrington11b}. This is illustrated in Figure~\ref{fig:2.64_sigma_lt_pol_2gamma} which separates the contribution of the TPE ($\Delta_{2\gamma}$) to the slope arising from $\gep$ for the E01-001 separations. In this analysis, the PT slope decreases rapidly with increasing $Q^2$, and almost all of the $\varepsilon$ dependence comes from TPE contributions for $Q^2>4$~GeV$^2$. While the assumptions in this extraction appear to be reasonable based on both TPE calculations and constraints from world data, e.g. constraints on non-linear contributions discussed in Sec.~\ref{sec:linearity}, such analyses have model dependence that can be challenging to quantify. We present below some global analysis using similar or modified assumptions to try and extract the TPE contributions from the data and, in some cases, propagate the uncertainty from these corrections to the extracted form factors. 

With the original E01-001 results~\cite{qattan05}, it was possible to better constrain the TPE contributions and use this information along with later Rosenbluth and polarization measurements to extract the proton form factors including estimated TPE corrections. Ref.~\cite{arrington07c} performed the first global analysis of the elastic e-p scattering data accounting for TPE corrections and an estimate of their uncertainties. The cross sections were corrected using the hadronic model of Ref.~\cite{blunden05a} for the nucleon intermediate state including improved form factors at the internal vertices, and then adding a small additional correction (with a 100\% uncertainty) to resolve the remaining discrepancy at larger $Q^2$ values. The $Q^2$ dependence used in this correction was guided by other calculations that go beyond the intermediate elastic state~\cite{kondratyuk07, afanasev05b}. A global fit was performed to the TPE-corrected cross sections and recoil polarization data, allowing  $\gep$ and $\gmp$ to be extracted up to $Q^2 \approx$ 6 (30)~GeV$^2$ for $\gep$ ($\gmp$). The TPE corrections to $\gep$ were significant for $Q^2 \gtorder 2$~GeV$^2$, bringing the ratio $\gegm$ from Rosenbluth and recoil polarization into agreement. The  TPE correction to $\gmp$ is smaller, on the few percent level at the larger $Q^2$ values. Similar analyses, including additional data sets and updated estimates of the form factor uncertainties were also performed in Refs.~\cite{venkat11, Ye:2017gyb}.

Other analyses used a parameterization of the TPE contribution and extracted both the TPE correction and the form factors from a global fit of the corrected cross section measurements and polarization extractions of $\mugegm$, typically assumed to be unaffected by TPE. 

Bernauer \etal~\cite{bernauer13} performed such an analysis, with an emphasis on the low-$Q^2$ high-precision measurements from Mainz. They apply radiative corrections according to Maximon and Tjon~\cite{maximon00}, rather than Mo and Tsai~\cite{tsai71}, and then apply the Feshbach correction to account for Coulomb distortion in the $Q^2=0$ limit~\cite{mckinley48}. They then fit $\gep$, $\gmp$, and an additional TPE contribution, $\delta_{TPE}=-(1-\varepsilon) a \ln{(b Q^2 +1)}$. From this, they extract a series of parameterizations of the TPE contribution based on different parameterizations of the proton form factors.

Alberico \etal~\cite{alberico09} performed a similar analysis, using somewhat different parameterizations of the TPE contribution and making different assumptions for the form factor parameterizations. Another analysis was performed by Qattan, \etal~\cite{Qattan:2012zf}, which combined high-$Q^2$ TPE-corrected proton form factors~\cite{qattan11a} with neutron form factor measurements to extract the up- and down-quark contributions to the charge and magnetic form factors. A later extension of this analysis was performed~\cite{Qattan:2015qxa}, making a simultaneous fit to the proton form factors and TPE correction, using the TPE parameterization of Ref.~\cite{borisyuk06b}, and including both low- and high-$Q^2$ data.

\subsection{Estimating TPE corrections}\label{sec:TPE_extractions}

The analyses discussed in the previous sections all required TPE corrections with an $\varepsilon$ dependence of several percent to resolve the discrepancy at large $Q^2$. However, the size of the high-$Q^2$ corrections varied by up to a factor of two between different extractions, and while they yielded smaller low-$Q^2$ corrections, even the sign of these corrections depended on the approach taken as illustrated in Refs.~\cite{Qattan:2012zf, Qattan:2015qxa}. These analyses were more focused on the extraction of the proton form factors, using calculations or parameterizations of the TPE contributions, and all took the simplified approach of applying a single TPE correction to the unpolarized cross section. Other analyses focused more on constraining the TPE contributions, or included additional observables to allow for model-dependent extractions of the different TPE amplitudes. We summarize these works below.

As described in Sec~\ref{sec:TPE_theory_Phenomen}, a complete extraction of elastic electron-proton scattering requires three complex form factors. In their initial examination of the discrepancy, Guichon and Vanderhaeghen~\cite{guichon03} estimated the TPE contributions for all three form factors and accounted for their impact on both polarization and cross-section measurements. But estimates based on only the cross section and polarization require significant assumptions, as the discrepancy can only constrain one of the three TPE amplitudes.

Ref.~\cite{arrington05} used the same formalism for the form factors~\cite{guichon03}, but a different set of assumptions about the impact of each of the TPE amplitudes. This analysis assumed that all of the beyond-Born contributions were of comparable size (order $\alpha$) and were $\varepsilon$ independent. Under these assumptions, the TPE contribution to $\gep$ has a much smaller effect on the cross section and is ignored. The TPE contribution to $\gmp$ yields a correction that depends only on $Q^2$, and the third amplitude, $Y_{2\gamma}$ in Ref.~\cite{guichon03} is extracted at the $Q^2$ value of each Rosenbluth experiment based on the difference between Rosenbluth extractions of $\mugegm$ and a fit to the polarization extractions, accounting for experimental uncertainties in both types of measurements. The TPE contribution to $\gmp$ is then used to ensure that the total TPE contribution to the cross section is zero at $\varepsilon=1$, based on earlier comparisons of positron-proton and electron-proton scattering~\cite{arrington04b} and consistent with the constraints from crossing symmetry~\cite{chen07}. The extracted TPE amplitudes and their estimated uncertainties are then parametrized as a function of $Q^2$, and used to apply TPE corrections to the form factors obtained from a global Rosenbluth analysis~\cite{arrington04a} and the new recoil polarization data. 

Additional measurements, in particular the $\varepsilon$ dependence of the recoil polarization extraction of $\mugegm$~\cite{meziane11, Puckett:2017flj} and of the longitudinal component of the polarization, provide additional information that  allows for estimates of the TPE amplitudes with fewer assumptions and reduced model dependence. Two such analyses were performed~\cite{guttmann11,borisyuk11}, providing extractions of the three TPE amplitudes at $Q^2=2.5$~GeV$^2$. While these were able to constrain the amplitudes with far weaker assumptions than in Refs.~\cite{guichon03, arrington05}, some assumptions on the functional form of the amplitudes are required, and these analyses take a somewhat different approach in extracting the amplitudes. Later analyses extended these separations to cover a range of $Q^2$ values~\cite{Qattan:2017zwt, Qattan:2018epw}, guided by the data at 2.5~GeV$^2$, but making additional assumptions to estimate the $Q^2$ dependence. 

From the existing calculations and observed discrepancy, it appears that a TPE contribution of 5--8\%, roughly linear in $\varepsilon$ is required to resolve the discrepancy at high $Q^2$. This is relatively consistent with the approximate size and $\varepsilon$ and $Q^2$ dependence of several calculations, making TPE the consensus explanation for the discrepancy. It has also been demonstrated~\cite{arrington07c} that the extraction of the proton form factors is not dominated by TPE contributions, as long as the entire discrepancy is due to TPE and the corrections are sufficiently linear.

We note that most of these analyses were based on cross sections based on the Mo and Tsai prescription~\cite{mo69, tsai71}. Other works have looked at modified approaches or additional corrections unrelated to two-photon exchange~\cite{maximon00, Kuraev:2013dra, Gramolin:2016hjt, Afanasev:2020ejr, Afanasev:2023gev}, and using these prescriptions could modify the TPE contribution needed to resolve the discrepancy. A recent analysis included newer high-precision cross-section measurements at large $Q^2$ values~\cite{Christy:2021snt} and previous Rosenbluth measurements using the updated Maximon and Tjon prescription~\cite{maximon00}. This work extended the constraints on TPE to larger $Q^2$ values, and demonstrated that the Maximon and Tjon procedure reduced the discrepancy between Rosenbluth and polarization measurements by one third. However, the updated prescription is only available for electron detection~\cite{Stefan:2024inp}, so the Super-Rosenbluth measurements cannot be updated to see the impact of the different prescription. It may be that the discrepancy as observed with the Super-Rosenbluth data would have a similar reduction, although the $\varepsilon$ dependence is smaller for proton detection.


\subsection{Search for nonlinearities}\label{sec:linearity}

\begin{figure}[!htbp]
\begin{center}
\includegraphics*[width=0.95\columnwidth]{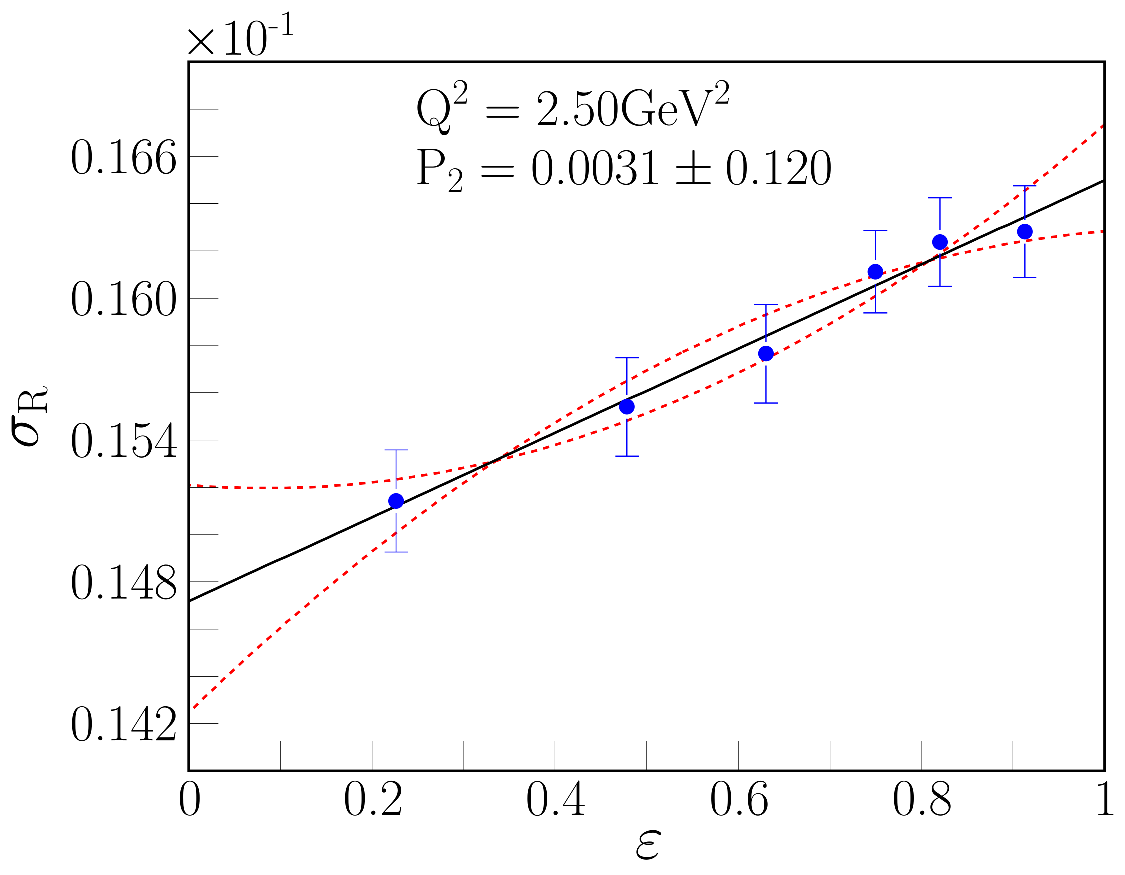} \\
\includegraphics*[width=0.95\columnwidth]{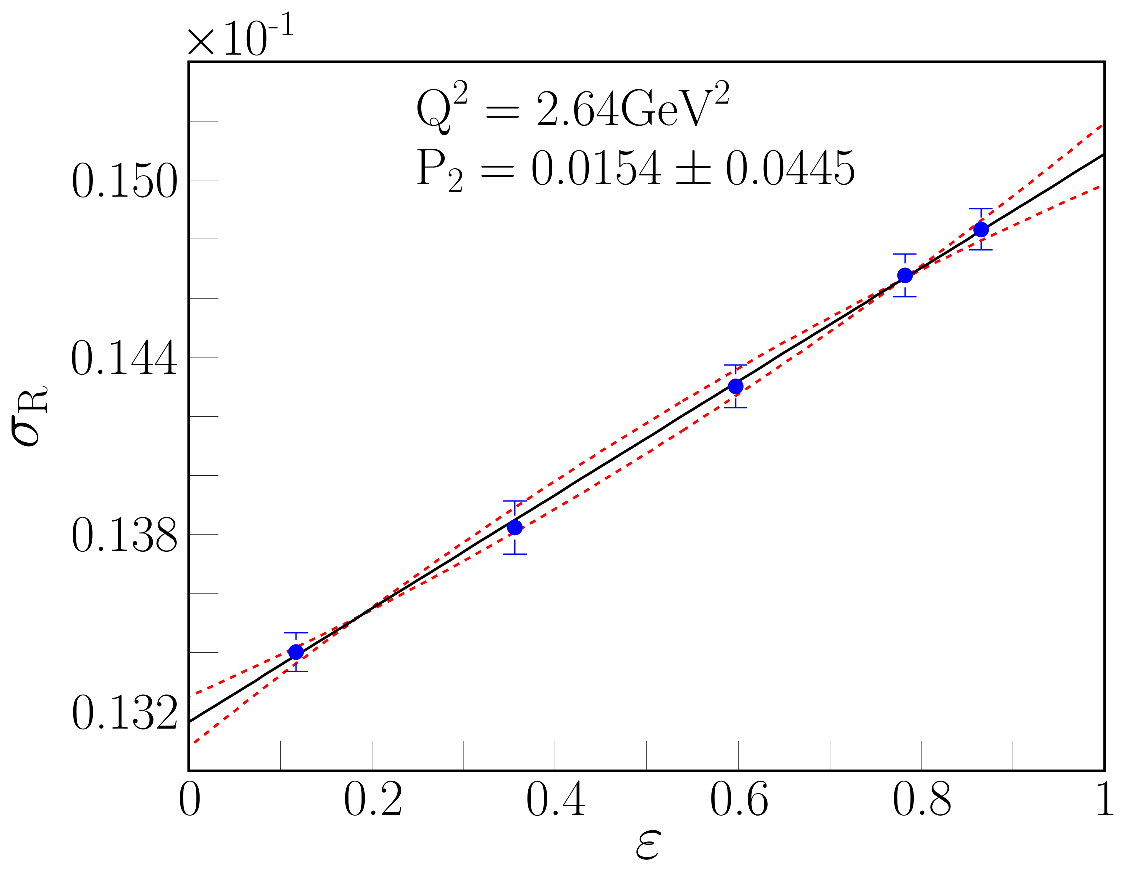}
\end{center}
\caption{(Color online) Nonlinearity constraints from SLAC NE11 at $Q^2$ = 2.50~GeV$^2$ (top) and E01-001 at $Q^2$ = 2.64~GeV$^2$. The solid black line is the linear fit, while the dashed red lines show the result of the quadratic fit (Eq.~\ref{linearity_test_equation} with $P_2$ increased (decreased) by 1$\sigma$ from the central $P_2$ value.}
\label{fig:lt_sigma_red_nonlinear}
\end{figure}

Due to the linearity of the reduced cross section with $\varepsilon$ in the Born approximation, any deviation from linearity would have to come from higher-order terms that are not included in standard radiative correction procedures. Observation of a nonlinearity would provide a clean signature of TPE and give information about the nonlinear component of TPE, assuming all other aspects of the radiative correction procedure are complete. To search for a deviation from linearity in the reduced cross section, we fit the measured cross sections to a second-order polynomial of the form
\begin{equation} \label{linearity_test_equation}
\sigma_{R} = P_0\Big(1 + P_1(\varepsilon-0.5) + P_2(\varepsilon-0.5)^2 \Big),
\end{equation}
where $P_2$ is the curvature parameter and provides a simple measure of the size of the nonlinear term relative to the cross section at $\varepsilon$ = 0.5. Figure~\ref{fig:lt_sigma_red_nonlinear} shows such fits done for the SLAC NE11~\cite{andivahis94} and the E01-001 experiments at $Q^2$ = 2.50 and 2.64~GeV$^2$, respectively. SLAC NE18 yields $P_2$=0.003$\pm$0.120, while the E01-001 yields $P_2$=0.015$\pm$0.045, providing a much better constraint on $P_2$. Moreover, as $\varepsilon \to 0$, the variation of $P_0$ between the linear and quadratic fits can help estimate the possible TPE contribution to $\delta (\tau G^2_{Mp})$, as seen in the spread of the red dotted curves as the reduced cross section is extrapolated to $\varepsilon$=0. Table~\ref{nonlinearity_values} shows the extracted curvature parameters and uncertainties, along with the estimated uncertainty on $G^2_{Mp}$ using this fit function for all three $Q^2$ values from E01-001. The initial E01-001 results~\cite{qattan05} were included in a global analysis of elastic e-p scattering Rosenbluth separations~\cite{tvaskis06} which concluded that such nonlinear effects are consistent with zero, with a global average value of $\langle P_2 \rangle$=0.019$\pm$0.027, and a 95\% confidence level upper limit  $|P_2|_{\mbox{max}}$ of 6.4\%.

\begin{table}[!htbp]
\begin{center}
\caption{The curvature parameter ($P_2$) and uncertainty ($\delta P_2$), and the uncertainty in
$\tau G^2_{Mp}$ as extracted for the SLAC NE11 data at $Q^2$ = 2.50~GeV$^2$
and for E01-001.}
\begin{tabular}{c c c c c}
\hline \hline
Parameter & NE11          & E01-001      & E01-001       &E01-001    \\
          &$Q^2$=2.50     & $Q^2$=2.64   & $Q^2$=3.20    &$Q^2$=4.10 \\
          &~GeV$^2$   & ~GeV$^2$ &~GeV$^2$   &~GeV$^2$               \\
\hline
$P_2$                     &0.003      &0.015        &0.013       &0.057          \\
$\delta P_2$            &$\pm$0.120 &$\pm$0.044   &$\pm$0.056  &$\pm$0.12       \\
$\delta (\tau G^2_{Mp})$ (\%) &3.27       &0.87         &1.10        &2.19             \\
\hline \hline
\end{tabular}
\label{nonlinearity_values}
\end{center}
\end{table}

For comparison, calculations in hadronic~\cite{blunden03} and partonic~\cite{chen04} frameworks give small nonlinear contributions at these $Q^2$ values, consistent with the limits we set, although the overall size of the TPE contributions in these calculations is insufficient to fully resolve the discrepancy. Chen~\etal~\cite{chen07} parametrized the TPE contributions based on crossing symmetry, yielding somewhat larger nonlinear contributions, although the predicted $P_2$ is particularly sensitive to the exact $\varepsilon$ range examined. Calculations of the curvature in partonic models using generalized parton distribution (GPDs) as input~\cite{abidin08} showed large sensitivity of the parameter $P_2$ to the type of GPDs used. Calculations based on modified Regge GPDs~\cite{abidin08} yielded negative values for $P_2$ and about 1.4$\sigma$ away from $\langle P_2 \rangle$ = 0.013$\pm$0.033 obtained for the points $Q^2$ = 2.64 and 3.20~GeV$^2$ from this work. 

\subsection{Test of Radiative Corrections} \label{rad_corrections}

In the Born approximation, $\ge$ is extracted from the slope of $\sigma_R$ vs $\varepsilon$. The slope becomes very small at large $Q^2$, making the $\varepsilon$ dependence of both conventional and higher order radiative corrections extremely important. For E01-001, we applied the radiative corrections procedure of Mo and Tsai~\cite{mo69, tsai71} as modified by Walker and Ent~\cite{walker94, walkerphd, ent01, makinsphd} and implemented in the elastic e-p simulation code SIMC. Note that Ref.~\cite{ent01} has errors in some of the equations, and SIMC uses the correct versions that appear in Ref.~\cite{makins94}. Full coincidence (e,e$'$p) simulations are performed using the prescription of Refs.~\cite{makinsphd,ent01} taking into account Bremsstrahlung from all three tails (Bremsstrahlung from the incident electron, scattered electron, and scattered proton).

Previous Rosenbluth measurements~\cite{walker94, andivahis94, christy04} estimated scale and random uncertainties in the radiative corrections of 1.0\% and 0.60\%, respectively. In these measurements, electrons rather than protons were detected. In our case, we assign the same scale uncertainty of 1.0\% for both arms and reduce the random uncertainty as quoted by the previous measurements to account for the reduced $\varepsilon$ dependence of the Bremsstrahlung correction, as shown in Fig.~\ref{fig:rad_corr_epsilon} and discussed below. For the left arm, we apply a slope contribution of 0.30\% and a random contribution of 0.20\%, while for the right arm measurement at lower $Q^2$, we apply a 0.2\% slope uncertainty and a 0.2\% random uncertainty.

The main $\varepsilon$ dependence in the radiative corrections comes from the internal and external Bremsstrahlung corrections. Bremsstrahlung yields significantly different corrections for electron and proton detection. For example, at $Q^2$ = 2.64~GeV$^2$, the $\varepsilon$ dependence over the range of the E01-001 measurement is roughly -8\% for proton detection, but +17\% for electron detection~\cite{afanasev01} as shown in Fig.~\ref{fig:rad_corr_epsilon}. Thus, one expects a 25\% difference between the slopes of the electron and proton detection cross sections before applying radiative corrections. The fact that the slopes are consistent after the corrections provides a unique test of the radiative correction procedures. Given that the slopes after radiative corrections agree at the 2-3\% level, we can estimate that the conventional Bremsstrahlung correction on the slope is good at the $\ltorder$10\% level. Assuming this 10\% applies to both electron and proton detection, this would yield a 0.8\% uncertainty on the $\varepsilon$ dependence for the case of proton detection. This constraint is not sufficient to reduce the radiative correction uncertainties applied based on conventional estimates of the uncertainties. However, it is a new consistency check on the Bremsstrahlung corrections and does provide a meaningful limit on potential $\varepsilon$-dependent errors associated with the assumptions used in evaluating conventional radiative corrections.

\section{Recent Experimental Two-Photon-Exchange Studies} \label{TPE_experimental}

After the E01-001 experiment confirmed and better quantified the discrepancy between the Rosenbluth separation and recoil polarization results, several experiments were proposed and/or carried out aimed at measuring the size of the TPE corrections at modest-to-large $Q^2$ values, both at Jefferson Lab and worldwide. These included measurements of the $\varepsilon$ dependence of the ratio $R = \gegm$ in polarization observables~\cite{meziane11}, searching for nonlinearities in $\sigma_R$ vs $\varepsilon$~\cite{e05017}, and measurements of the ratio, $R_{e^+e^-}$, of positron-proton to electron-proton elastic-scattering cross sections~\cite{e07005, vepp_proposal, kohl09, nikolenko10a, nikolenko10b, kohl09, e12+23-008, e12+23-012}.

JLab experiment E05-017~\cite{e05017} carried out a high-precision Rosenbluth separation measurement similar to the Super-Rosenbluth E01-001 experiment reported in this work. The experiment ran in Hall C at Jefferson Lab and detected elastic protons over a wider $\varepsilon$ and $Q^2$ range. Such measurements will extend  precise extraction of TPE effects from the difference between Rosenbluth and recoil polarization measurements to larger $Q^2$. For two $Q^2$ values, many $\varepsilon$ values were measured to provide improved constraints on nonlinearities in the $\varepsilon$ dependence of the cross sections. 

The GEp--2$\gamma$ collaboration~\cite{meziane11} searched for effects beyond the OPE by measuring the $\varepsilon$ dependence of the ratio $R = \gegm$ and the longitudinal polarization transfer component $P_l$ in the elastic ($\overrightarrow{e},e'\overrightarrow{p}$) reaction for three different beam energies of 1.87, 2.84, and 3.63~GeV at a fixed $Q^2$ value of 2.5~GeV$^2$. The experiment was carried out in Hall C at Jefferson Lab, where a longitudinally polarized electron beam was scattered elastically off a 20-cm liquid hydrogen target. The measured ratio $R$ was found to be essentially independent of $\varepsilon$ at the 1.5\% level, suggesting that the TPE amplitudes are either small or cancel in the ratio. On the other hand, the ratio $P_l/P^{Born}_l$ showed an enhancement at large $\varepsilon$ at the (1.4$\pm$0.8)\% level~\cite{Puckett:2017flj}.  These data were used to constrain the TPE amplitudes with a reduced set of assumptions about the $\varepsilon$ dependence of the amplitudes~\cite{borisyuk11, guttmann11, Qattan:2017zwt, Qattan:2018epw} as discussed in Section~\ref{sec:TPE_extractions}

The effects of TPE corrections have the opposite sign for electrons and positrons, i.e. $\sigma(e^{\pm}) = \sigma_{Born}(1 \mp \delta_{2\gamma})$, where $\delta_{2\gamma}$ is the TPE correction yielding a charge asymmetry $R_{e^+e^-} \approx 1 - 2\delta_{2\gamma}$, so any deviation of $R$ from unity is a model-independent indication of TPE in elastic e-p scattering. Several such measurements were made before 1980 (See~\cite{arrington04b} and references therein), but were generally limited to low $Q^2$ and/or small angle (large $\varepsilon$ values), where the TPE contributions are small. Three modern experiments have made such measurements, expanding the $Q^2$ and $\varepsilon$ coverage into the range of interest to explain the form factor discrepancy.

The first is the CLAS Collaboration E07-005 experiment~\cite{e07005}, which used the CLAS detector at Jefferson Lab to make novel measurements of $R_{e^+e^-}$~\cite{moteabbed13, CLAS:2014xso, CLAS:2016fvy}. They measured the $Q^2$ and $\varepsilon$ dependencies of the charge asymmetry using a mixed beam of $e^{+}$ and $e^{-}$ produced via pair production from a secondary photon beam. Detection of both the struck proton and the scattered lepton was used to separate and simultaneously measure $\sigma(e^{+}p \rightarrow e^{+}p)$ and $\sigma(e^{-}p \rightarrow e^{-}p)$ elastic scattering cross sections. Cross-section ratios were extracted as a function of $\varepsilon$ for $Q^2=0.85$ and 1.45~GeV$^2$, and the $Q^2$ dependence for $\varepsilon=0.45$ and 0.88. The results showed a systematic increase in $R_{e^+e^-}$ at the largest $Q^2$ and favored hadronic TPE calculations, but the precision was insufficient to strongly exclude the no-TPE hypothesis.

The second is the VEPP-3 experiment~\cite{vepp_proposal, nikolenko10b, gramolin12, Rachek:2014fam, Nikolenko:2015xsa}. The internal target at the VEPP-3 electron-positron storage ring at Novosibirsk was used to extract the ratio $R_{e^+ e^-}$ using beams of 1.0 and 1.6 GeV. Small-angle detectors were used to normalize the relative luminosity by requiring $R_{e^+ e^-}=1$ at the largest $\varepsilon$ (and lowest $Q^2$) measurement at each energy. Measurements were made at $Q^2=1.60$, 1.0, and 0.8~GeV$^2$, with $\varepsilon \approx 0.4$, 0.25, and 0.4, respectively. In all cases, the lower-$\varepsilon$ values of $R_{e^+e^-}$ were 2-3$\sigma$ above unity.

The third is the OLYMPUS experiment~\cite{kohl09}, where the DORIS lepton storage ring at DESY was used to extract the ratio $R_{e^{+} e^{-}}$ at a fixed beam energy of 2~GeV for scattering angles from 25-75 degrees. This corresponds to $\varepsilon$ values from 0.45-0.9, $Q^2$ from 0.6-2.0~GeV$^2$, with the largest $Q^2$ values corresponding to the lowest $\varepsilon$ values. The OLYMPUS results also showed a systematic $\varepsilon$ dependence, but had limited statistics in the high-$Q^2$, low-$\varepsilon$ region where TPE contributions are believed to be most important.

Reference~\cite{Afanasev:2017gsk} provides a detailed summary of all three of these experiments and their results, as well as a combined analysis of the data. The measurements are consistent with small TPE contributions at low $Q^2$ and large $\varepsilon$, as seen in nearly all TPE calculations~\cite{arrington11b}, with a ratio that is larger for lower $\varepsilon$ values. For different treatments of the normalization uncertainties, they find that the no-TPE hypothesis is excluded at the 98\% or 99.5\% confidence level. 

Finally, two experiments have been proposed to make additional measurements using positron and electron beams at JLab. One of these makes direct measurements of the positron-proton and electron-proton cross sections in CLAS12~\cite{e12+23-008} over a wide kinematic range, while the other will perform Super-Rosenbluth extractions similar to this work and E05-017~\cite{e05017} with electrons and, separately, with positrons~\cite{e12+23-012, Arrington:2021kdp}. While JLab does not currently provide positron beams, there is an ongoing effort to develop positron beams~\cite{Afanasev:2019xmr, Accardi:2020swt, Arrington:2021alx} for a range of measurements, including direct and dramatically expanded TPE studies.

\section{Conclusions}

High precision measurements of the elastic e-p scattering cross sections were made at $Q^2$ = 0.50, 2.64, 3.20, and 4.10~GeV$^2$ at Hall A of the Thomas Jefferson National Accelerator
Facility. Protons were detected, in contrast to previous measurements where the scattered electrons were detected, to significantly decrease any $\varepsilon$-dependent corrections and systematic uncertainties.  A single spectrometer, HRS-L, measured the scattered protons of interest at $Q^2$ = 2.64, 3.20, and 4.10~GeV$^2$, while simultaneous measurements at $Q^2$ = 0.5~GeV$^2$ were carried out using the HRS-R. For all of the right arm measurements, $\varepsilon$ was above 0.9, and so the expected $\varepsilon$ dependence over that small range was precisely known based on previous Rosenbluth measurements. This allowed the HRS-R data to be used as a luminosity monitor, checking the corrections and uncertainties associated with the beam current measurements and target density fluctuation corrections. While the absolute uncertainty on the cross sections is at the 3\% level, the relative uncertainties which go into the determination of $\gegm$ are below 1\%.

The results of this work are in agreement with the previous Rosenbluth data and are inconsistent with the high-$Q^2$ recoil polarization results. The E01-001 experiment provided systematic uncertainties much smaller than the best previous Rosenbluth measurements~\cite{andivahis94,christy04}, and comparable to those of the recoil polarization, clearly establishing the discrepancy between the Rosenbluth separations and recoil polarization results. Furthermore, the high precision of the results confirmed that the discrepancy is not an experimental error in the Rosenbluth measurements or technique, confirmed the reliability of the elastic e-p scattering cross sections extracted from previous Rosenbluth separations, provided a unique test of the conventional radiative corrections used, and constituted a precise measurement of the discrepancy.


\begin{acknowledgments}

This work was supported by the U. S. Department of Energy, Office of Nuclear Physics, under contracts DE-AC02-05CH11231 and DE-AC02-06CH11357 and contract DE-AC05-06OR23177 under which Jefferson Science Associates, LLC operates the Thomas Jefferson National Accelerator Facility.

\textit{Data availability.}
The data that support the findings of this article are not publicly
available.  The data are available from the authors upon reasonable
request.

\end{acknowledgments}

\bibliographystyle{apsrev4-2}
\bibliography{longpaper_e01001}

\end{document}